\documentclass[preprint,prd,aps,showpacs,showkeys,nofootinbib]{revtex4}
\begin{document}


\title{Residue theorem and summing over Kaluza-Klein excitations}

\author{Tai-Fu Feng$^{a,b}$\footnote{email:fengtf@dlut.edu.cn}, Jian-Bin Chen$^a$, Tie-Jun Gao$^a$,
Ke-Sheng Sun$^a$}

\affiliation{$^a$Department of Physics, Dalian University of Technology,
Dalian, 116024, China\\
$^b$Center for High Energy Physics, Peking University, Beijing 100871, China}

\begin{abstract}
Applying the equations of motion together with corresponding boundary conditions of bulk profiles
at infrared and ultraviolet branes, we verify some lemmas on the
eigenvalues of Kaluze-Klein modes in framework of warped extra dimension with the custodial symmetry
$SU(3)_c\times SU(2)_L\times SU(2)_R\times U(1)_X\times P_{LR}$. Using the lemmas and
performing properly analytic extensions of bulk profiles, we present
the sufficient condition for a convergent series of Kaluze-Klein excitations
and sum over the series through the residue theorem. The method can also be
applied to sum over the infinite series of Kaluze-Klein excitations
in unified extra dimension. Additional, we analyze the possible connection
between the propagators in five dimensional full theory and the product of
bulk profiles with corresponding propagators of exciting Kaluze-Klein modes in
four dimensional effective theory, and recover some relations presented in literature for
warped and unified extra dimensions respectively. As an example, we demonstrate
that the corrections from neutral Higgs to the Wilson coefficients of relevant
operators for $B\rightarrow X_s\gamma$ contain the suppression factor
$m_b^3m_s/m_{_{\rm w}}^4$ comparing with that from other sectors, thus can be neglected safely.
\end{abstract}

\keywords{residue theorem, extra dimension, Kaluza-Klein mode}
\pacs{11.10Kk, 04.50.-h}

\maketitle

\section{Introduction\label{sec1}}
\indent\indent
Extensions of the standard model (SM) with a warped dimension\cite{RS,Chang,Csaki,
Gherghetta1}, where all SM fields are propagating in the bulk, provide a naturally
geometrical solution to the hierarchy problem regarding the huge difference between
the Planck scale and the electroweak one. The small mixing between zero modes
and heavy Kaluza-Klein (KK) excitations can induce the observed fermion masses
and corresponding weak mixing angles\cite{Grossman,Gherghetta2}, and suppress
flavor changing neutral current (FCNC) couplings\cite{Huber1,Agashe1}. Additional,
realistic models of electroweak symmetry breaking in warped extra dimension are constructed
in literature\cite{Agashe2,Csaki2,Agashe3,Csaki3,Contino,Carena}, and the gauge coupling
unification with a warped extra dimension is also analyzed in Ref.\cite{Agashe4,Agashe5}.

If the SM gauge group $SU(2)_L\times U(1)_Y$ is chosen in the bulk for extensions
of the SM with a warped extra dimension, the electroweak precision observables, for example
the experimental data on the $S,\;T$ parameters and the well-measured
$Z\bar{b}_Lb_L$ coupling\cite{Agashe6,Casagrande,Bauer1,Bauer2}, generally require
that the exciting KK modes are heavier than 10 TeV and exceed the reach of
colliders running now. In order to accomodate light exciting KK modes with
${\cal O}(1{\rm TeV})$ masses in warped extra dimension,
literature\cite{Agashe7,Santiago1} enlarges the gauge group in the bulk to
$SU(3)_c\times SU(2)_L\times SU(2)_R\times U(1)_X\times P_{LR}$. With an appropriate
choice of quark bulk masses, one indeed obtains the agreement with
the electroweak precision data in the presence of light KK excitations\cite{Djouadi,Bouchart}.
Actually, the electroweak precision observables are consistent with the light
fermion KK modes with masses even below $1{\rm Tev}$ while the masses of the KK
gauge bosons are forced to be at least $(2-3){\rm TeV}$ to be consistent with
the data on the parameter $S$.

However, the large FCNC transitions are aroused by light exciting KK modes if we assume
that the hierarchy of fermion masses together with corresponding weak mixings solely originate
from geometry and the fundamental Yukawa couplings in five dimension are anarchic\cite{Csaki4,
Buras1,Agashe8,Davidson,Iltan,Agashe9}. To suppress the large FCNC
processes mediated by light exciting KK modes, literature\cite{Cacciapaglia} introduces
the bulk and brane flavor symmetries where the naturally geometric explanation
of fermion hierarchies is abandoned. The author of Ref.\cite{Santiago2} proposes
the minimal flavor protection where a global $U(3)$ bulk flavor symmetry is imposed
on the triplet reprsentations of the local gauge symmetry $SU(2)_L\times SU(2)_R$
in the quark sector. In a similar way, literature\cite{Chen1} chooses the global flavor
symmetry $U(3)_L\times U(3)_R$ on lepton sector to control relevant
FCNC transitions. Another approach to suppress FCNC processes in warped
extra dimension introduces two horizontal $U(1)$ symmetries which guarantee
bulk masses in alignment with Yukawa couplings for charge $-1/3$ quarks and charge
$-1$ leptons respectively\cite{Csaki5}.
An analogous strategy to solve the FCNC problems in warped geometry is based on
$A_4$ flavor symmetry\cite{Csaki6,Kadosh}. A very different approach has been
presented in Ref.\cite{Perez1,Perez2} where the bulk mass matrices are expressed
in terms of five dimensional Yukawa couplings, thus flavor violation at low energy can be suppressed
rationally. Comparing with the choices mentioned above,
the warped extra dimension with a soft-wall\cite{Falkowski1,Gherghetta3,
Delgado,Santiago3,Gherghetta4,Cabrer1} perhaps provides a natural solution
to accomodate FCNC transitions at acceptable level and the lightest KK excitations
with ${\cal O}(1{\rm TeV})$ masses simultaneously\cite{Huber2}.

In a warped extra dimension with the custodial symmetry
$SU(3)_c\times SU(2)_L\times SU(2)_R\times U(1)_X\times P_{LR}$, a meticulous
analysis on the electroweak and flavor structure is provided in Ref.\cite{Buras2},
a complete study of rare $K$ and $B$ meson decays is presented in Ref.\cite{Buras3},
the impact from KK excitations of fermions on the couplings among
the SM particles is given in Ref.\cite{Buras4}. Assuming all fields are
propagating in the bulk, the authors of literature\cite{Azatov} analyze the FCNC processes
mediated by a light Higgs, the authors of literature\cite{Agashe10}
present an analysis on production and decay of KK gravitons at the LHC,
and the authors of literature\cite{Agashe11} study signals for KK
excitations of electroweak and strong gauge bosons in the LHC. Additional, an analysis of
loop-induced rare lepton FCNC transition $\mu\rightarrow e\gamma$ in a warped extra
dimension with anarchic Yukawa couplings and IR brane localized Higgs is presented
in literature\cite{Csaki7} through the five dimensional mixed position/momentum space
formalism\cite{Carena1} and mass insertion approach.

It is well known that all virtual KK excitations contribute their corrections
to theoretical predictions on the physical quantities at electroweak scale,
and those theoretical corrections should be summed over infinite KK modes in principle\cite{Hirn,Azatov}.
In this work, we verify some lemmas on the eigenvalues of exciting KK modes in
extension of the SM with a warped extra dimension and the custodial symmetry
$SU(3)_c\times SU(2)_L\times SU(2)_R\times U(1)_X\times P_{LR}$\cite{Agashe1,Agashe12}.
Performing properly analytic extensions of the bulk profiles, we sum over the infinite series of
KK modes through the residue theorem. Additional, we also present the sufficient
condition for a convergent series of infinite KK modes in extensions of the SM
with a warped extra dimension. The method can also be applied to sum over
the infinite series of KK modes in unified extra dimension.
The emphasized point here is that the authors of literature\cite{Hirn} also propose
to sum over infinite series of KK modes applying the equations of motion and
corresponding completeness relations of bulk profiles. Nevertheless, the method
proposed there can only be applied to sum over the infinite series of KK modes
for the five dimensional fields with zero bulk mass and zero modes in extensions
of the SM with a warped extra dimension.

Our presentation is organized as follows. In section \ref{sec2}, the main ingredients
of a warped extra dimension with custodial symmetry are summarized briefly, the KK decompositions
of all five dimensional fields and relevant bulk profiles are given here also.
In section \ref{sec3}, we present the verification of relevant lemmas on the eigenvalues
of exciting KK modes in detail. In order to sum over the infinite series of KK modes properly, we also
discuss how to extend bulk profiles analytically. We discuss the possible relation
between the perturbative expansions in four dimensional effective theory and five
dimensional full theory in section \ref{sec4}. Furthermore, we also recover the equations
presented in Ref.\cite{Casagrande} through the residue theorem. In section \ref{sec5}, we
show how to sum over the infinite series of KK modes in unified extra dimension using residue theorem
through recover an important relation applied extensively in literature.
As an example, we demonstrate in section \ref{sec6} that the corrections from neutral Higgs to the
Wilson coefficients of relevant operators for $B\rightarrow X_s\gamma$ contain the suppression factor
$m_b^3m_s/m_{_{\rm w}}^4$ in comparison with that
from other sectors, thus can be neglected safely. Our conclusions are summarized
in Section \ref{sec7}.

\section{A warped extra dimension with custodial symmetry\label{sec2}}
\indent\indent
In the Randall-Sundrum (RS) scenario, four dimensional Minkowskian space-time
is embedded into a slice of five dimensional anti de-Sitter (ADS$_5$)
space with curvature $k$. The fifth dimension is a $S^1/Z_2\times Z_2^\prime$
orbifold of size r labeled by a coordinate $\phi\in[-\pi,\pi]$, thus
the points $(x^\mu,\phi)$, $(x^\mu,\pi-\phi)$, $(x^\mu,\pi+\phi)$ and $(x^\mu,-\phi)$ are
identified all. The corresponding metric of non-factorizable RS geometry is written as
\begin{eqnarray}
&&ds^2=e^{-2\sigma(\phi)}\eta_{\mu\nu}dx^\mu dx^\nu-r^2d\phi^2,
\;\sigma(\phi)=kr|\phi|,
\label{metric}
\end{eqnarray}
where $x^\mu\;(\mu=0,\;1,\;2,\;3)$ are the coordinates on the four dimensional
hyper-surfaces of constant $\phi$ with metric $\eta_{\mu\nu}=(1,-1,-1,-1)$,
and $e^\sigma$ is called the warp factor. Two branes are located on the
orbifold fixed points $\phi=0$ and $\phi=\pi/2$, respectively. The brane
on $\phi=0$ is called Planck or ultra-violet (UV) brane, and the brane on
$\phi=\pi/2$ is called TeV or infra-red (IR) brane. Assuming the parameters
$k$ and $1/r$ to be of order the fundamental Planck scale $M_{\rm PI}$ and choosing
the product $kr\simeq24$, one gets the inverse warp factor
\begin{eqnarray}
&&\epsilon={\Lambda_{\rm IR}\over\Lambda_{\rm UV}}\equiv e^{-kr\pi/2}\simeq10^{-16},
\label{warped-factor}
\end{eqnarray}
which explains the hierarchy between the electroweak and Planck scale naturally.
Meanwhile, the mass scale of low-lying KK excitations is set as
\begin{eqnarray}
&&\Lambda_{_{KK}}\equiv k\epsilon=ke^{-kr\pi/2}={\cal O}(1{\rm TeV}).
\label{KK-scale}
\end{eqnarray}
In the gauge symmetry $SU(2)_L\times SU(2)_R\times U(1)_X\times P_{LR}$,
the discrete symmetry $P_{LR}$ interchanging the local groups $SU(2)_L$ and
$SU(2)_R$ implies that the five dimensional gauge couplings satisfy the relation
$g_{5L}=g_{5R}=g_5$. The local gauge group $SU(2)_L\times SU(2)_R\times U(1)_X$
is broken to the SM gauge group by the boundary conditions (BCs) on the UV brane:
\begin{eqnarray}
&&W_{L,\mu}^{1,2,3}(++),\;B_\mu(++),\;W_{R,\mu}^{1,2}(-+),\;Z_{X,\mu}(-+),\;\;
(\mu=0,\;1,\;2,\;3),
\nonumber\\
&&W_{L,5}^{1,2,3}(--),\;B_5(--),\;W_{R,5}^{1,2}(+-),\;Z_{X,5}(+-).
\label{BCs-gauge}
\end{eqnarray}
Here, the first (second) sign is the BC on the UV (IR) brane: $+$ denotes a Neumann
BC and $-$ denotes a Dirichlet BC. The zero modes of fields with (++)BCs are massless
before the electroweak symmetry is spontaneously broken, and are
identified as the SM gauge bosons. The fields with other BCs only contain
massive KK modes. The third component of $SU(2)_R$ gauge fields
$W_{R,M}^3$ and the $U(1)_X$ gauge field $\tilde{B}_M$ are expressed in
terms of the neutral gauge fields $Z_{X,M}$ and $B_M$ as
\begin{eqnarray}
&&W_{R,M}^3={g_5Z_{X,M}+g_{5X}B_M\over\sqrt{g_5^2+g_{5X}^2}},\;
\tilde{B}_M=-{g_{5X}Z_{X,M}-g_5B_M\over\sqrt{g_5^2+g_{5X}^2}},\;(M=0,\;1,\;2,\;3,\;5),
\label{gauge1}
\end{eqnarray}
where $g_{5X}$ is the five dimensional gauge coupling of $U(1)_X$. The Lagrangian
for gauge sector is written as
\begin{eqnarray}
&&{\cal L}_{gauge}={\sqrt{\cal G}\over r}{\cal G}^{KM}{\cal G}^{LN}\Big(-{1\over4}W_{KL}^iW_{MN}^i
-{1\over4}\tilde{W}_{KL}^i\tilde{W}_{MN}^i-{1\over4}\tilde{B}_{KL}\tilde{B}_{MN}
-{1\over4}G_{KL}^aG_{MN}^a\Big)\;.
\label{L_gauge}
\end{eqnarray}

For convenience in our discussion further, we define the following five dimensional
fields\cite{Agashe11}
\begin{eqnarray}
&&A_A={\sqrt{g_5^2+g_{5X}^2}B_A+g_{5X}W_{L,A}^3\over\sqrt{g_5^2+2g_{5X}^2}},
\nonumber\\
&&Z_{A}={-g_{5X}B_A+\sqrt{g_5^2+g_{5X}^2}W_{L,A}^3\over\sqrt{g_5^2+2g_{5X}^2}},
\nonumber\\
&&W_{L,A}^{\pm}={1\over\sqrt{2}}\Big(W_{L,A}^1\mp iW_{L,A}^2\Big)\;,
\nonumber\\
&&W_{R,A}^{\pm}={1\over\sqrt{2}}\Big(W_{R,A}^1\mp iW_{R,A}^2\Big)\;.
\label{gauge2}
\end{eqnarray}

As regards the matter fields, the quarks of one generation are embedded into the multiplets\cite{Buras5}:
\begin{eqnarray}
&&Q_L^i=\left(\begin{array}{ll}\chi_{u_L}^i(-+)_{5/3}&q_{u_L}^i(++)_{2/3}\\
\chi_{d_L}^i(-+)_{2/3}&q_{d_L}^i(++)_{-1/3}\end{array}\right)\;,\;\;
Q_{u_R}^i=u_R^i(++)_{2/3}
\nonumber\\
&&\tilde{Q}_{d_R}^i=\left(\begin{array}{l}X_R^i(-+)_{5/3}\\U_R^i(-+)_{2/3}
\\D_R^i(-+)_{-1/3}\end{array}\right)\;,\;\;
Q_{d_R}^i=\left(\begin{array}{l}\tilde{X}_R^i(-+)_{5/3}\\ \tilde{U}_R^i(-+)_{2/3}
\\d_R^i(++)_{-1/3}\end{array}\right)\;,
\label{SM-quarks}
\end{eqnarray}
and the states with opposite chirality are written as
\begin{eqnarray}
&&Q_R^i=\left(\begin{array}{ll}\chi_{u_R}^i(+-)_{5/3}&q_{u_R}^i(--)_{2/3}\\
\chi_{d_R}^i(+-)_{2/3}&q_{d_R}^i(--)_{-1/3}\end{array}\right)\;,\;\;
Q_{u_L}^i=u_L^i(--)_{2/3}
\nonumber\\
&&\tilde{Q}_{d_L}^i=\left(\begin{array}{l}X_L^i(+-)_{5/3}\\U_L^i(+-)_{2/3}
\\D_L^i(+-)_{-1/3}\end{array}\right)\;,\;\;
Q_{d_L}^i=\left(\begin{array}{l}\tilde{X}_L^i(+-)_{5/3}\\ \tilde{U}_L^i(+-)_{2/3}
\\d_L^i(--)_{-1/3}\end{array}\right)\;.
\label{opposite-quarks}
\end{eqnarray}
Here, $i=1,\;2,\;3$ denote the indices of generation, the $U(1)_X$ charges are all assigned as
\begin{eqnarray}
&&Y_{Q^i}=Y_{u^i}=Y_{Q_d^i}={2\over3}\;.
\label{U1_charges}
\end{eqnarray}

In order to give the kinetic terms of triplets, we redefine the quarks in
triplet as
\begin{eqnarray}
&&\tilde{T}_{Q_R}^i=\left(\begin{array}{c}{1\over\sqrt{2}}(X_R^i+D_R^i)
\\{i\over\sqrt{2}}(X_R^i-D_R^i)\\U_R^i\end{array}\right)\;,\;\;
T_{Q_R}^i=\left(\begin{array}{c}{1\over\sqrt{2}}(\tilde{X}_R^i+d_R^i)
\\{i\over\sqrt{2}}(\tilde{X}_R^i-d_R^i)\\ \tilde{U}_R^i\end{array}\right)\;,
\nonumber\\
&&\tilde{T}_{Q_L}^i=\left(\begin{array}{c}{1\over\sqrt{2}}(X_L^i+D_L^i)
\\{i\over\sqrt{2}}(X_L^i-D_L^i)\\U_L^i\end{array}\right)\;,\;\;
T_{Q_L}^i=\left(\begin{array}{c}{1\over\sqrt{2}}(\tilde{X}_L^i+d_L^i)
\\{i\over\sqrt{2}}(\tilde{X}_L^i-d_L^i)\\ \tilde{U}_L^i\end{array}\right)\;.
\label{redefine-triplet}
\end{eqnarray}

Correspondingly, the Lagrangian for kinetic terms of quarks can be written as
\begin{eqnarray}
&&{\cal L}_{Q}
={\sqrt{\cal G}\over2r}\sum\limits_{i=1}^3
\Big\{(\overline{Q}^i)_{a_1a_2}iE_A^M\gamma^A\Big[\Big({1\over2}(\partial_M
-\overleftarrow{\partial}_M)+ig_{5s}T^aG^a_M+ig_{5X}Y_{Q^i}\tilde{B}_M\Big)\delta_{a_1b_1}
\delta_{a_2b_2}
\nonumber\\
&&\hspace{1.2cm}
+ig_5({\sigma^{c_1}\over2})_{a_1b_1}W_{L,M}^{c_1}\delta_{a_2b_2}
+ig_5({\sigma^{c_2}\over2})_{a_2b_2}W_{R,M}^{c_2}\delta_{a_1b_1}\Big](Q^i)_{b_1b_2}
\nonumber\\
&&\hspace{1.2cm}
+(\overline{Q}^i)_{a_1a_2}\Big[iE_A^M\gamma^A\omega_M-{\rm sgn}(\phi)(c_{_B})_{ij}\Big]
(Q^j)_{a_1a_2}
\nonumber\\
&&\hspace{1.2cm}
+\overline{u}^i\Big[iE_A^M\gamma^A\Big({1\over2}(\partial_M
-\overleftarrow{\partial}_M)+ig_{5s}T^aG^a_M+ig_{5X}Y_{u^i}\tilde{B}_M\Big)\delta_{ij}
\nonumber\\
&&\hspace{1.2cm}
+iE_A^M\gamma^A\omega_M-{\rm sgn}(\phi)(c_{_S})_{ij}\Big]u^j
\nonumber\\
&&\hspace{1.2cm}
+(\overline{\tilde{T}}_Q^i)_{a1}iE_A^M\gamma^A\Big[\Big({1\over2}(\partial_M
-\overleftarrow{\partial}_M)+ig_{5s}T^aG^a_M+ig_{5X}Y_{Q_d^i}\tilde{B}_M\Big)
\delta_{a_1b_1}
\nonumber\\
&&\hspace{1.2cm}
+g_5\varepsilon_{a_1b_1c_1}W_{L,M}^{c_1}\Big](\tilde{T}_Q^i)_{b_1}
+(\overline{\tilde{T}}_Q^i)_{a_1}\Big[iE_A^M\gamma^A\omega_M-{\rm sgn}(\phi)(\eta_3)_{ij}\Big]
(\tilde{T}_Q^j)_{a_1}
\nonumber\\
&&\hspace{1.2cm}
+(\overline{T}_Q^i)_{a_1}iE_A^M\gamma^A\Big[\Big({1\over2}(\partial_M
-\overleftarrow{\partial}_M)+ig_{5s}T^aG^a_M+ig_{5X}Y_{Q_d^i}\tilde{B}_M\Big)
\delta_{a_1b_1}
\nonumber\\
&&\hspace{1.2cm}
+g_5\varepsilon_{a_1b_1c_1}W_{R,M}^{c_1}\Big](T_Q^i)_{b_1}
+(\overline{T}_Q^i)_{a_1}\Big[iE_A^M\gamma^A\omega_M-{\rm sgn}(\phi)(c_{_T})_{ij}
\Big](T_Q^j)_{a_1}+h.c.\Big\}\;,
\label{kinetic_quark}
\end{eqnarray}
with $\gamma^A=(\gamma^\mu,\;-i\gamma^5)$, the inverse vielbein
$E_B^A={\it diag}(e^{\sigma(\phi)},\;e^{\sigma(\phi)},\;e^{\sigma(\phi)},\;
e^{\sigma(\phi)},{1\over r})$, and the spin connection
$\omega_A=({\rm sgn}(\phi){i\over2}ke^{-\sigma(\phi)}\gamma_\mu\gamma^5,0)$.
Generally, three bulk mass matrices
$c_{_B},\;c_{_S},\;c_{_T}$ are arbitrarily hermitian $3\times3$ matrices.

To break down the electroweak symmetry, we introduce an IR brane located
Higgs which transforms as a self-dual bidoublet under the gauge group
$SU(2)_L\times SU(2)_R$, and transforms as a singlet with charge $Y_H=0$
under the gauge group $U(1)_X$:
\begin{eqnarray}
&&H=\left(\begin{array}{cc}-i\pi^+/\sqrt{2}&-(h^0-i\pi^0)/2\\
(h^0+i\pi^0)/2&i\pi^-/\sqrt{2}
\end{array}\right)\;.
\label{Higgs-bidoublet}
\end{eqnarray}
After normalizing the kinetic term of Higgs in five dimension, we write the
corresponding Lagrangian as
\begin{eqnarray}
&&{\cal L}_{H}={\rm Tr}\Big[\Big(D_\mu
\Phi(x)\Big)^\dagger\Big(D^\mu\Phi(x)\Big)\Big]
-\mu^2{\rm Tr}\Big(\Phi^\dagger(x)\Phi(x)\Big)
+{\lambda\over2}\Big[{\rm Tr}\Big(\Phi^\dagger(x)\Phi(x)\Big)\Big]^2\;.
\label{IR-Brane-Higgs0}
\end{eqnarray}
Accordingly, the Yukawa couplings between quarks and Higgs can be formulated as
\begin{eqnarray}
&&{\cal L}_{Y}^Q=
e^{kr\pi/2}\sqrt{-{\cal G}_{_{IR}}}\sum\limits_{i,j=1}^3\Big\{\sqrt{2}
\lambda_{ij}^{u}\overline{Q}_{a\alpha}^iH_{a\alpha}u^j-2\lambda_{ij}^{d}\Big[
\overline{Q}_{a\alpha}^i(\tau^c)_{ab}(\tilde{T}_d^j)_cH_{b\alpha}
\nonumber\\
&&\hspace{1.2cm}
+\overline{Q}_{a\alpha}^i(\tau^c)_{\alpha\beta}(T_d^j)_cH_{a\beta}\Big]+h.c.\Big\}\;,
\label{Yukawa-quarks}
\end{eqnarray}
here the metric on the IR brane ${\cal G}_{_{IR}}^{\mu\nu}=e^{kr\pi/2}\eta^{\mu\nu}$.
In the following we choose to work in the background gauge
with $\xi=1$\cite{Randall1}. Furthermore,
we also assume that three bulk mass matrices $c_{_B},\;c_{_S},\;c_{_T}$ are real and
diagonal, i.e. each of them is described by three real parameters. This can
always be obtained through some appropriate field redefinitions.

The KK decompositions of five dimensional gauge fields are extensively studied
in literature, and can be written in our notations as
\begin{eqnarray}
&&A_\mu(x,\phi)={1\over\sqrt{r}}\sum\limits_{n=0}^\infty A_\mu^{(n)}(x)
\chi_{_{(++)}}^A(y_{_{(++)}}^{A(n)},\phi),
\nonumber\\
&&A_\phi(x,\phi)={1\over \Lambda_{_{KK}}\sqrt{r}}\sum\limits_{n=1}^\infty{1\over y_{_A}^{(n)}}
\varphi_{_A}^{(n)}(x)\partial_\phi\chi_{_{(++)}}^A(y_{_{(++)}}^{A(n)},\phi),
\nonumber\\
&&Z_\mu(x,\phi)={1\over\sqrt{r}}\sum\limits_{n=0}^\infty Z_\mu^{(n)}(x)
\chi_{_{(++)}}^Z(y_{_{(++)}}^{Z(n)},\phi),
\nonumber\\
&&Z_\phi(x,\phi)={1\over \Lambda_{_{KK}}\sqrt{r}}\sum\limits_{n=1}^\infty{1\over y_{_Z}^{(n)}}\varphi_{_Z}^{(n)}(x)
\partial_\phi\chi_{_{(++)}}^Z(y_{_{(++)}}^{Z(n)},\phi),
\nonumber\\
&&Z_{X,\mu}(x,\phi)={1\over\sqrt{r}}\sum\limits_{n=1}^\infty Z_{X,\mu}^{(n)}(x)
\chi_{_{(-+)}}^{Z_X}(y_{_{(-+)}}^{Z_X(n)},\phi),
\nonumber\\
&&Z_{X,\phi}(x,\phi)={1\over \Lambda_{_{KK}}\sqrt{r}}\sum\limits_{n=1}^\infty
{1\over y_{_{Z_X}}^{(n)}}\varphi_{_{Z_X}}^{(n)}(x)\partial_\phi
\chi_{_{(-+)}}^{Z_X}(y_{_{(-+)}}^{Z_X(n)},\phi),
\nonumber\\
&&W_{L,\mu}^\pm(x,\phi)={1\over\sqrt{r}}\sum\limits_{n=0}^\infty W_{L,\mu}^{\pm(n)}(x)
\chi_{_{(++)}}^{W_L}(y_{_{(++)}}^{W_L(n)},\phi),
\nonumber\\
&&W_{L,\phi}^\pm(x,\phi)={1\over \Lambda_{_{KK}}\sqrt{r}}\sum\limits_{n=1}^\infty{1\over y_{_{W_L}}^{(n)}}
\varphi_{_{W_L}}^{\pm(n)}(x)\partial_\phi\chi_{_{(++)}}^{W_L}(y_{_{(++)}}^{W_L(n)},\phi),
\nonumber\\
&&W_{R,\mu}^\pm(x,\phi)={1\over\sqrt{r}}\sum\limits_{n=1}^\infty W_{R,\mu}^{\pm(n)}(x)
\chi_{_{(-+)}}^{W_R}(y_{_{(-+)}}^{W_R(n)},\phi),
\nonumber\\
&&W_{R,\phi}^\pm(x,\phi)={1\over \Lambda_{_{KK}}\sqrt{r}}\sum\limits_{n=1}^\infty{1\over y_{_{W_R}}^{(n)}}
\varphi_{_{W_R}}^{\pm(n)}(x)\partial_\phi\chi_{_{(-+)}}^{W_R}(y_{_{(-+)}}^{W_R(n)},\phi),
\nonumber\\
&&G^\pm(x,\phi)=\sum\limits_n\Big\{b_n^{W_L}(\phi)\varphi_{W_L}^{\pm,(n)}(x)
+b_n^{W_R}(\phi)\varphi_{W_R}^{\pm,(n)}(x)\Big\},
\nonumber\\
&&G^0(x,\phi)=\sum\limits_n\Big\{b_n^Z(\phi)\varphi_Z^{(n)}(x)
+b_n^{Z_X}(\phi)\varphi_{Z_X}^{(n)}(x)\Big\},
\nonumber\\
&&G_\mu^a(x,\phi)={1\over\sqrt{r}}\sum\limits_{n=0}^\infty G_\mu^{a,(n)}(x)
\chi_{_{(++)}}^g(y_{_{(++)}}^{g(n)},\phi),
\nonumber\\
&&G_\phi^a(x,\phi)={1\over \Lambda_{_{KK}}\sqrt{r}}\sum\limits_{n=1}^\infty{1\over y_{_g}^{(n)}}
\varphi_{_g}^{a,(n)}(x)\partial_\phi\chi_{_{(++)}}^g(y_{_{(++)}}^{g(n)},\phi).
\label{KK-decomposition-Gauge-Higgs}
\end{eqnarray}
As $G=A,\;Z,\;W_L^\pm,\;g$, $y_{_{(++)}}^{G(n)}\;(n=0,\;1,\;\cdots,\;\infty)$ denote
the roots of equation $z^2R_{_{(++)}}^{G,\epsilon}(z)\equiv0$ with
\begin{eqnarray}
&&R_{_{(++)}}^{G,\epsilon}(z)=Y_0(z)J_0(z\epsilon)-J_0(z)Y_0(z\epsilon)\;.
\label{root1}
\end{eqnarray}
When $G=Z_{_X},\;W_R^\pm$, $y_{_{(-+)}}^{G(n)}\;(n=1,\;2,\;\cdots,\;\infty)$ denote
the roots of equation $R_{_{(-+)}}^{G,\epsilon}(z)\equiv0$ with
\begin{eqnarray}
R_{_{(-+)}}^{G,\epsilon}(z)=Y_0(z)J_1(z\epsilon)-J_0(z)Y_1(z\epsilon)\;.
\label{root2}
\end{eqnarray}
In order to give the bulk profiles of those five dimensional fields conveniently, we introduce
a coordinate $t=\epsilon\exp(\sigma(\phi))$ which takes values between $t=\epsilon$
(UV brane) and $t=1$ (IR brane). In terms of the variable $t$, we write the bulk files
for those gauge fields as
\begin{eqnarray}
&&\chi_{_{(++)}}^G(y_{_{(++)}}^{G(n)},\phi)={t\Phi_{_{(uv)}}^{G(+)}(y_{_{(++)}}^{G(n)},t)\over
N_{_{(uv)}}^{G(+)}(y_{_{(++)}}^{G(n)})}={t\Phi_{_{(ir)}}^{G(+)}(y_{_{(++)}}^{G(n)},t)\over
N_{_{(ir)}}^{G(+)}(y_{_{(++)}}^{G(n)})}\;,
\nonumber\\
&&\chi_{_{(-+)}}^G(y_{_{(-+)}}^{G(n)},\phi)={t\Phi_{_{(uv)}}^{G(-)}(y_{_{(-+)}}^{G(n)},t)\over
N_{_{(uv)}}^{G(-)}(y_{_{(-+)}}^{G(n)})}={t\Phi_{_{(ir)}}^{G(+)}(y_{_{(-+)}}^{G(n)},t)\over
N_{_{(ir)}}^{G(+)}(y_{_{(-+)}}^{G(n)})}\;,
\label{gauge-5th}
\end{eqnarray}
with
\begin{eqnarray}
&&\Phi_{_{(uv)}}^{G(+)}(y,t)=Y_0(y\epsilon)J_1(yt)-J_0(y\epsilon)Y_1(yt)\;,
\nonumber\\
&&\Phi_{_{(ir)}}^{G(+)}(y,t)=Y_0(y)J_1(yt)-J_0(y)Y_1(yt)\;,
\nonumber\\
&&\Phi_{_{(uv)}}^{G(-)}(y,t)=Y_1(y\epsilon)J_1(yt)-J_1(y\epsilon)Y_1(yt)\;,
\nonumber\\
&&\Psi_{_{(uv)}}^{G(-)}(y,t)=Y_0(y\epsilon)J_0(yt)-J_0(y\epsilon)Y_0(yt)\;,
\nonumber\\
&&\Psi_{_{(ir)}}^{G(-)}(y,t)=Y_0(y)J_0(yt)-J_0(y)Y_0(yt)\;,
\nonumber\\
&&\Psi_{_{(uv)}}^{G(+)}(y,t)=Y_1(y\epsilon)J_0(yt)-J_1(y\epsilon)Y_0(yt)\;.
\label{Phi}
\end{eqnarray}
It is easy to check Eq.(\ref{Phi}) satisfying the equation of motions
\begin{eqnarray}
&&t{\partial\Phi_{_{(uv)}}^{G(+)}\over\partial t}(y,t)+\Phi_{_{(uv)}}^{G(+)}(y,t)
=yt\Psi_{_{(uv)}}^{G(-)}(y,t)\;,
\nonumber\\
&&t{\partial\Phi_{_{(ir)}}^{G(+)}\over\partial t}(y,t)+\Phi_{_{(ir)}}^{G(+)}(y,t)
=yt\Psi_{_{(ir)}}^{G(-)}(y,t)\;,
\nonumber\\
&&t{\partial\Phi_{_{(uv)}}^{G(-)}\over\partial t}(y,t)+\Phi_{_{(uv)}}^{G(-)}(y,t)
=yt\Psi_{_{(uv)}}^{G(+)}(y,t)\;,
\nonumber\\
&&t{\partial\Psi_{_{(uv)}}^{G(-)}\over\partial t}(y,t)
=-yt\Phi_{_{(uv)}}^{G(+)}(y,t)\;,
\nonumber\\
&&t{\partial\Psi_{_{(ir)}}^{G(-)}\over\partial t}(y,t)
=-yt\Phi_{_{(ir)}}^{G(+)}(y,t)\;,
\nonumber\\
&&t{\partial\Psi_{_{(uv)}}^{G(+)}\over\partial t}(y,t)
=-yt\Phi_{_{(uv)}}^{G(-)}(y,t)\;.
\label{EM-gauge}
\end{eqnarray}
The corresponding normalization constants are formulated as
\begin{eqnarray}
&&\Big[N_{_{(uv)}}^{G(+)}(y)\Big]^2={2\over kr}\Big\{\Big(\Phi_{_{(uv)}}^{G(+)}(y,1)\Big)^2
-\epsilon^2\Big(\Phi_{_{(uv)}}^{G(+)}(y,\epsilon)\Big)^2+\Big(\Psi_{_{(uv)}}^{G(-)}(y,1)\Big)^2
\nonumber\\
&&\hspace{2.6cm}
-{2\over y}\Phi_{_{(uv)}}^{G(+)}(y,1)\Psi_{_{(uv)}}^{G(-)}(y,1)\Big\}\;,
\nonumber\\
&&\Big[N_{_{(ir)}}^{G(+)}(y)\Big]^2={2\over kr}\Big\{\Big(\Phi_{_{(ir)}}^{G(+)}(y,1)\Big)^2
-\epsilon^2\Big(\Phi_{_{(ir)}}^{G(+)}(y,\epsilon)\Big)^2-\epsilon^2\Big(\Psi_{_{(ir)}}^{G(-)}(y,\epsilon)\Big)^2
\nonumber\\
&&\hspace{2.6cm}
+{2\epsilon\over y}\Phi_{_{(ir)}}^{G(+)}(y,\epsilon)\Psi_{_{(ir)}}^{G(-)}(y,\epsilon)\Big\}\;,
\nonumber\\
&&\Big[N_{_{(uv)}}^{G(-)}(y)\Big]^2={2\over kr}\Big\{\Big(\Psi_{_{(uv)}}^{G(+)}(y,1)\Big)^2
-\epsilon^2\Big(\Psi_{_{(uv)}}^{G(+)}(y,\epsilon)\Big)^2+\Big(\Phi_{_{(uv)}}^{G(-)}(y,1)\Big)^2
\nonumber\\
&&\hspace{2.6cm}
-{2\over y}\Psi_{_{(uv)}}^{G(+)}(y,1)\Phi_{_{(uv)}}^{G(-)}(y,1)\Big\}\;.
\label{normalization-gauge}
\end{eqnarray}

Similarly, the KK decompositions of five dimensional quark fields are written as
\begin{eqnarray}
&&\chi_{u_L}^i(x,\phi)={e^{2\sigma(\phi)}\over\sqrt{r}}\sum\limits_n
\chi_{u_L}^{i,(n)}(x)f_{_{(-+)}}^{L,c_B^i}(y_{_{(\mp\pm)}}^{c_B^i(n)},\phi),
\nonumber\\
&&\chi_{d_L}^i(x,\phi)={e^{2\sigma(\phi)}\over\sqrt{r}}\sum\limits_n
\chi_{d_L}^{i,(n)}(x)f_{_{(-+)}}^{L,c_B^i}(y_{_{(\mp\pm)}}^{c_B^i(n)},\phi),
\nonumber\\
&&q_{u_L}^i(x,\phi)={e^{2\sigma(\phi)}\over\sqrt{r}}\sum\limits_n
q_{u_L}^{i,(n)}(x)f_{_{(++)}}^{L,c_B^i}(y_{_{(\pm\pm)}}^{c_B^i(n)},\phi),
\nonumber\\
&&q_{d_L}^i(x,\phi)={e^{2\sigma(\phi)}\over\sqrt{r}}\sum\limits_n
q_{d_L}^{i,(n)}(x)f_{_{(++)}}^{L,c_B^i}(y_{_{(\pm\pm)}}^{c_B^i(n)},\phi),
\nonumber\\
&&u_R^i(x,\phi)={e^{2\sigma(\phi)}\over\sqrt{r}}\sum\limits_n
u_R^{i,(n)}(x)f_{_{(++)}}^{R,c_S^i}(y_{_{(\mp\mp)}}^{c_S^i(n)},\phi),
\nonumber\\
&&X_R^i(x,\phi)={e^{2\sigma(\phi)}\over\sqrt{r}}\sum\limits_n
X_R^{i,(n)}(x)f_{_{(-+)}}^{R,c_T^i}(y_{_{(\pm\mp)}}^{c_T^i(n)},\phi),
\nonumber\\
&&U_R^i(x,\phi)={e^{2\sigma(\phi)}\over\sqrt{r}}\sum\limits_n
U_R^{i,(n)}(x)f_{_{(-+)}}^{R,c_T^i}(y_{_{(\pm\mp)}}^{c_T^i(n)},\phi),
\nonumber\\
&&D_R^i(x,\phi)={e^{2\sigma(\phi)}\over\sqrt{r}}\sum\limits_n
D_R^{i,(n)}(x)f_{_{(-+)}}^{R,c_T^i}(y_{_{(\pm\mp)}}^{c_T^i(n)},\phi),
\nonumber\\
&&\tilde{X}_R^i(x,\phi)={e^{2\sigma(\phi)}\over\sqrt{r}}\sum\limits_n
\tilde{X}_R^{i,(n)}(x)f_{_{(-+)}}^{R,c_T^i}(y_{_{(\pm\mp)}}^{c_T^i(n)},\phi),
\nonumber\\
&&\tilde{U}_R^i(x,\phi)={e^{2\sigma(\phi)}\over\sqrt{r}}\sum\limits_n
\tilde{U}_R^{i,(n)}(x)f_{_{(-+)}}^{R,c_T^i}(y_{_{(\pm\mp)}}^{c_T^i(n)},\phi),
\nonumber\\
&&d_R^i(x,\phi)={e^{2\sigma(\phi)}\over\sqrt{r}}\sum\limits_n
d_R^{i,(n)}(x)f_{_{(++)}}^{R,c_T^i}(y_{_{(\mp\mp)}}^{c_T^i(n)},\phi),
\nonumber\\
&&X_L^i(x,\phi)={e^{2\sigma(\phi)}\over\sqrt{r}}\sum\limits_n
X_L^{i,(n)}(x)f_{_{(+-)}}^{L,c_T^i}(y_{_{(\pm\mp)}}^{c_T^i(n)},\phi),
\nonumber\\
&&U_L^i(x,\phi)={e^{2\sigma(\phi)}\over\sqrt{r}}\sum\limits_n
U_L^{i,(n)}(x)f_{_{(+-)}}^{L,c_T^i}(y_{_{(\pm\mp)}}^{c_T^i(n)},\phi),
\nonumber\\
&&D_L^i(x,\phi)={e^{2\sigma(\phi)}\over\sqrt{r}}\sum\limits_n
D_L^{i,(n)}(x)f_{_{(+-)}}^{L,c_T^i}(y_{_{(\pm\mp)}}^{c_T^i(n)},\phi),
\nonumber\\
&&\tilde{X}_L^i(x,\phi)={e^{2\sigma(\phi)}\over\sqrt{r}}\sum\limits_n
\tilde{X}_L^{i,(n)}(x)f_{_{(+-)}}^{L,c_T^i}(y_{_{(\pm\mp)}}^{c_T^i(n)},\phi),
\nonumber\\
&&\tilde{U}_L^i(x,\phi)={e^{2\sigma(\phi)}\over\sqrt{r}}\sum\limits_n
\tilde{U}_L^{i,(n)}(x)f_{_{(+-)}}^{L,c_T^i}(y_{_{(\pm\mp)}}^{c_T^i(n)},\phi),
\nonumber\\
&&d_L^i(x,\phi)={e^{2\sigma(\phi)}\over\sqrt{r}}\sum\limits_n
d_L^{i,(n)}(x)f_{_{(--)}}^{L,c_T^i}(y_{_{(\mp\mp)}}^{c_T^i(n)},\phi),
\nonumber\\
&&u_L^i(x,\phi)={e^{2\sigma(\phi)}\over\sqrt{r}}\sum\limits_n
u_L^{i,(n)}(x)f_{_{(--)}}^{L,c_S^i}(y_{_{(\mp\mp)}}^{c_S^i(n)},\phi),
\nonumber\\
&&\chi_{u_R}^i(x,\phi)={e^{2\sigma(\phi)}\over\sqrt{r}}\sum\limits_n
\chi_{u_R}^{i,(n)}(x)f_{_{(+-)}}^{R,c_B^i}(y_{_{(\mp\pm)}}^{c_B^i(n)},\phi),
\nonumber\\
&&\chi_{d_R}^i(x,\phi)={e^{2\sigma(\phi)}\over\sqrt{r}}\sum\limits_n
\chi_{d_R}^{i,(n)}(x)f_{_{(+-)}}^{R,c_B^i}(y_{_{(\mp\pm)}}^{c_B^i(n)},\phi),
\nonumber\\
&&q_{u_R}^i(x,\phi)={e^{2\sigma(\phi)}\over\sqrt{r}}\sum\limits_n
q_{u_R}^{i,(n)}(x)f_{_{(--)}}^{R,c_B^i}(y_{_{(\pm\pm)}}^{c_B^i(n)},\phi),
\nonumber\\
&&q_{d_R}^i(x,\phi)={e^{2\sigma(\phi)}\over\sqrt{r}}\sum\limits_n
q_{d_R}^{i,(n)}(x)f_{_{(--)}}^{R,c_B^i}(y_{_{(\pm\pm)}}^{c_B^i(n)},\phi).
\label{KK-decomposition-quark}
\end{eqnarray}
In Eq.(\ref{KK-decomposition-quark}), the eigenvalues $y_{_{(\pm\pm)}}^{c(n)}\;(n\ge1)$
satisfy the equation $R_{_{(\pm\pm)}}^{c,\epsilon}(z)\equiv0$,
$y_{_{(\mp\mp)}}^{c(n)}\;(n\ge1)$ satisfy the equation $R_{_{(\mp\mp)}}^{c,\epsilon}(z)\equiv0$,
$y_{_{(\pm\mp)}}^{c(n)}\;(n\ge1)$ satisfy the equation
$R_{_{(\pm\mp)}}^{c,\epsilon}(z)\equiv0$, and the eigenvalues $y_{_{(\mp\pm)}}^{c(n)}\;(n\ge1)$
satisfy the equation $R_{_{(\mp\pm)}}^{c,\epsilon}(z)\equiv0$, respectively.
Here, the concrete expressions of $R_{_{(\pm\pm)}}^{c,\epsilon}(z),\;R_{_{(\pm\mp)}}^{c,\epsilon}(z)
\;,R_{_{(\mp\pm)}}^{c,\epsilon}(z),\;R_{_{(\mp\mp)}}^{c,\epsilon}(z)$ are
\begin{eqnarray}
&&R_{_{(\pm\pm)}}^{c,\epsilon}(z)=\left\{\begin{array}{ll}Y_{N}(z)J_{N}(z\epsilon)-J_{N}(z)Y_{N}(z\epsilon),
&c=N+{1\over2}\\
J_{-c+{1\over2}}(z)J_{c-{1\over2}}(z\epsilon)-J_{c-{1\over2}}(z)J_{-c+{1\over2}}(z\epsilon),
&c\neq N+{1\over2}\end{array}\right.\;,
\nonumber\\
&&R_{_{(\pm\mp)}}^{c,\epsilon}(z)=\left\{\begin{array}{ll}J_{N+1}(z)Y_{N}(z\epsilon)-Y_{N+1}(z)J_{N}(z\epsilon),
&c=N+{1\over2}\\
J_{c+{1\over2}}(z)J_{-c+{1\over2}}(z\epsilon)+J_{-c-{1\over2}}(z)J_{c-{1\over2}}(z\epsilon),
&c\neq N+{1\over2}\end{array}\right.\;,
\nonumber\\
&&R_{_{(\mp\pm)}}^{c,\epsilon}(z)=\left\{\begin{array}{ll}Y_{N}(z)J_{N+1}(z\epsilon)-J_{N}(z)Y_{N+1}(z\epsilon),
&c=N+{1\over2}\\
J_{-c+{1\over2}}(z)J_{c+{1\over2}}(z\epsilon)+J_{c-{1\over2}}(z)J_{-c-{1\over2}}(z\epsilon),
&c\neq N+{1\over2}\end{array}\right.\;,
\nonumber\\
&&R_{_{(\mp\mp)}}^{c,\epsilon}(z)=\left\{\begin{array}{ll}J_{N+1}(z)Y_{N+1}(z\epsilon)
-Y_{N+1}(z)J_{N+1}(z\epsilon),&c=N+{1\over2}\\
J_{c+{1\over2}}(z)J_{-c-{1\over2}}(z\epsilon)-J_{-c-{1\over2}}(z)J_{c+{1\over2}}(z\epsilon),
&c\neq N+{1\over2}\end{array}\right.\;.
\label{root-quark}
\end{eqnarray}
In terms of the variable $t$, those bulk profiles for five dimensional fermions are formulated as
\begin{eqnarray}
&&f_{_{(++)}}^{L,c}(y_{_{(\pm\pm)}}^{c(n)},\phi)={\sqrt{t}\;\varphi_{_{L_{(uv)}}}^{c(+)}
(y_{_{(\pm\pm)}}^{c(n)},t)\over N_{_{L_{(uv)}}}^{c(+)}(y_{_{(\pm\pm)}}^{c(n)})}
={\sqrt{t}\;\varphi_{_{L_{(ir)}}}^{c(+)}(y_{_{(\pm\pm)}}^{c(n)},t)
\over N_{_{L_{(ir)}}}^{c(+)}(y_{_{(\pm\pm)}}^{c(n)})}\;,
\nonumber\\
&&f_{_{(+-)}}^{L,c}(y_{_{(\pm\mp)}}^{c(n)},\phi)={\sqrt{t}\;\varphi_{_{L_{(uv)}}}^{c(+)}
(y_{_{(\pm\mp)}}^{c(n)},t)\over N_{_{L_{(uv)}}}^{c(+)}(y_{_{(\pm\mp)}}^{c(n)})}
={\sqrt{t}\;\varphi_{_{L_{(ir)}}}^{c(-)}(y_{_{(\pm\mp)}}^{c(n)},t)\over
N_{_{L_{(ir)}}}^{c(-)}(y_{_{(\pm\mp)}}^{c(n)})}\;,
\nonumber\\
&&f_{_{(-+)}}^{L,c}(y_{_{(\mp\pm)}}^{c(n)},\phi)={\sqrt{t}\;\varphi_{_{L_{(uv)}}}^{c(-)}
(y_{_{(\mp\pm)}}^{c(n)},t)\over N_{_{L_{(uv)}}}^{c(-)}(y_{_{(\mp\pm)}}^{c(n)})}
={\sqrt{t}\;\varphi_{_{L_{(ir)}}}^{c(+)}(y_{_{(\mp\pm)}}^{c(n)},t)\over
N_{_{L_{(ir)}}}^{c(+)}(y_{_{(\mp\pm)}}^{c(n)})}\;,
\nonumber\\
&&f_{_{(--)}}^{L,c}(y_{_{(\mp\mp)}}^{c(n)},\phi)={\sqrt{t}\;\varphi_{_{L_{(uv)}}}^{c(-)}
(y_{_{(\mp\mp)}}^{c(n)},t)\over N_{_{L_{(uv)}}}^{c(-)}(y_{_{(\mp\mp)}}^{c(n)})}
={\sqrt{t}\;\varphi_{_{L_{(ir)}}}^{c(-)}(y_{_{(\mp\mp)}}^{c(n)},t)\over
N_{_{L_{(ir)}}}^{c(-)}(y_{_{(\mp\mp)}}^{c(n)})}\;,
\nonumber\\
&&f_{_{(++)}}^{R,c}(y_{_{(\mp\mp)}}^{c(n)},\phi)={\sqrt{t}\;\varphi_{_{R_{(uv)}}}^{c(+)}
(y_{_{(\mp\mp)}}^{c(n)},t)\over N_{_{R_{(uv)}}}^{c(+)}(y_{_{(\mp\mp)}}^{c(n)})}
={\sqrt{t}\;\varphi_{_{R_{(ir)}}}^{c(+)}(y_{_{(\mp\mp)}}^{c(n)},t)\over
N_{_{R_{(ir)}}}^{c(+)}(y_{_{(\mp\mp)}}^{c(n)})}\;,
\nonumber\\
&&f_{_{(+-)}}^{R,c}(y_{_{(\mp\pm)}}^{c(n)},\phi)={\sqrt{t}\;\varphi_{_{R_{(uv)}}}^{c(+)}
(y_{_{(\mp\pm)}}^{c(n)},t)\over N_{_{R_{(uv)}}}^{c(+)}(y_{_{(\mp\pm)}}^{c(n)})}
={\sqrt{t}\;\varphi_{_{R_{(ir)}}}^{c(-)}(y_{_{(\mp\pm)}}^{c(n)},t)\over
N_{_{R_{(ir)}}}^{c(-)}(y_{_{(\mp\pm)}}^{c(n)})}\;,
\nonumber\\
&&f_{_{(-+)}}^{R,c}(y_{_{(\pm\mp)}}^{c(n)},\phi)={\sqrt{t}\;\varphi_{_{R_{(uv)}}}^{c(-)}
(y_{_{(\pm\mp)}}^{c(n)},t)\over N_{_{R_{(uv)}}}^{c(-)}(y_{_{(\pm\mp)}}^{c(n)})}
={\sqrt{t}\;\varphi_{_{R_{(ir)}}}^{c(+)}(y_{_{(\pm\mp)}}^{c(n)},t)\over
N_{_{R_{(ir)}}}^{c(+)}(y_{_{(\pm\mp)}}^{c(n)})}\;,
\nonumber\\
&&f_{_{(--)}}^{R,c}(y_{_{(\pm\pm)}}^{c(n)},\phi)={\sqrt{t}\;\varphi_{_{R_{(uv)}}}^{c(-)}
(y_{_{(\pm\pm)}}^{c(n)},t)\over N_{_{R_{(uv)}}}^{c(-)}(y_{_{(\pm\pm)}}^{c(n)})}
={\sqrt{t}\;\varphi_{_{R_{(ir)}}}^{c(-)}(y_{_{(\pm\pm)}}^{c(n)},t)\over
N_{_{R_{(ir)}}}^{c(-)}(y_{_{(\pm\pm)}}^{c(n)})}
\label{quark-5th}
\end{eqnarray}
with
\begin{eqnarray}
&&\varphi_{_{L_{(ir)}}}^{c(+)}(y,t)=\left\{\begin{array}{ll}Y_{N}(y)J_{N+1}(yt)-J_{N}(y)Y_{N+1}(yt),
&c=N+{1\over2}\\
J_{-c+{1\over2}}(y)J_{c+{1\over2}}(yt)+J_{c-{1\over2}}(y)J_{-c-{1\over2}}(yt),
&c\neq N+{1\over2}\end{array}\right.\;,
\nonumber\\
&&\varphi_{_{L_{(uv)}}}^{c(+)}(y,t)=\left\{\begin{array}{ll}Y_{N}(y\epsilon)J_{N+1}(yt)-J_{N}(y\epsilon)Y_{N+1}(yt),
&c=N+{1\over2}\\
J_{-c+{1\over2}}(y\epsilon)J_{c+{1\over2}}(yt)+J_{c-{1\over2}}(y\epsilon)J_{-c-{1\over2}}(yt),
&c\neq N+{1\over2}\end{array}\right.\;,
\nonumber\\
&&\varphi_{_{L_{(ir)}}}^{c(-)}(y,t)=\left\{\begin{array}{ll}Y_{N+1}(y)J_{N+1}(yt)-J_{N+1}(y)Y_{N+1}(yt),
&c=N+{1\over2}\\
J_{-c-{1\over2}}(y)J_{c+{1\over2}}(yt)-J_{c+{1\over2}}(y)J_{-c-{1\over2}}(yt),
&c\neq N+{1\over2}\end{array}\right.\;,
\nonumber\\
&&\varphi_{_{L_{(uv)}}}^{c(-)}(y,t)=\left\{\begin{array}{ll}Y_{N+1}(y\epsilon)J_{N+1}(yt)-J_{N+1}(y\epsilon)Y_{N+1}(yt),
&c=N+{1\over2}\\
J_{-c-{1\over2}}(y\epsilon)J_{c+{1\over2}}(yt)-J_{c+{1\over2}}(y\epsilon)J_{-c-{1\over2}}(yt),
&c\neq N+{1\over2}\end{array}\right.\;,
\nonumber\\
&&\varphi_{_{R_{(ir)}}}^{c(-)}(y,t)=\left\{\begin{array}{ll}Y_{N}(y)J_{N}(yt)-J_{N}(y)Y_{N}(yt),
&c=N+{1\over2}\\
J_{-c+{1\over2}}(y)J_{c-{1\over2}}(yt)-J_{c-{1\over2}}(y)J_{-c+{1\over2}}(yt),
&c\neq N+{1\over2}\end{array}\right.\;,
\nonumber\\
&&\varphi_{_{R_{(uv)}}}^{c(-)}(y,t)=\left\{\begin{array}{ll}Y_{N}(y\epsilon)J_{N}(yt)-J_{N}(y\epsilon)Y_{N}(yt),
&c=N+{1\over2}\\
J_{-c+{1\over2}}(y\epsilon)J_{c-{1\over2}}(yt)-J_{c-{1\over2}}(y\epsilon)J_{-c+{1\over2}}(yt),
&c\neq N+{1\over2}\end{array}\right.\;,
\nonumber\\
&&\varphi_{_{R_{(ir)}}}^{c(+)}(y,t)=\left\{\begin{array}{ll}Y_{N+1}(y)J_{N}(yt)-J_{N+1}(y)Y_{N}(yt),
&c=N+{1\over2}\\
J_{-c-{1\over2}}(y)J_{c-{1\over2}}(yt)+J_{c+{1\over2}}(y)J_{-c+{1\over2}}(yt),
&c\neq N+{1\over2}\end{array}\right.\;,
\nonumber\\
&&\varphi_{_{R_{(uv)}}}^{c(+)}(y,t)=\left\{\begin{array}{ll}Y_{N+1}(y\epsilon)J_{N}(yt)-J_{N+1}(y\epsilon)Y_{N}(yt),
&c=N+{1\over2}\\
J_{-c-{1\over2}}(y\epsilon)J_{c-{1\over2}}(yt)+J_{c+{1\over2}}(y\epsilon)J_{-c+{1\over2}}(yt),
&c\neq N+{1\over2}\end{array}\right.\;.
\label{varphi-i}
\end{eqnarray}
It should pointed out that those bulk profiles satisfy the following equations of motion
\begin{eqnarray}
&&t{\partial\varphi_{_{L_{(ir)}}}^{c(\pm)}\over\partial t}(y,t)+\Big(c+{1\over2}\Big)
\varphi_{_{L_{(ir)}}}^{c(\pm)}(y,t)=yt\varphi_{_{R_{(ir)}}}^{c(\mp)}(y,t)\;,
\nonumber\\
&&t{\partial\varphi_{_{R_{(ir)}}}^{c(\pm)}\over\partial t}(z,t)-\Big(c-{1\over2}\Big)
\varphi_{_{R_{(ir)}}}^{c(\pm)}(y,t)=-yt\varphi_{_{L_{(ir)}}}^{c(\mp)}(y,t)\;,
\nonumber\\
&&t{\partial\varphi_{_{L_{(uv)}}}^{c(\pm)}\over\partial t}(y,t)+\Big(c+{1\over2}\Big)
\varphi_{_{L_{(uv)}}}^{c(\pm)}(y,t)=yt\varphi_{_{R_{(uv)}}}^{c(\mp)}(y,t)\;,
\nonumber\\
&&t{\partial\varphi_{_{R_{(uv)}}}^{c(\pm)}\over\partial t}(y,t)-\Big(c-{1\over2}\Big)
\varphi_{_{R_{(uv)}}}^{c(\pm)}(y,t)=-yt\varphi_{_{L_{(uv)}}}^{c(\mp)}(y,t)\;.
\label{EM-fermion}
\end{eqnarray}

Meanwhile, the normalization factors can be written as
\begin{eqnarray}
&&\Big[N_{_{L_{(ir)}}}^{c(+)}(y)\Big]^2={2\over kr\epsilon}\Big\{\Big(\varphi_{_{L_{(ir)}}}^{c(+)}(y,1)\Big)^2
-\epsilon^2\Big(\varphi_{_{L_{(ir)}}}^{c(+)}(y,\epsilon)\Big)^2-\epsilon^2\Big(\varphi_{_{R_{(ir)}}}^{c(-)}(y,\epsilon)\Big)^2
\nonumber\\
&&\hspace{2.5cm}
+(2c+1){\epsilon\over y}\varphi_{_{L_{(ir)}}}^{c(+)}(y,\epsilon)\varphi_{_{R_{(ir)}}}^{c(-)}(y,\epsilon)\Big\}\;,
\nonumber\\
&&\Big[N_{_{R_{(ir)}}}^{c(-)}(y)\Big]^2={2\over kr\epsilon}\Big\{\Big(\varphi_{_{L_{(ir)}}}^{c(+)}(y,1)\Big)^2
-\epsilon^2\Big(\varphi_{_{L_{(ir)}}}^{c(+)}(y,\epsilon)\Big)^2-\epsilon^2\Big(\varphi_{_{R_{(ir)}}}^{c(-)}(y,\epsilon)\Big)^2
\nonumber\\
&&\hspace{2.5cm}
+(2c-1){\epsilon\over y}\varphi_{_{L_{(ir)}}}^{c(+)}(y,\epsilon)\varphi_{_{R_{(ir)}}}^{c(-)}(y,\epsilon)\Big\}\;,
\nonumber\\
&&\Big[N_{_{L_{(ir)}}}^{c(-)}(y)\Big]^2={2\over kr\epsilon}\Big\{
-\epsilon^2\Big(\varphi_{_{L_{(ir)}}}^{c(-)}(y,\epsilon)\Big)^2
+\Big(\varphi_{_{R_{(ir)}}}^{c(+)}(y,1)\Big)^2-\epsilon^2\Big(\varphi_{_{R_{(ir)}}}^{c(+)}(y,\epsilon)\Big)^2
\nonumber\\
&&\hspace{2.5cm}
+(2c+1){\epsilon\over y}\varphi_{_{L_{(ir)}}}^{c(-)}(y,\epsilon)\varphi_{_{R_{(ir)}}}^{c(+)}(y,\epsilon)\Big\}\;,
\nonumber\\
&&\Big[N_{_{R_{(ir)}}}^{c(+)}(y)\Big]^2={2\over kr\epsilon}\Big\{
-\epsilon^2\Big(\varphi_{_{L_{(ir)}}}^{c(-)}(y,\epsilon)\Big)^2
+\Big(\varphi_{_{R_{(ir)}}}^{c(+)}(y,1)\Big)^2-\epsilon^2\Big(\varphi_{_{R_{(ir)}}}^{c(+)}(y,\epsilon)\Big)^2
\nonumber\\
&&\hspace{2.5cm}
+(2c-1){\epsilon\over y}\varphi_{_{L_{(ir)}}}^{c(-)}(y,\epsilon)\varphi_{_{R_{(ir)}}}^{c(+)}(y,\epsilon)\Big\}\;,
\nonumber\\
&&\Big[N_{_{L_{(uv)}}}^{c(+)}(y)\Big]^2={2\over kr\epsilon}\Big\{\Big(\varphi_{_{L_{(uv)}}}^{c(+)}(y,1)\Big)^2
-\epsilon^2\Big(\varphi_{_{L_{(uv)}}}^{c(+)}(y,\epsilon)\Big)^2+\Big(\varphi_{_{R_{(uv)}}}^{c(-)}(y,1)\Big)^2
\nonumber\\
&&\hspace{2.5cm}
-{2c+1\over y}\varphi_{_{L_{(uv)}}}^{c(+)}(y,1)\varphi_{_{R_{(uv)}}}^{c(-)}(y,1)\Big\}\;,
\nonumber\\
&&\Big[N_{_{R_{(uv)}}}^{c(-)}(y)\Big]^2={2\over kr\epsilon}\Big\{\Big(\varphi_{_{L_{(uv)}}}^{c(+)}(y,1)\Big)^2
-\epsilon^2\Big(\varphi_{_{L_{(uv)}}}^{c(+)}(y,\epsilon)\Big)^2+\Big(\varphi_{_{R_{(uv)}}}^{c(-)}(y,1)\Big)^2
\nonumber\\
&&\hspace{2.5cm}
-{2c-1\over y}\varphi_{_{L_{(uv)}}}^{c(+)}(y,1)\varphi_{_{R_{(uv)}}}^{c(-)}(y,1)\Big\}\;,
\nonumber\\
&&\Big[N_{_{L_{(uv)}}}^{c(-)}(y)\Big]^2={2\over kr\epsilon}\Big\{
\Big(\varphi_{_{L_{(uv)}}}^{c(-)}(y,1)\Big)^2
+\Big(\varphi_{_{R_{(uv)}}}^{c(+)}(y,1)\Big)^2-\epsilon^2\Big(\varphi_{_{R_{(uv)}}}^{c(+)}(y,\epsilon)\Big)^2
\nonumber\\
&&\hspace{2.5cm}
-{2c+1\over y}\varphi_{_{L_{(uv)}}}^{c(-)}(y,1)\varphi_{_{R_{(uv)}}}^{c(+)}(y,1)\Big\}\;,
\nonumber\\
&&\Big[N_{_{R_{(uv)}}}^{c(+)}(y)\Big]^2={2\over kr\epsilon}\Big\{
\Big(\varphi_{_{L_{(uv)}}}^{c(-)}(y,1)\Big)^2
+\Big(\varphi_{_{R_{(uv)}}}^{c(+)}(y,1)\Big)^2-\epsilon^2\Big(\varphi_{_{R_{(uv)}}}^{c(+)}(y,\epsilon)\Big)^2
\nonumber\\
&&\hspace{2.5cm}
-{2c-1\over y}\varphi_{_{L_{(uv)}}}^{c(-)}(y,1)\varphi_{_{R_{(uv)}}}^{c(+)}(y,1)\Big\}\;.
\label{normalization-quark}
\end{eqnarray}
With the preparations above, we verify some lemmas on the
eigenvalues of KK modes in the framework of warped extra dimension,
then show how to sum over the infinite series of KK modes using the residue theorem.

\section{Summing over infinite series of KK modes\label{sec3}}
\indent\indent
As mentioned above, the radiative corrections from all virtual KK modes to the physics
quantities at electroweak scale should be summed over in principle in order to obtain the
theoretical predictions in extensions of the SM with a warped or unified extra dimension.
To sum over the infinite series $\sum\limits_{n=1}^\infty(f(y_{_{(BCs)}}^{c(n)})+f(y_{_{(BCs)}}^{c(-n)}))$
where $f(z)$ is analytic function except some limited in number nonzero poles
$z_1,\;z_2,\;\cdots,\;z_{n_0}$ and the possible pole $z=0$, we construct another function $G(z)$ which
is analytic except its poles of order one $y_{_{(BCs)}}^{c(\pm1)},\;\cdots,\;
y_{_{(BCs)}}^{c(\pm n)},\cdots$, and the possible pole of order $m$: $z=0\;\;(m\ge1)$. Additional,
the residues of $G(z)$ are uniform one at nonzero poles $z=\;y_{_{(BCs)}}^{c(\pm1)},\;\cdots,\;
y_{_{(BCs)}}^{c(\pm n)},\cdots$. If the closed rectifiable contour
$C_{_{r_{(n)}}}$ does not pass through any of the points $y_{_{(BCs)}}^{c(\pm k)}\;(k=1,\;2,\;\cdots,\;n)$,
and the region with the boundary $C_{_{r_{(n)}}}$ contains the points
$0,\;y_{_{(BCs)}}^{c(\pm1)},\;\cdots,\;y_{_{(BCs)}}^{c(\pm n)}$ and $z_1,\;z_2,\;\cdots,\;z_{n_0}$.
According the residue theorem\cite{textbook}, we obtain
\begin{eqnarray}
&&\oint_{C_{_{r_{(n)}}}}G(z)f(z)dz=i2\pi\Big\{\sum\limits_{i=1}^n\Big[
f(y_{_{(BCs)}}^{c(i)})+f(y_{_{(BCs)}}^{c(-i)})\Big]
\nonumber\\
&&\hspace{3.6cm}
+{\rm Res}\Big(G(z)f(z),z=0\Big)+\sum\limits_{i=1}^{n_0}{\rm Res}\Big(G(z)f(z),z=z_i\Big)\Big\}\;.
\label{residue1}
\end{eqnarray}
If the limit $$\lim\limits_{n\rightarrow\infty}\oint_{C_{_{r_{(n)}}}}G(z)f(z)dz=0\;,$$
then we can sum over the infinite series
\begin{eqnarray}
\sum\limits_{i=1}^\infty\Big[f(y_{_{(BCs)}}^{c(i)})
+f(y_{_{(BCs)}}^{c(-i)})\Big]=-{\rm Res}\Big(G(z)f(z),z=0\Big)-\sum\limits_{i=1}^{n_0}{\rm Res}\Big(G(z)f(z),z=z_i\Big)\;.
\label{residue1aa}
\end{eqnarray}
To construct the function $G(z)$ and find the suitable contour $C_{_{r_{(n)}}}$,
we verify some lemmas on the eigenvalues of KK modes firstly.

{\it Lemma1: If $y_{_{\rm(BCs)}}^{c(n)}(n=1,\;2,\;\cdots,\;\infty)$ satisfy the equation
$R_{_{\rm(BCs)}}^{c,\epsilon}(y_{_{\rm(BCs)}}^{c(n)})=0$ where $c$ denotes the bulk mass
of five dimensional fermions, then $y_{_{\rm(BCs)}}^{c(n)}$ is real.}

Proof. Taking ${\rm(BCs)}=(\pm\pm)$ as an example, we show how to demonstrate
that the roots of $R_{_{\pm\pm}}^{c,\epsilon}(y_{_{\pm\pm}}^{c(n)})=0$ are real.
If the left-handed five dimensional fermion satisfies the BCs $(++)$ and the dual
right-handed five dimensional fermion satisfies the BCs $(--)$, the corresponding bulk files
are written as
\begin{eqnarray}
&&f_{_{(++)}}^{L,c}(y_{_{(\pm\pm)}}^{c(n)},\phi)={\sqrt{t}\varphi_{_{L_{(ir)}}}^{c(+)}
(y_{_{(\pm\pm)}}^{c(n)},t)\over N_{_{L_{(ir)}}}^{c(+)}(y_{_{(\pm\pm)}}^{c(n)})}
={\sqrt{t}\varphi_{_{L_{(uv)}}}^{c(+)}
(y_{_{(\pm\pm)}}^{c(n)},t)\over N_{_{L_{(uv)}}}^{c(+)}(y_{_{(\pm\pm)}}^{c(n)})}\;,
\nonumber\\
&&f_{_{(--)}}^{R,c}(y_{_{(\pm\pm)}}^{c(n)},\phi)={\sqrt{t}\varphi_{_{R_{(ir)}}}^{c(-)}
(y_{_{(\pm\pm)}}^{c(n)},t)\over N_{_{R_{(ir)}}}^{c(-)}(y_{_{(\pm\pm)}}^{c(n)})}
={\sqrt{t}\varphi_{_{R_{(uv)}}}^{c(-)}
(y_{_{(\pm\pm)}}^{c(n)},t)\over N_{_{R_{(uv)}}}^{c(-)}(y_{_{(\pm\pm)}}^{c(n)})}\;,
\label{lemma1a}
\end{eqnarray}
where $\varphi_{_{L_{(ir)}}}^{c(+)}(y_{_{(\pm\pm)}}^{c(n)},t),\;\varphi_{_{L_{(uv)}}}^{c(+)}(y_{_{(\pm\pm)}}^{c(n)},t),
\;\varphi_{_{R_{(ir)}}}^{c(-)}(y_{_{(\pm\pm)}}^{c(n)},t),\;\varphi_{_{R_{(uv)}}}^{c(-)}(y_{_{(\pm\pm)}}^{c(n)},t)$
satisfy the equations of motion in Eq.(\ref{EM-fermion}).
Using Eq.(\ref{EM-fermion}), we can derive easily the following equations
\begin{eqnarray}
&&{1\over t}{\partial\over\partial t}\Big(t{\partial\varphi_{_{L_{(ir)}}}^{c(+)}\over\partial t}\Big)
(y_{_{(\pm\pm)}}^{c(n)},t)+\bigg(\Big[y_{_{(\pm\pm)}}^{c(n)}\Big]^2
-{(c+{1\over2})^2\over t^2}\bigg)\varphi_{_{L_{(ir)}}}^{c(+)}(y_{_{(\pm\pm)}}^{c(n)},t)=0
\;,\nonumber\\
&&{1\over t}{\partial\over\partial t}\Big(t{\partial\varphi_{_{R_{(ir)}}}^{c(-)}\over\partial t}\Big)
(y_{_{(\pm\pm)}}^{c(n)},t)+\bigg(\Big[y_{_{(\pm\pm)}}^{c(n)}\Big]^2
-{(c-{1\over2})^2\over t^2}\bigg)\varphi_{_{R_{(ir)}}}^{c(-)}(y_{_{(\pm\pm)}}^{c(n)},t)=0\;.
\label{lemma1c}
\end{eqnarray}
If $y_{_{(\pm\pm)}}^{c(n)}=r+is$ ($r,s$ are real and $s\neq0$) satisfies
$R_{_{(\pm\pm)}}^{c,\epsilon}(y_{_{(\pm\pm)}}^{c(n)})=0$, then its conjugate
$\overline{y}_{_{(\pm\pm)}}^{c(n)}=r-is$ also satisfies $R_{_{(\pm\pm)}}^{c,\epsilon}
(\overline{y}_{_{(\pm\pm)}}^{c(n)})=0$. This implies
\begin{eqnarray}
&&{1\over t}{\partial\over\partial t}\Big(t{\partial\varphi_{_{L_{(ir)}}}^{c(+)}\over\partial t}\Big)
(\overline{y}_{_{(\pm\pm)}}^{c(n)},t)+\bigg(\Big[\overline{y}_{_{(\pm\pm)}}^{c(n)}\Big]^2
-{(c+{1\over2})^2\over t^2}\bigg)\varphi_{_{L_{(ir)}}}^{c(+)}(\overline{y}_{_{(\pm\pm)}}^{c(n)},t)=0\;.
\label{lemma1d}
\end{eqnarray}
Using the first equation in Eq(\ref{lemma1c}) and that in Eq(\ref{lemma1d}), we find
\begin{eqnarray}
&&\bigg(\Big[y_{_{(\pm\pm)}}^{c(n)}\Big]^2-\Big[\overline{y}_{_{(\pm\pm)}}^{c(n)}\Big]^2\bigg)
\int_\epsilon^1dtt\varphi_{_{L_{(ir)}}}^{c(+)}(y_{_{(\pm\pm)}}^{c(n)},t)
\varphi_{_{L_{(ir)}}}^{c(+)}(\overline{y}_{_{(\pm\pm)}}^{c(n)},t)
\nonumber\\
&&\hspace{-0.6cm}=
\bigg\{t\varphi_{_{L_{(ir)}}}^{c(+)}(y_{_{(\pm\pm)}}^{c(n)},t){\partial\varphi_{_{L_{(ir)}}}^{c(+)}\over\partial t}
(\overline{y}_{_{(\pm\pm)}}^{c(n)},t)-t\varphi_{_{L_{(ir)}}}^{c(+)}(\overline{y}_{_{(\pm\pm)}}^{c(n)},t)
{\partial\varphi_{_{L_{(ir)}}}^{c(+)}\over\partial t}(y_{_{(\pm\pm)}}^{c(n)},t)\bigg\}_\epsilon^1
\nonumber\\
&&\hspace{-0.6cm}=
\bigg\{t\overline{y}_{_{(\pm\pm)}}^{c(n)}\varphi_{_{L_{(ir)}}}^{c(+)}(y_{_{(\pm\pm)}}^{c(n)},t)
\varphi_{_{R_{(ir)}}}^{c(-)}
(\overline{y}_{_{(\pm\pm)}}^{c(n)},t)-t\overline{y}_{_{(\pm\pm)}}^{c(n)}
\varphi_{_{L_{(ir)}}}^{c(+)}(\overline{y}_{_{(\pm\pm)}}^{c(n)},t)
\varphi_{_{R_{(ir)}}}^{c(-)}(y_{_{(\pm\pm)}}^{c(n)},t)\bigg\}_\epsilon^1\;.
\label{lemma1e}
\end{eqnarray}
Similarly, we also obtain
\begin{eqnarray}
&&\bigg(\Big[y_{_{(\pm\pm)}}^{c(n)}\Big]^2-\Big[\overline{y}_{_{(\pm\pm)}}^{c(n)}\Big]^2\bigg)
\int_\epsilon^1dtt\varphi_{_{R_{(ir)}}}^{c(-)}(y_{_{(\pm\pm)}}^{c(n)},t)
\varphi_{_{R_{(ir)}}}^{c(-)}(\overline{y}_{_{(\pm\pm)}}^{c(n)},t)
\nonumber\\
&&\hspace{-0.6cm}=
\bigg\{-t\overline{y}_{_{(\pm\pm)}}^{c(n)}\varphi_{_{R_{(ir)}}}^{c(-)}(y_{_{(\pm\pm)}}^{c(n)},t)
\varphi_{_{L_{(ir)}}}^{c(+)}(\overline{y}_{_{(\pm\pm)}}^{c(n)},t)+t\overline{y}_{_{(\pm\pm)}}^{c(n)}
\varphi_{_{R_{(ir)}}}^{c(-)}(\overline{y}_{_{(\pm\pm)}}^{c(n)},t)
\varphi_{_{L_{(ir)}}}^{c(+)}(y_{_{(\pm\pm)}}^{c(n)},t)\bigg\}_\epsilon^1\;.
\label{lemma1f}
\end{eqnarray}
Since $t\in(\epsilon,1)$ is real, $\varphi_{_{L_{(ir)}}}^{c(+)}(\overline{y}_{_{(\pm\pm)}}^{c(n)},t)
=\Big[\varphi_{_{L_{(ir)}}}^{c(+)}(y_{_{(\pm\pm)}}^{c(n)},t)\Big]^*$,
$\varphi_{_{R_{(ir)}}}^{c(-)}(\overline{y}_{_{(\pm\pm)}}^{c(n)},t)
=\Big[\varphi_{_{R_{(ir)}}}^{c(-)}(y_{_{(\pm\pm)}}^{c(n)},t)\Big]^*$, then
\begin{eqnarray}
&&\int_\epsilon^1dtt\varphi_{_{L_{(ir)}}}^{c(+)}(y_{_{(\pm\pm)}}^{c(n)},t)
\varphi_{_{L_{(ir)}}}^{c(+)}(\overline{y}_{_{(\pm\pm)}}^{c(n)},t)
=\int_\epsilon^1dtt\Big|\varphi_{_{L_{(ir)}}}^{c(+)}(y_{_{(\pm\pm)}}^{c(n)},t)\Big|^2>0\;,
\nonumber\\
&&\int_\epsilon^1dtt\varphi_{_{R_{(ir)}}}^{c(-)}(y_{_{(\pm\pm)}}^{c(n)},t)
\varphi_{_{R_{(ir)}}}^{c(-)}(\overline{y}_{_{(\pm\pm)}}^{c(n)},t)
=\int_\epsilon^1dtt\Big|\varphi_{_{R_{(ir)}}}^{c(-)}(y_{_{(\pm\pm)}}^{c(n)},t)\Big|^2>0
\label{lemma1g}
\end{eqnarray}
for the nontrivial functions $\varphi_{_{L_{(ir)}}}^{c(+)}(y_{_{(\pm\pm)}}^{c(n)},t),
\varphi_{_{R_{(ir)}}}^{c(-)}(y_{_{(\pm\pm)}}^{c(n)},t)$. If $r\neq0$, then
$\Big[y_{_{(\pm\pm)}}^{c(n)}\Big]^2-\Big[\overline{y}_{_{(\pm\pm)}}^{c(n)}\Big]^2=i4rs\neq0$.
Applying $(--)$ BCs satisfied by the bulk profiles of right-handed fermions, we
derive the following equations from Eq(\ref{lemma1e}) and Eq(\ref{lemma1f}):
\begin{eqnarray}
&&\int_\epsilon^1dtt\Big|\varphi_{_{L_{(ir)}}}^{c(+)}(y_{_{(\pm\pm)}}^{c(n)},t)\Big|^2=0\;,
\nonumber\\
&&\int_\epsilon^1dtt\Big|\varphi_{_{R_{(ir)}}}^{c(-)}(y_{_{(\pm\pm)}}^{c(n)},t)\Big|^2=0\;,
\label{lemma1h}
\end{eqnarray}
which are contrary to the inequalities in Eq(\ref{lemma1g}). For $r=0$,
\begin{eqnarray}
&&\int_\epsilon^1dtt\Big|\varphi_{_{L_{(ir)}}}^{c(+)}(y_{_{(\pm\pm)}}^{c(n)},t)\Big|^2
\nonumber\\
&&\hspace{-0.6cm}=
\lim\limits_{r\rightarrow0}{1\over i4rs}
\bigg\{t\overline{y}_{_{(\pm\pm)}}^{c(n)}\varphi_{_{L_{(ir)}}}^{c(+)}(y_{_{(\pm\pm)}}^{c(n)},t)
\varphi_{_{R_{(ir)}}}^{c(-)}(\overline{y}_{_{(\pm\pm)}}^{c(n)},t)-t\overline{y}_{_{(\pm\pm)}}^{c(n)}
\varphi_{_{L_{(ir)}}}^{c(+)}(\overline{y}_{_{(\pm\pm)}}^{c(n)},t)
\varphi_{_{R_{(ir)}}}^{c(-)}(y_{_{(\pm\pm)}}^{c(n)},t)\bigg\}_\epsilon^1
\nonumber\\
&&\hspace{-0.6cm}=
{1\over i4s}\bigg\{t\varphi_{_{L_{(ir)}}}^{c(+)}(y_{_{(\pm\pm)}}^{c(n)},t)\varphi_{_{R_{(ir)}}}^{c(-)}
(\overline{y}_{_{(\pm\pm)}}^{c(n)},t)+{\overline{y}_{_{(\pm\pm)}}^{c(n)}\over y_{_{(\pm\pm)}}^{c(n)}}
t^2{\partial\varphi_{_{L_{(ir)}}}^{c(+)}\over\partial t}(y_{_{(\pm\pm)}}^{c(n)},t)\varphi_{_{R_{(ir)}}}^{c(-)}
(\overline{y}_{_{(\pm\pm)}}^{c(n)},t)
\nonumber\\&&
+t^2\varphi_{_{L_{(ir)}}}^{c(+)}(y_{_{(\pm\pm)}}^{c(n)},t){\partial\varphi_{_{R_{(ir)}}}^{c(-)}\over\partial t}
(\overline{y}_{_{(\pm\pm)}}^{c(n)},t)-t\varphi_{_{L_{(ir)}}}^{c(+)}(\overline{y}_{_{(\pm\pm)}}^{c(n)},t)
\varphi_{_{R_{(ir)}}}^{c(-)}(y_{_{(\pm\pm)}}^{c(n)},t)
\nonumber\\&&
-{y_{_{(\pm\pm)}}^{c(n)}\over \overline{y}_{_{(\pm\pm)}}^{c(n)}}
t^2{\partial\varphi_{_{L_{(ir)}}}^{c(+)}\over\partial t}(\overline{y}_{_{(\pm\pm)}}^{c(n)},t)\varphi_{_{R_{(ir)}}}^{c(-)}
(y_{_{(\pm\pm)}}^{c(n)},t)-t^2\varphi_{_{L_{(ir)}}}^{c(+)}(\overline{y}_{_{(\pm\pm)}}^{c(n)},t)
{\partial\varphi_{_{R_{(ir)}}}^{c(-)}\over\partial t}(y_{_{(\pm\pm)}}^{c(n)},t)
\bigg\}_\epsilon^1
\nonumber\\
&&\hspace{-0.6cm}=
{1\over2}\Big\{\Big|\varphi_{_{L_{(ir)}}}^{c(+)}(y_{_{(\pm\pm)}}^{c(n)},1)\Big|^2
-\epsilon^2\Big|\varphi_{_{L_{(ir)}}}^{c(+)}(y_{_{(\pm\pm)}}^{c(n)},\epsilon)\Big|^2\Big\}\;.
\label{lemma1i}
\end{eqnarray}
In the last step, we apply the equation of motion for the bulk profiles of fermions
in Eq(\ref{EM-fermion}) and $(--)$ BCs satisfied by the bulk files of
right-handed fermions. Similarly, we can derive
\begin{eqnarray}
&&\int_\epsilon^1dtt\Big|\varphi_{_{R_{(ir)}}}^{c(-)}(y_{_{(\pm\pm)}}^{c(n)},t)\Big|^2
=-{1\over2}\Big\{\Big|\varphi_{_{L_{(ir)}}}^{c(+)}(y_{_{(\pm\pm)}}^{c(n)},1)\Big|^2
-\epsilon^2\Big|\varphi_{_{L_{(ir)}}}^{c(+)}(y_{_{(\pm\pm)}}^{c(n)},\epsilon)\Big|^2\Big\}\;.
\label{lemma1j}
\end{eqnarray}
Eq(\ref{lemma1i}) and Eq(\ref{lemma1j}) are also contrary to the inequalities in Eq(\ref{lemma1g}).
In other words, $R_{_{(\pm\pm)}}^{c,\epsilon}(z)=0$ only has the real roots.
Analogously, we can verify that the equations $R_{_{(\pm\mp)}}^{c,\epsilon}(z)=0$,
$R_{_{(\mp\pm)}}^{c,\epsilon}(z)=0$, $R_{_{(\mp\mp)}}^{c,\epsilon}(z)=0$,
$R_{_{(++)}}^{G,\epsilon}(z)=0$ and $R_{_{(-+)}}^{G,\epsilon}(z)=0$ only have
real roots.

{\it Lemma2: If $y_{_{\rm(BCs)}}^{c(n)}(n=1,\;2,\;\cdots,\;\infty)$ satisfy
$R_{_{\rm(BCs)}}^{c,\epsilon}(y_{_{\rm(BCs)}}^{c(n)})=0$, then
$y_{_{\rm(BCs)}}^{c(-n)}=-y_{_{\rm(BCs)}}^{c(n)}$ and $\pm y_{_{\rm(BCs)}}^{c(1)},\;
\cdots,\;\pm y_{_{\rm(BCs)}}^{c(n)},\;\cdots$ are the zeros of order one of the function
$R_{_{\rm(BCs)}}^{c,\epsilon}(z)$.}

Proof. Assuming ${\rm(BCs)}=(\pm\pm)$, we firstly demonstrate
that $y_{_{(\pm\pm)}}^{c(1)},\;\cdots,\;y_{_{(\pm\pm)}}^{c(n)},\;\cdots$ are
the zeros of order one of the function $R_{_{(\pm\pm)}}^{c,\epsilon}(z)$.
For $c\neq N+{1\over2}$
\begin{eqnarray}
&&{dR_{_{(\pm\pm)}}^{c,\epsilon}\over dz}(z)=\Big[J_{-c-{1\over2}}(z)J_{c-{1\over2}}(z\epsilon)
+J_{c+{1\over2}}(z)J_{-c+{1\over2}}(z\epsilon)\Big]
\nonumber\\
&&\hspace{2.3cm}
-\epsilon\Big[J_{-c+{1\over2}}(z)J_{c+{1\over2}}(z\epsilon)+J_{c-{1\over2}}(z)
J_{-c-{1\over2}}(z\epsilon)\Big]
\nonumber\\
&&\hspace{2.3cm}
+{2c-1\over z}\Big[J_{-c+{1\over2}}(z)J_{c-{1\over2}}(z\epsilon)
-J_{c-{1\over2}}(z)J_{-c+{1\over2}}(z\epsilon)\Big]
\nonumber\\
&&\hspace{2.0cm}=
\varphi_{_{R_{(ir)}}}^{c(+)}(z,\epsilon)-\epsilon\varphi_{_{L_{(ir)}}}^{c(+)}(z,\epsilon)
+{2c-1\over z}\varphi_{_{R_{(ir)}}}^{c(-)}(z,\epsilon)\;.
\label{lemma2e}
\end{eqnarray}
As $c=N+1/2$,
\begin{eqnarray}
&&{dR_{_{(\pm\pm)}}^{c,\epsilon}\over dz}(z)=-\Big[
Y_{N+1}(z)J_{N}(z\epsilon)-J_{N+1}(z)Y_{N}(z\epsilon)\Big]
\nonumber\\
&&\hspace{2.3cm}
-\epsilon\Big[
Y_{N}(z)J_{N+1}(z\epsilon)-J_{N}(z)Y_{N+1}(z\epsilon)\Big]
\nonumber\\
&&\hspace{2.3cm}
+{2N\over z}\Big[Y_{N}(z)J_{N}(z\epsilon)-J_{N}(z)Y_{N}(z\epsilon)\Big]
\nonumber\\
&&\hspace{2.0cm}=
\varphi_{_{R_{(ir)}}}^{c(+)}(z,\epsilon)-\epsilon\varphi_{_{L_{(ir)}}}^{c(+)}(z,\epsilon)
+{2c-1\over z}\varphi_{_{R_{(ir)}}}^{c(-)}(z,\epsilon)\;.
\label{lemma2f}
\end{eqnarray}
When $y_{_{(\pm\pm)}}^{c(n)}(n=1,\;2,\;\cdots,\;\infty)$ satisfy the equation
$R_{_{(\pm\pm)}}^{c,\epsilon}(y_{_{(\pm\pm)}}^{c(n)})=0$, then
\begin{eqnarray}
&&{dR_{_{(\pm\pm)}}^{c,\epsilon}\over dz}(z)\bigg|_{z=y_{_{(\pm\pm)}}^{c(n)}}=
\varphi_{_{R_{(ir)}}}^{c(+)}(y_{_{(\pm\pm)}}^{c(n)},\epsilon)-\epsilon\varphi_{_{L_{(ir)}}}^{c(+)}
(y_{_{(\pm\pm)}}^{c(n)},\epsilon)\neq0\;.
\label{lemma2g}
\end{eqnarray}
In other words, $y_{_{(\pm\pm)}}^{c(n)}(n=1,\;2,\;\cdots,\;\infty)$ are the zeros of order one
for the function $R_{_{(\pm\pm)}}^{c,\epsilon}(z)$. Using concrete expressions
of the Bessel functions $J_{\nu}$ and $Y_{_\nu}$, we can verify directly $R_{_{(\pm\pm)}}^{c,\epsilon}(
-y_{_{(\pm\pm)}}^{c(n)})=0$ if $R_{_{(\pm\pm)}}^{c,\epsilon}(y_{_{(\pm\pm)}}^{c(n)})=0$.
Furthermore, we can obtain those similar results on the zeros of the functions
$R_{_{(\pm\mp)}}^{c,\epsilon}(z)$, $R_{_{(\mp\pm)}}^{c,\epsilon}(z)$, $R_{_{(\mp\mp)}}^{c,\epsilon}(z)$,
$R_{_{(++)}}^{G,\epsilon}(z)$ and $R_{_{(-+)}}^{G,\epsilon}(z)$.

When $z\rightarrow0$ and $c\neq N+{1\over2}$, we expand the function $R_{_{(\pm\pm)}}^{c,\epsilon}(z)$
according $z$ as
\begin{eqnarray}
&&R_{_{(\pm\pm)}}^{c,\epsilon}(z)={2(\epsilon^{c-{1\over2}}-\epsilon^{{1\over2}-c})\over(1-2c)
\Gamma({1\over2}-c)\Gamma({1\over2}+c)}\Big\{1+{\cal O}(z^2)\Big\}\;.
\label{lemma2b}
\end{eqnarray}
When $z\rightarrow0$ and $c=N+{1\over2}$, the function $R_{_{(\pm\pm)}}^{c,\epsilon}(z)$
is approximated as
\begin{eqnarray}
&&R_{_{(\pm\pm)}}^{c,\epsilon}(z)=\left\{\begin{array}{ll}
{1-\epsilon^{2N}\over N\pi\epsilon^N}\Big\{1+{\cal O}(z^2)\Big\}, & N\neq 0\\
-{2\ln\epsilon\over\pi}\Big\{1+{\cal O}(z^2)\Big\}, & N=0\end{array}\right.\;.
\label{lemma2c}
\end{eqnarray}
In other words, $z=0$ is not the zero of $R_{_{(\pm\pm)}}^{c,\epsilon}(z)$. Similarly,
we find that $z=0$ is not the zero of the functions $R_{_{(\mp\mp)}}^{c,\epsilon}(z)$ as well as
$R_{_{(++)}}^{G,\epsilon}(z)$ also, and is the pole of order one of the functions
$R_{_{(\pm\mp)}}^{c,\epsilon}(z)$, $R_{_{(\mp\pm)}}^{c,\epsilon}(z)$ together with
$R_{_{(-+)}}^{G,\epsilon}(z)$.

{\it Lemma3: Let function $f(z)$ be analytic except for limited in number isolated
singularities on the complex plane. If there are two constants ${\cal M}>0$ and ${\cal R}>0$,
we have $|zf(z)|\leq{\cal M}$ when $|z|>{\cal R}$. Then
\begin{eqnarray}
&&\lim\limits_{n\rightarrow\infty}\oint_{C_{_{r_{(n)}}}}\Big\{{2\over z}
+{1\over R_{_{\rm(BCs)}}^{c,\epsilon}(z)}{dR_{_{\rm(BCs)}}^{c,\epsilon}\over dz}(z)
\Big\}f(z)dz=0\;,
\label{lemma3a}
\end{eqnarray}
where the path $C_{_{r_{(n)}}}$ is the rectangular contour with four vertices
$(1\pm i)r_{(n)}$ and $(-1\pm i)r_{(n)}$ with
$y_{_{\rm(BCs)}}^{c(n)}<r_{(n)}<y_{_{\rm(BCs)}}^{c(n+1)}$.

Proof. Firstly, we illustrate how to demonstrate the lemma for the case ${\rm(BCs)}=(\pm\pm)$.
Since $|zf(z)|\leq{\cal M}$ when $|z|>{\cal R}$, all singularities of the function $f(z)$
all distribute within the region $|z|\le{\cal R}$. This implies that $zf(z)$ is analytic at
$z=\infty$:
\begin{eqnarray}
&&zf(z)=a_0+{a_1\over z}+{a_2\over z^2}+\cdots\;,\;\;|z|>{\cal R}\;,
\label{lemma3b}
\end{eqnarray}
or
\begin{eqnarray}
&&f(z)={a_0\over z}+{a_1\over z^2}+{a_2\over z^3}+\cdots\;,\;\;|z|>{\cal R}\;.
\label{lemma3c}
\end{eqnarray}
Applying residue theorem, we have
\begin{eqnarray}
&&\oint_{C_{_{r_{(n)}}}}\Big\{{2\over z^2}+{1\over zR_{_{(\pm\pm)}}^{c,\epsilon}(z)}
{dR_{_{(\pm\pm)}}^{c,\epsilon}\over dz}(z)\Big\}dz
\nonumber\\
&&\hspace{-0.6cm}=
i2\pi\Big\{{\rm Res}\Big({2\over z^2}+{1\over zR_{_{(\pm\pm)}}^{c,\epsilon}(z)}
{dR_{_{(\pm\pm)}}^{c,\epsilon}\over dz}(z),z=0\Big)
\nonumber\\&&
+\sum\limits_{i=1}^n\Big[{\rm Res}\Big({1\over zR_{_{(\pm\pm)}}^{c,\epsilon}(z)}
{dR_{_{(\pm\pm)}}^{c,\epsilon}\over dz}(z),z=y_{_{(\pm\pm)}}^{c(i)}\Big)
\nonumber\\&&
+{\rm Res}\Big({1\over zR_{_{(\pm\pm)}}^{c,\epsilon}(z)}
{dR_{_{(\pm\pm)}}^{c,\epsilon}\over dz}(z),z=-y_{_{(\pm\pm)}}^{c(i)}\Big)\Big]\Big\}
\label{lemma3d}
\end{eqnarray}
Because $2/z^2+(dR_{_{(\pm\pm)}}^{c,\epsilon}(z)/dz)/(zR_{_{(\pm\pm)}}^{c,\epsilon}(z))$
is even function of $z$, its Laurent series at the point $z=0$ does not contain
the term which is proportional to $1/z$. One directly has
\begin{eqnarray}
&&\oint_{C_{_{r_{(n)}}}}\Big\{{2\over z^2}+{1\over zR_{_{(\pm\pm)}}^{c,\epsilon}(z)}
{dR_{_{(\pm\pm)}}^{c,\epsilon}\over dz}(z)\Big\}dz
\nonumber\\
&&\hspace{-0.6cm}=
i2\pi\sum\limits_{i=1}^n\Big[{\rm Res}\Big({1\over zR_{_{(\pm\pm)}}^{c,\epsilon}(z)}
{dR_{_{(\pm\pm)}}^{c,\epsilon}\over dz}(z),z=y_{_{(\pm\pm)}}^{c(i)}\Big)
\nonumber\\&&
+{\rm Res}\Big({1\over zR_{_{(\pm\pm)}}^{c,\epsilon}(z)}
{dR_{_{(\pm\pm)}}^{c,\epsilon}\over dz}(z),z=-y_{_{(\pm\pm)}}^{c(i)}\Big)\Big]
\nonumber\\
&&\hspace{-0.6cm}=
i2\pi\sum\limits_{i=1}^n\Big[{1\over y_{_{(\pm\pm)}}^{c(i)}}
+{1\over -y_{_{(\pm\pm)}}^{c(i)}}\Big]=0\;,
\label{lemma3e}
\end{eqnarray}
then gets
\begin{eqnarray}
&&\oint_{C_{_{r_{(n)}}}}\Big\{{2\over z}+{1\over R_{_{(\pm\pm)}}^{c,\epsilon}(z)}
{dR_{_{(\pm\pm)}}^{c,\epsilon}\over dz}(z)\Big\}f(z)dz
\nonumber\\
&&\hspace{-0.6cm}=
\oint_{C_{_{r_{(n)}}}}\Big\{{2\over z}+{1\over R_{_{(\pm\pm)}}^{c,\epsilon}(z)}
{dR_{_{(\pm\pm)}}^{c,\epsilon}\over dz}(z)\Big\}\Big\{f(z)-{a_0\over z}\Big\}dz\;.
\label{lemma3f}
\end{eqnarray}
For $f(z)-a_0/z=1/z^2\Big\{a_1+a_2/z+\cdots+a_{n+1}/z^n+\cdots\Big\}$,
$a_1+a_2/z+\cdots+a_{n+1}/z^n+\cdots$ is the analytic function and its absolute
value has an upper limit in the region $|z|\ge{\cal R}^\prime,\;{\cal R}^\prime>{\cal R}$.
Assuming
\begin{eqnarray}
&&\Big|a_1+{a_2\over z}+\cdots+{a_{n+1}\over z^n}+\cdots\Big|\le {\cal M}^\prime,\;
as\;|z|\ge{\cal R}^\prime\;,
\label{lemma3g}
\end{eqnarray}
then one gets
\begin{eqnarray}
&&\Big|f(z)-{a_0\over z}\Big|\le {{\cal M}^\prime\over|z|^2},\;
as\;|z|\ge{\cal R}^\prime\;.
\label{lemma3h}
\end{eqnarray}
As $n$ is sufficiently large, we have $|z|\ge{\cal R}^\prime$ for $z\in
C_{_{r_{(n)}}}$, then get
\begin{eqnarray}
&&\bigg|\oint_{C_{_{r_{(n)}}}}\Big\{{2\over z}+{1\over R_{_{(\pm\pm)}}^{c,\epsilon}(z)}
{dR_{_{(\pm\pm)}}^{c,\epsilon}\over dz}(z)\Big\}\Big\{f(z)-{a_0\over z}\Big\}dz\bigg|
\nonumber\\
&&\hspace{-0.6cm}
\le{4r_{(n)}{\cal M}^\prime\over r_{(n)}^2}\times\Big\{the\;upper\;bound\;of\;
\Big|{1\over R_{_{(\pm\pm)}}^{c,\epsilon}(z)}{dR_{_{(\pm\pm)}}^{c,\epsilon}\over dz}(z)\Big|
\;for\;z\in C_{_{r_{(n)}}}\Big\}\;.
\label{lemma3i}
\end{eqnarray}
In order to obtain the upper bound of $\Big|(dR_{_{(\pm\pm)}}^{c,\epsilon}(z)/dz)
/R_{_{(\pm\pm)}}^{c,\epsilon}(z)\Big|$ for $z\in C_{_{r_{(n)}}}$, we express the
Bessel functions as \cite{Wangzx}
\begin{eqnarray}
&&J_\nu(z)={1\over\sqrt{2\pi z}}\Big\{\Big[e^{i(z-{\nu\pi\over2}-{\pi\over4})}
+e^{-i(z-{\nu\pi\over2}-{\pi\over4})}\Big]\Big(1+{\cal O}({1\over z^2})\Big)
\nonumber\\
&&\hspace{1.6cm}
+{i\over2\nu}(\nu^2-{1\over4})\Big[e^{i(z-{\nu\pi\over2}-{\pi\over4})}
-e^{-i(z-{\nu\pi\over2}-{\pi\over4})}\Big]\Big(1+{\cal O}({1\over z^2})\Big)\Big\}
\;,\nonumber\\
&&Y_\nu(z)={1\over\sqrt{2\pi z}}\Big\{-i\Big[e^{i(z-{\nu\pi\over2}-{\pi\over4})}
-e^{-i(z-{\nu\pi\over2}-{\pi\over4})}\Big]\Big(1+{\cal O}({1\over z^2})\Big)
\nonumber\\
&&\hspace{1.6cm}
+{1\over2\nu}(\nu^2-{1\over4})\Big[e^{i(z-{\nu\pi\over2}-{\pi\over4})}
+e^{-i(z-{\nu\pi\over2}-{\pi\over4})}\Big]\Big(1+{\cal O}({1\over z^2})\Big)\Big\}
\label{lemma3j}
\end{eqnarray}
for $|z|\rightarrow\infty$. Using the above equations, we approximate
$R_{_{(\pm\pm)}}^{c,\epsilon}(z)$ and $dR_{_{(\pm\pm)}}^{c,\epsilon}(z)/dz$ as
\begin{eqnarray}
&&R_{_{(\pm\pm)}}^{c,\epsilon}(z)=-{i\cos c\pi\over\pi\sqrt{\epsilon}z}
\Big\{\Big[e^{i(1-\epsilon)z}-e^{-i(1-\epsilon)z}\Big]\Big(1+{\cal O}({1\over z^2})\Big)
\nonumber\\
&&\hspace{2.1cm}
+{ic(c-1)\over2z}\Big({1\over\epsilon}-1\Big)\Big[e^{i(1-\epsilon)z}
+e^{-i(1-\epsilon)z}\Big]\Big(1+{\cal O}({1\over z^2})\Big)\Big\}
\;,\nonumber\\
&&{dR_{_{(\pm\pm)}}^{c,\epsilon}\over dz}(z)={\cos c\pi\over\pi\sqrt{\epsilon}z}
\Big\{(1-\epsilon)\Big[e^{i(1-\epsilon)z}+e^{-i(1-\epsilon)z}\Big]\Big(1+{\cal O}({1\over z^2})\Big)
+{i\over z}\Big[1
\nonumber\\
&&\hspace{2.3cm}
+\Big(1-{(1+\epsilon^2)\over2\epsilon}\Big)c(c-1)\Big]\Big[e^{i(1-\epsilon)z}
-e^{-i(1-\epsilon)z}\Big]\Big(1+{\cal O}({1\over z^2})\Big)\Big\}
\label{lemma3k}
\end{eqnarray}
for $c\neq N+1/2$ and $|z|\rightarrow\infty$. When $c=N+1/2$, the functions
$R_{_{(\pm\pm)}}^{c,\epsilon}(z)$ and $dR_{_{(\pm\pm)}}^{c,\epsilon}(z)/dz$ at
$|z|\rightarrow\infty$ can be similarly approximated as
\begin{eqnarray}
&&R_{_{(\pm\pm)}}^{c,\epsilon}(z)=-{i\over\pi\sqrt{\epsilon}z}
\Big\{\Big[e^{i(1-\epsilon)z}-e^{-i(1-\epsilon)z}\Big]\Big(1+{\cal O}({1\over z^2})\Big)
\nonumber\\
&&\hspace{2.1cm}
+{i\over2z}\Big(N^2-{1\over4}\Big)\Big({1\over\epsilon}-1\Big)\Big[e^{i(1-\epsilon)z}
+e^{-i(1-\epsilon)z}\Big]\Big(1+{\cal O}({1\over z^2})\Big)\Big\}
\;,\nonumber\\
&&{dR_{_{(\pm\pm)}}^{c,\epsilon}\over dz}(z)={1\over\pi\sqrt{\epsilon}z}
\Big\{(1-\epsilon)\Big[e^{i(1-\epsilon)z}+e^{-i(1-\epsilon)z}\Big]\Big(1+{\cal O}({1\over z^2})\Big)
+{i\over z}\Big[1
\nonumber\\
&&\hspace{2.3cm}
+\Big(1-{(1+\epsilon^2)\over2\epsilon}\Big)(N^2-{1\over4})\Big]\Big[e^{i(1-\epsilon)z}
-e^{-i(1-\epsilon)z}\Big]\Big(1+{\cal O}({1\over z^2})\Big)\Big\}\;.
\label{lemma3l}
\end{eqnarray}
Using Eq(\ref{lemma3k}) and Eq(\ref{lemma3l}), one obtains
\begin{eqnarray}
&&{1\over R_{_{(\pm\pm)}}^{c,\epsilon}(z)}{dR_{_{(\pm\pm)}}^{c,\epsilon}\over dz}(z)
\nonumber\\
&&\hspace{-0.6cm}=
i(1-\epsilon){e^{i(1-\epsilon)z}+e^{-i(1-\epsilon)z}
+{i\over z}\Big[{1\over1-\epsilon}-{(1-\epsilon)c(c-1)\over2\epsilon}\Big]\Big[e^{i(1-\epsilon)z}
-e^{-i(1-\epsilon)z}\Big]\over e^{i(1-\epsilon)z}-e^{-i(1-\epsilon)z}
+{ic(c-1)\over2z}\Big({1\over\epsilon}-1\Big)\Big[e^{i(1-\epsilon)z}
+e^{-i(1-\epsilon)z}\Big]}\Big[1+{\cal O}({1\over z^2})\Big]\;.
\label{lemma3m}
\end{eqnarray}
If $n$ is sufficiently large, $y_{_{(\pm\pm)}}^{c(n)}$ is approximately
given by \cite{Buras5}
\begin{eqnarray}
&&y_{_{(\pm\pm)}}^{c(n)}\simeq\Big[n+{1\over2}\Big(\Big|c+{1\over2}\Big|-1\Big)-{1\over4}\Big]\pi\;.
\label{lemma3n}
\end{eqnarray}
The fact implies that the interval between $y_{_{(\pm\pm)}}^{c(n)}$ and $y_{_{(\pm\pm)}}^{c(n+1)}$
is about $\pi$ as $n\gg1$. When
\begin{eqnarray}
&&y_{_{(\pm\pm)}}^{c(n)}<{\pi\over1-\epsilon}\Big(N_0+{1\over4}\Big)
<{\pi\over1-\epsilon}\Big(N_0+{3\over4}\Big)\leq y_{_{(\pm\pm)}}^{c(n+1)}
\label{lemma3o}
\end{eqnarray}
or
\begin{eqnarray}
&&y_{_{(\pm\pm)}}^{c(n)}<{\pi\over1-\epsilon}\Big(N_0+{1\over4}\Big)
< y_{_{(\pm\pm)}}^{c(n+1)}\leq{\pi\over1-\epsilon}\Big(N_0+{3\over4}\Big)
\label{lemma3o1}
\end{eqnarray}
where the positive integer $N_0$ obviously turns large along with increasing of the number $n$,
one can choose
\begin{eqnarray}
&&r_{(n)}={\pi\over1-\epsilon}\Big(N_0+{1\over4}\Big)\;.
\label{lemma3p}
\end{eqnarray}
When the point $z$ belongs to the left- and right- borders of $C_{_{r_{(n)}}}$, i.e.
$z=\mp\Big(N_0+1/4\Big)\pi/(1-\epsilon)+iy$ with
$-\Big(N_0+1/4\Big)\pi/(1-\epsilon)\leq y\leq\Big(N_0+1/4\Big)\pi/(1-\epsilon)$,
we have
\begin{eqnarray}
&&\bigg|{1\over R_{_{(\pm\pm)}}^{c,\epsilon}(z)}{dR_{_{(\pm\pm)}}^{c,\epsilon}\over dz}(z)\bigg|
\nonumber\\
&&\hspace{-0.6cm}\leq
(1-\epsilon){\Big|e^{i(1-\epsilon)z}+e^{-i(1-\epsilon)z}\Big|
+{1\over|z|}\Big|{1\over1-\epsilon}-{(1-\epsilon)c(c-1)\over2\epsilon}\Big|\Big|e^{i(1-\epsilon)z}
-e^{-i(1-\epsilon)z}\Big|\over\Big|e^{i(1-\epsilon)z}-e^{-i(1-\epsilon)z}\Big|
-{|c(c-1)|\over2|z|}\Big({1\over\epsilon}-1\Big)\Big|e^{i(1-\epsilon)z}
+e^{-i(1-\epsilon)z}\Big|}\Big[1+{\cal O}({1\over|z|^2})\Big|
\nonumber\\
&&\hspace{-0.6cm}
\leq(1-\epsilon)\bigg\{1+{1\over r_{(n)}}\Big[{1\over1-\epsilon}
+|c(c-1)|\Big({1\over\epsilon}-1\Big)\Big]\Big\}
\Big[1+{\cal O}({1\over r_{(n)}^2})\Big]\;.
\label{lemma3x}
\end{eqnarray}
When the point $z$ belongs to the upper- and down- borders of $C_{_{r_{(n)}}}$, i.e.
$z=x\pm i\Big(N_0+1/4\Big)\pi/(1-\epsilon)$ with
$-\Big(N_0+1/4\Big)\pi/(1-\epsilon)\leq x\leq\Big(N_0+1/4\Big)\pi/(1-\epsilon)$,
we similarly obtain
\begin{eqnarray}
&&\bigg|{1\over R_{_{(\pm\pm)}}^{c,\epsilon}(z)}{dR_{_{(\pm\pm)}}^{c,\epsilon}\over dz}(z)\bigg|
\nonumber\\
&&\hspace{-0.6cm}\leq
(1-\epsilon){\Big|e^{i2(1-\epsilon)x}+e^{(2N_0+{1\over2})\pi}\Big|
+{1\over|z|}\Big|{1\over1-\epsilon}-{(1-\epsilon)c(c-1)\over2\epsilon}\Big|\Big|e^{i2(1-\epsilon)x}
-e^{(2N_0+{1\over2})\pi}\Big|\over\Big|e^{i2(1-\epsilon)x}-e^{(2N_0+{1\over2})\pi}\Big|
-{|c(c-1)|\over2|z|}\Big({1\over\epsilon}-1\Big)\Big|e^{i2(1-\epsilon)x}
+e^{(2N_0+{1\over2})\pi}\Big|}\Big[1+{\cal O}({1\over|z|^2})\Big|
\nonumber\\
&&\hspace{-0.6cm}\leq
(1-\epsilon)\Big\{1+{1\over|z|}\Big|{1\over1-\epsilon}-{(1-\epsilon)c(c-1)\over2\epsilon}\Big|
+{|c(c-1)|\over2|z|}\Big({1\over\epsilon}-1\Big)\Big\}
\nonumber\\
&&\hspace{0.0cm}\times
{e^{(2N_0+{1\over2})\pi}+1\over e^{(2N_0+{1\over2})\pi}
-1-{|c(c-1)|\over|z|}\Big({1\over\epsilon}-1\Big)}
\Big[1+{\cal O}({1\over |z|^2})\Big]
\nonumber\\
&&\hspace{-0.6cm}\leq
2(1-\epsilon)\bigg\{1+{1\over r_{(n)}}\Big[{1\over1-\epsilon}
+|c(c-1)|\Big({1\over\epsilon}-1\Big)\Big]\Big\}
\Big[1+{\cal O}({1\over r_{(n)}^2})\Big]\;.
\label{lemma3y}
\end{eqnarray}
As
\begin{eqnarray}
&&{\pi\over1-\epsilon}\Big(N_0+{1\over4}\Big)\leq y_{_{(\pm\pm)}}^{c(n)}
<{\pi\over1-\epsilon}\Big(N_0+{3\over4}\Big)< y_{_{(\pm\pm)}}^{c(n+1)}\;,
\label{lemma3z}
\end{eqnarray}
we choose
\begin{eqnarray}
&&r_{(n)}={\pi\over1-\epsilon}\Big(N_0+{3\over4}\Big)\;,
\label{lemma3A}
\end{eqnarray}
and similarly get the upper bounds in Eq(\ref{lemma3x}) and Eq(\ref{lemma3y}).
Applying Eq(\ref{lemma3i}), we have
\begin{eqnarray}
&&\lim\limits_{r_{(n)}\rightarrow\infty}
\bigg|\oint_{C_{_{r_{(n)}}}}\Big\{{2\over z}+{1\over R_{_{(\pm\pm)}}^{c,\epsilon}(z)}
{dR_{_{(\pm\pm)}}^{c,\epsilon}\over dz}(z)\Big\}\Big\{f(z)-{a_0\over z}\Big\}dz\bigg|
\nonumber\\
&&\hspace{-0.6cm}
\le\lim\limits_{r_{(n)}\rightarrow\infty}{4(1-\epsilon){\cal M}^\prime\over r_{(n)}}
\bigg\{1+{1\over r_{(n)}}\Big[{1\over1-\epsilon}+|c(c-1)|
\Big({1\over\epsilon}-1\Big)\Big]\bigg\}\Big[1+{\cal O}({1\over r_{(n)}^2})\Big]
\nonumber\\
&&\hspace{-0.6cm}=0\;.
\label{lemma3B}
\end{eqnarray}
In other words, the integral
\begin{eqnarray}
&&\lim\limits_{r_{(n)}\rightarrow\infty}
\oint_{C_{_{r_{(n)}}}}\Big\{{2\over z}+{1\over R_{_{(\pm\pm)}}^{c,\epsilon}(z)}
{dR_{_{(\pm\pm)}}^{c,\epsilon}\over dz}(z)\Big\}f(z)=0\;.
\label{lemma3C}
\end{eqnarray}

Using the equation in Eq(\ref{root-quark}) and Eq(\ref{lemma3j}), we have
\begin{eqnarray}
&&{1\over R_{_{(\pm\mp)}}^{c,\epsilon}(z)}{dR_{_{(\pm\mp)}}^{c,\epsilon}\over dz}(z)=
i(1-\epsilon){e^{i(1-\epsilon)z}-e^{-i(1-\epsilon)z}\over e^{i(1-\epsilon)z}+e^{-i(1-\epsilon)z}}
\Big\{1-{i\over2z}\Big[{c(c-1)\over\epsilon}
\nonumber\\
&&\hspace{3.9cm}
-c-1+{(c^2-1)\epsilon^2\over1-\epsilon}\Big]{e^{i(1-\epsilon)z}+e^{-i(1-\epsilon)z}\over
e^{i(1-\epsilon)z}-e^{-i(1-\epsilon)z}}
\nonumber\\
&&\hspace{3.9cm}
+{ic\over2z}\Big({c-1\over\epsilon}-c-1\Big){e^{i(1-\epsilon)z}-e^{-i(1-\epsilon)z}
\over e^{i(1-\epsilon)z}+e^{-i(1-\epsilon)z}}\Big\}\Big[1+{\cal O}({1\over z^2})\Big]\;,
\label{lemma3D}
\end{eqnarray}
for the case ${\rm BCs}=(\pm\mp)$. If $n$ is sufficiently large, $y_{_{(\pm\mp)}}^{c(n)}$
is approximately given by \cite{Buras5}
\begin{eqnarray}
&&y_{_{(\pm\mp)}}^{c(n)}\simeq\Big[n+{1\over2}\Big(\Big|c+{1\over2}\Big|-1\Big)+{1\over4}\Big]\pi\;.
\label{lemma3E}
\end{eqnarray}
The fact also implies that the interval between $y_{_{(\pm\mp)}}^{c(n)}$ and $y_{_{(\pm\mp)}}^{c(n+1)}$
is about $\pi$ as $n\gg1$. When
\begin{eqnarray}
&&y_{_{(\pm\pm)}}^{c(n)}<{\pi\over1-\epsilon}\Big(N_0-{1\over4}\Big)
<{\pi\over1-\epsilon}\Big(N_0+{1\over4}\Big)\leq y_{_{(\pm\pm)}}^{c(n+1)}
\label{lemma3F}
\end{eqnarray}
or
\begin{eqnarray}
&&y_{_{(\pm\pm)}}^{c(n)}<{\pi\over1-\epsilon}\Big(N_0-{1\over4}\Big)
< y_{_{(\pm\pm)}}^{c(n+1)}\leq{\pi\over1-\epsilon}\Big(N_0+{1\over4}\Big)
\label{lemma3F1}
\end{eqnarray}
where the positive integer $N_0$ obviously turns large along with increasing of the number $n$.
One can obviously choose
\begin{eqnarray}
&&r_{(n)}={\pi\over1-\epsilon}\Big(N_0-{1\over4}\Big)\;.
\label{lemma3G}
\end{eqnarray}
As
\begin{eqnarray}
&&{\pi\over1-\epsilon}\Big(N_0-{1\over4}\Big)\leq y_{_{(\pm\pm)}}^{c(n)}
<{\pi\over1-\epsilon}\Big(N_0+{1\over4}\Big)< y_{_{(\pm\pm)}}^{c(n+1)}\;,
\label{lemma3G1}
\end{eqnarray}
one can choose $r_{(n)}=(N_0+1/4)\pi/(1-\epsilon)$.
Then performing the similar analysis above, we finally get
\begin{eqnarray}
&&\lim\limits_{r_{(n)}\rightarrow\infty}
\oint_{C_{_{r_{(n)}}}}\Big\{{2\over z}+{1\over R_{_{(\pm\mp)}}^{c,\epsilon}(z)}
{dR_{_{(\pm\mp)}}^{c,\epsilon}\over dz}(z)\Big\}f(z)=0\;.
\label{lemma3H}
\end{eqnarray}
As for the function $R_{_{(\mp\pm)}}^{c,\epsilon}(z),\;R_{_{(\mp\mp)}}^{c,\epsilon}(z),\;
R_{_{(++)}}^{G,\epsilon}(z)$ and $R_{_{(-+)}}^{G,\epsilon}(z)$, we derive the similar equations.

Using the lemmas verified above and Eq(\ref{residue1}), we summarize the summing over
infinite series of KK modes as
\begin{eqnarray}
&&\sum\limits_{i=1}^\infty\Big[f(y_{_{\rm(BCs)}}^{c(i)})+f(-y_{_{\rm(BCs)}}^{c(i)})\Big]
=-{\rm Res}\Big[\Big({2\over z}+{1\over R_{_{\rm(BCs)}}^{c,\epsilon}(z)}
{dR_{_{\rm(BCs)}}^{c,\epsilon}\over dz}(z)\Big)f(z),z=0\Big]
\nonumber\\
&&\hspace{5.2cm}
-\sum\limits_{i=1}^{n_0}{\rm Res}\Big[\Big({2\over z}
+{1\over R_{_{\rm(BCs)}}^{c,\epsilon}(z)}
{dR_{_{\rm(BCs)}}^{c,\epsilon}\over dz}(z)\Big)f(z),z=z_i\Big]\;,
\nonumber\\
&&\sum\limits_{i=1}^\infty\Big[f(y_{_{\rm(BCs)}}^{G(i)})+f(-y_{_{\rm(BCs)}}^{G(i)})\Big]
=-{\rm Res}\Big[\Big({2\over z}+{1\over R_{_{\rm(BCs)}}^{G,\epsilon}(z)}
{dR_{_{\rm(BCs)}}^{G,\epsilon}\over dz}(z)\Big)f(z),z=0\Big]
\nonumber\\
&&\hspace{5.2cm}
-\sum\limits_{i=1}^{n_0}{\rm Res}\Big[\Big({2\over z}
+{1\over R_{_{\rm(BCs)}}^{G,\epsilon}(z)}
{dR_{_{\rm(BCs)}}^{G,\epsilon}\over dz}(z)\Big)f(z),z=z_i\Big]
\label{lemma3I}
\end{eqnarray}
where
\begin{eqnarray}
&&\lim\limits_{|z|\rightarrow\infty}|zf(z)|\le{\cal M},\;\;0<{\cal M}<\infty\;.
\label{lemma3J}
\end{eqnarray}
Actually, Eq(\ref{lemma3J}) is the sufficient condition to judge if
the infinite series $\sum\limits_{i=1}^\infty\Big[f(y_{_{(BCs)}}^{c(i)})+f(-y_{_{(BCs)}}^{c(i)})\Big]$
is convergent.

In extensions of the SM with a warped extra dimension, the bulk profiles in Eq.(\ref{quark-5th}) affect
amplitudes for relevant processes in terms of $\Big[\chi_{_{(BCs)}}^G(y_{_{(BCs)}}^{G(n)},\phi)\Big]
\Big[\chi_{_{(BCs)}}^G(y_{_{(BCs)}}^{G(n)},\phi^\prime)\Big]$, $\Big[f_{_{(BCs)}}^{L,c}(y_{_{(BCs)}}^{c(n)},\phi)\Big]
\Big[f_{_{(BCs)}}^{L,c}(y_{_{(BCs)}}^{c(n)},\phi^\prime)\Big]$ and $\Big[f_{_{(BCs)}}^{R,c}(y_{_{(BCs)}}^{c(n)},\phi)\Big]
\Big[f_{_{(BCs)}}^{R,c}(y_{_{(BCs)}}^{c(n)},\phi^\prime)\Big]$. In order to sum over the
infinite series of KK modes properly, one should analytically extend the above combinations
of bulk profiles to the complex plane. Here, we illustrate how to extend analytically the
combinations of bulk profiles for gauge fields satisfying $(++)$ ($(-+)$) BCs
in the complex plane. When $y=y_{_{(++)}}^{G(n)}$ satisfies the equation
$R_{_{(++)}}^{G,\epsilon}(y_{_{(++)}}^{G(n)})=0$, the combination of bulk profiles for
gauge fields satisfying $(++)$ BCs can be formulated as
\begin{eqnarray}
&&\Big[\chi_{_{(++)}}^G(y,\phi)\Big]\Big[\chi_{_{(++)}}^G(y,\phi^\prime)\Big]
={tt^\prime\Phi_{_{(uv)}}^{G(+)}(y,t)\Phi_{_{(uv)}}^{G(+)}
(y,t^\prime)\over\Big[N_{_{(uv)}}^{G(+)}(y)\Big]^2}
\nonumber\\
&&\hspace{4.6cm}
={tt^\prime\Phi_{_{(ir)}}^{G(+)}(y,t)\Phi_{_{(ir)}}^{G(+)}
(y,t^\prime)\over\Big[N_{_{(ir)}}^{G(+)}(y)\Big]^2}
\nonumber\\
&&\hspace{4.6cm}
={tt^\prime\Phi_{_{(uv)}}^{G(+)}(y,t)\Phi_{_{(ir)}}^{G(+)}
(y,t^\prime)\over\Big[N_{_{(uv)}}^{G(+)}(y)\Big]\Big[N_{_{(ir)}}^{G(+)}(y)\Big]}
\nonumber\\
&&\hspace{4.6cm}
={tt^\prime\Phi_{_{(ir)}}^{G(+)}(y,t)\Phi_{_{(uv)}}^{G(+)}
(y,t^\prime)\over\Big[N_{_{(uv)}}^{G(+)}(y)\Big]\Big[N_{_{(ir)}}^{G(+)}(y)\Big]}\;,
\label{bulkfiles-extension1}
\end{eqnarray}
with $t=\epsilon e^{\sigma(\phi)},\;t^\prime=\epsilon e^{\sigma(\phi^\prime)}$.
When $y=y_{_{(-+)}}^{G(n)}$ satisfies the equation
$R_{_{(-+)}}^{G,\epsilon}(y_{_{(-+)}}^{G(n)})=0$, the combination of bulk profiles for
gauge fields satisfying the BCs $(-+)$ can be written as
\begin{eqnarray}
&&\Big[\chi_{_{(-+)}}^G(y,\phi)\Big]\Big[\chi_{_{(-+)}}^G(y,\phi^\prime)\Big]
={tt^\prime\Phi_{_{(uv)}}^{G(-)}(y,t)\Phi_{_{(uv)}}^{G(-)}
(y,t^\prime)\over\Big[N_{_{(uv)}}^{G(-)}(y)\Big]^2}
\nonumber\\
&&\hspace{4.6cm}
={tt^\prime\Phi_{_{(ir)}}^{G(+)}(y,t)\Phi_{_{(ir)}}^{G(+)}
(y,t^\prime)\over\Big[N_{_{(ir)}}^{G(+)}(y)\Big]^2}
\nonumber\\
&&\hspace{4.6cm}
={tt^\prime\Phi_{_{(uv)}}^{G(-)}(y,t)\Phi_{_{(ir)}}^{G(+)}
(y,t^\prime)\over\Big[N_{_{(uv)}}^{G(-)}(y)\Big]\Big[N_{_{(ir)}}^{G(+)}(y)\Big]}
\nonumber\\
&&\hspace{4.6cm}
={tt^\prime\Phi_{_{(ir)}}^{G(+)}(y,t)\Phi_{_{(uv)}}^{G(-)}
(y,t^\prime)\over\Big[N_{_{(uv)}}^{G(-)}(y)\Big]\Big[N_{_{(ir)}}^{G(+)}(y)\Big]}\;.
\label{bulkfiles-extension1a}
\end{eqnarray}
The  combinations of bulk profiles for gauge fields certainly satisfy the corresponding BCs:
\begin{eqnarray}
&&{\partial\over\partial\phi_{_{(uv)}}}
\Big[\chi_{_{(++)}}^G(y,\phi)\Big]\Big[\chi_{_{(++)}}^G(y,\phi^\prime)\Big]
\bigg|_{\phi_{_{(uv)}}=0}=0\;,
\nonumber\\
&&{\partial\over\partial\phi_{_{(ir)}}}
\Big[\chi_{_{(++)}}^G(y,\phi)\Big]\Big[\chi_{_{(++)}}^G(y,\phi^\prime)\Big]
\bigg|_{\phi_{_{(ir)}}=\pi/2}=0\;,
\nonumber\\
&&\Big[\chi_{_{(-+)}}^G(y,\phi)\Big]\Big[\chi_{_{(-+)}}^G(y,\phi^\prime)\Big]
\bigg|_{\phi_{_{(uv)}}=0}=0\;,
\nonumber\\
&&{\partial\over\partial\phi_{_{(ir)}}}
\Big[\chi_{_{(-+)}}^G(y,\phi)\Big]\Big[\chi_{_{(-+)}}^G(y,\phi^\prime)\Big]
\bigg|_{\phi_{_{(ir)}}=\pi/2}=0\;,
\label{bulkfiles-extension2}
\end{eqnarray}
with $\phi_{_{(uv)}}=\min(\phi,\;\phi^\prime)$, $\phi_{_{(ir)}}=\max(\phi,\;\phi^\prime)$.
Considering Eq.(\ref{bulkfiles-extension2}), we analytically extend
the combinations of bulk profiles from Eq.(\ref{bulkfiles-extension1}) and Eq.(\ref{bulkfiles-extension1a})
in the complex plane as
\begin{eqnarray}
&&\Big[\chi_{_{(++)}}^G(z,\phi)\Big]\Big[\chi_{_{(++)}}^G(z,\phi^\prime)\Big]
={tt^\prime\over\Big[N_{_{(uv)}}^{G(+)}(z)\Big]\Big[N_{_{(ir)}}^{G(+)}(z)\Big]}
\Big\{\theta(t-t^\prime)\Phi_{_{(uv)}}^{G(+)}(z,t^\prime)\Phi_{_{(ir)}}^{G(+)}(z,t)
\nonumber\\
&&\hspace{4.8cm}
+\theta(t^\prime-t)\Phi_{_{(uv)}}^{G(+)}(z,t)\Phi_{_{(ir)}}^{G(+)}(z,t^\prime)\Big\}\;,
\nonumber\\
&&\Big[\chi_{_{(-+)}}^G(z,\phi)\Big]\Big[\chi_{_{(-+)}}^G(z,\phi^\prime)\Big]
={tt^\prime\over\Big[N_{_{(uv)}}^{G(-)}(z)\Big]\Big[N_{_{(ir)}}^{G(+)}(z)\Big]}
\Big\{\theta(t-t^\prime)\Phi_{_{(uv)}}^{G(-)}(z,t^\prime)\Phi_{_{(ir)}}^{G(+)}(z,t)
\nonumber\\
&&\hspace{4.8cm}
+\theta(t^\prime-t)\Phi_{_{(uv)}}^{G(-)}(z,t)\Phi_{_{(ir)}}^{G(+)}(z,t^\prime)\Big\}\;.
\label{bulkfiles-extension3}
\end{eqnarray}
Here, the step function $\theta(x)$ is defined as
\begin{eqnarray}
&&\theta(x)=\left\{\begin{array}{ll}1,&x>0\;;\\{1\over2},&x=0\;;\\0,&x<0\;.\end{array}\right.
\label{step-function}
\end{eqnarray}
To guarantee that the combinations of bulk profiles are
uniformly bounded in the complex plane, we analytically extend the corresponding
normalization factors in Eq.(\ref{bulkfiles-extension3}) as
\begin{eqnarray}
&&\Big|N_{_{(uv)}}^{G(+)}(z)\Big|^2={2\over kr(z^2-\bar{z}^2)}
\Big\{\bar{z}\Phi_{_{(uv)}}^{G(+)}(z,1)\Psi_{_{(uv)}}^{G(-)}(\bar{z},1)
-z\Phi_{_{(uv)}}^{G(+)}(\bar{z},1)\Psi_{_{(uv)}}^{G(-)}(z,1)\Big\}
\nonumber\\
&&\hspace{2.5cm}
+{2\over kr}\Upsilon(z)\;,
\nonumber\\
&&\Big|N_{_{(uv)}}^{G(-)}(z)\Big|^2={2\over kr(z^2-\bar{z}^2)}
\Big\{\bar{z}\Phi_{_{(uv)}}^{G(-)}(z,1)\Psi_{_{(uv)}}^{G(+)}(\bar{z},1)
-z\Phi_{_{(uv)}}^{G(-)}(\bar{z},1)\Psi_{_{(uv)}}^{G(+)}(z,1)\Big\}
\nonumber\\
&&\hspace{2.5cm}
+{2\over kr}\Upsilon(z)\;,
\nonumber\\
&&\Big|N_{_{(ir)}}^{G(+)}(z)\Big|^2={2\epsilon\over kr(z^2-\bar{z}^2)}
\Big\{z\Phi_{_{(ir)}}^{G(+)}(\bar{z},\epsilon)\Psi_{_{(ir)}}^{G(-)}(z,\epsilon)
-\bar{z}\Phi_{_{(ir)}}^{G(+)}(z,\epsilon)\Psi_{_{(ir)}}^{G(-)}(\bar{z},\epsilon)\Big\}
\nonumber\\
&&\hspace{2.5cm}
+{2\over kr}\Upsilon(z)\;.
\label{normalization-z-plane}
\end{eqnarray}
Here, $\bar{z}$ represents the conjugate of $z$, and the non-negative function $\Upsilon(z)$
is defined as
\begin{eqnarray}
&&\Upsilon(z)={1\over\pi^2|z|^2}\bigg\{(1-\epsilon)\Big(e^{-i(1-\epsilon)(z-\bar{z})}
+e^{i(1-\epsilon)(z-\bar{z})}\Big)
+{e^{-i(1-\epsilon)(z-\bar{z})}-e^{i(1-\epsilon)(z-\bar{z})}\over i(z-\bar{z})}\bigg\}.
\label{upsilon}
\end{eqnarray}
In the limit of $\bar{z}=z$ (i.e. $z$ is real), one easily gets
\begin{eqnarray}
&&\lim\limits_{\bar{z}\rightarrow z}\Upsilon(z)=0
\label{upsilon-limit}
\end{eqnarray}
and the normalization factors in Eq.(\ref{normalization-z-plane}) recover the corresponding
expressions in Eq.(\ref{normalization-gauge}). Similarly, we can analytically generalize
the normalization constants of bulk profiles for fermions as
\begin{eqnarray}
&&\Big|N_{_{L_{(ir)}}}^{c(+)}(z)\Big|^2={2\over kr(z^2-\bar{z}^2)}
\Big\{z\varphi_{_{L_{(ir)}}}^{c(+)}(\bar{z},\epsilon)\varphi_{_{R_{(ir)}}}^{c(-)}(z,\epsilon)
-\bar{z}\varphi_{_{L_{(ir)}}}^{c(+)}(z,\epsilon)\varphi_{_{R_{(ir)}}}^{c(-)}(\bar{z},\epsilon)\Big\}
\nonumber\\
&&\hspace{2.5cm}
+{2\cos^2c\pi\over kr\epsilon}\Upsilon(z)\;,
\nonumber\\
&&\Big|N_{_{L_{(uv)}}}^{c(+)}(z)\Big|^2={2\over kr\epsilon(z^2-\bar{z}^2)}
\Big\{\bar{z}\varphi_{_{L_{(uv)}}}^{c(+)}(z,1)\varphi_{_{R_{(uv)}}}^{c(-)}(\bar{z},1)
-z\varphi_{_{L_{(uv)}}}^{c(+)}(\bar{z},1)\varphi_{_{R_{(uv)}}}^{c(-)}(z,1)\Big\}
\nonumber\\
&&\hspace{2.5cm}
+{2\cos^2c\pi\over kr\epsilon}\Upsilon(z)\;,
\nonumber\\
&&\Big|N_{_{R_{(ir)}}}^{c(-)}(z)\Big|^2={2\over kr(z^2-\bar{z}^2)}
\Big\{\bar{z}\varphi_{_{L_{(ir)}}}^{c(+)}(\bar{z},\epsilon)\varphi_{_{R_{(ir)}}}^{c(-)}(z,\epsilon)
-z\varphi_{_{L_{(ir)}}}^{c(+)}(z,\epsilon)\varphi_{_{R_{(ir)}}}^{c(-)}(\bar{z},\epsilon)\Big\}
\nonumber\\
&&\hspace{2.5cm}
+{2\cos^2c\pi\over kr\epsilon}\Upsilon(z)\;,
\nonumber\\
&&\Big|N_{_{R_{(uv)}}}^{c(-)}(z)\Big|^2={2\over kr\epsilon(z^2-\bar{z}^2)}
\Big\{z\varphi_{_{L_{(uv)}}}^{c(+)}(z,1)\varphi_{_{R_{(uv)}}}^{c(-)}(\bar{z},1)
-\bar{z}\varphi_{_{L_{(uv)}}}^{c(+)}(\bar{z},1)\varphi_{_{R_{(uv)}}}^{c(-)}(z,1)\Big\}
\nonumber\\
&&\hspace{2.5cm}
+{2\cos^2c\pi\over kr\epsilon}\Upsilon(z)\;,
\nonumber\\
&&\Big|N_{_{L_{(ir)}}}^{c(-)}(z)\Big|^2={2\over kr(z^2-\bar{z}^2)}
\Big\{z\varphi_{_{L_{(ir)}}}^{c(-)}(\bar{z},\epsilon)\varphi_{_{R_{(ir)}}}^{c(+)}(z,\epsilon)
-\bar{z}\varphi_{_{L_{(ir)}}}^{c(-)}(z,\epsilon)\varphi_{_{R_{(ir)}}}^{c(+)}(\bar{z},\epsilon)\Big\}
\nonumber\\
&&\hspace{2.5cm}
+{2\cos^2c\pi\over kr\epsilon}\Upsilon(z)\;,
\nonumber\\
&&\Big|N_{_{L_{(uv)}}}^{c(-)}(z)\Big|^2={2\over kr\epsilon(z^2-\bar{z}^2)}
\Big\{\bar{z}\varphi_{_{L_{(uv)}}}^{c(-)}(z,1)\varphi_{_{R_{(uv)}}}^{c(+)}(\bar{z},1)
-z\varphi_{_{L_{(uv)}}}^{c(-)}(\bar{z},1)\varphi_{_{R_{(uv)}}}^{c(+)}(z,1)\Big\}
\nonumber\\
&&\hspace{2.5cm}
+{2\cos^2c\pi\over kr\epsilon}\Upsilon(z)\;,
\nonumber\\
&&\Big|N_{_{R_{(ir)}}}^{c(+)}(z)\Big|^2={2\over kr(z^2-\bar{z}^2)}
\Big\{\bar{z}\varphi_{_{L_{(ir)}}}^{c(-)}(\bar{z},\epsilon)\varphi_{_{R_{(ir)}}}^{c(+)}(z,\epsilon)
-z\varphi_{_{L_{(ir)}}}^{c(-)}(z,\epsilon)\varphi_{_{R_{(ir)}}}^{c(+)}(\bar{z},\epsilon)\Big\}
\nonumber\\
&&\hspace{2.5cm}
+{2\cos^2c\pi\over kr\epsilon}\Upsilon(z)\;,
\nonumber\\
&&\Big|N_{_{R_{(uv)}}}^{c(+)}(z)\Big|^2={2\over kr\epsilon(z^2-\bar{z}^2)}
\Big\{z\varphi_{_{L_{(uv)}}}^{c(-)}(z,1)\varphi_{_{R_{(uv)}}}^{c(+)}(\bar{z},1)
-\bar{z}\varphi_{_{L_{(uv)}}}^{c(-)}(\bar{z},1)\varphi_{_{R_{(uv)}}}^{c(+)}(z,1)\Big\}
\nonumber\\
&&\hspace{2.5cm}
+{2\cos^2c\pi\over kr\epsilon}\Upsilon(z)\;.
\label{normalization-z-plane1}
\end{eqnarray}
Using the above normalization constant defined in
Eq.(\ref{normalization-z-plane1}), we write the uniformly bounded combinations
of bulk profiles for fermion fields in the complex plane as
\begin{eqnarray}
&&\Big[f_{_{(++)}}^{L,c}(z,\phi)\Big]\Big[f_{_{(++)}}^{L,c}(z,\phi^\prime)\Big]
={\sqrt{tt^\prime}\over N_{_{L_{(uv)}}}^{c(+)}(z)N_{_{L_{(ir)}}}^{c(+)}(z)}
\Big\{\theta(t-t^\prime)\varphi_{_{L_{(uv)}}}^{c(+)}(z,t^\prime)\varphi_{_{L_{(ir)}}}^{c(+)}(z,t)
\nonumber\\
&&\hspace{4.8cm}
+\theta(t^\prime-t)\varphi_{_{L_{(uv)}}}^{c(+)}(z,t)\varphi_{_{L_{(ir)}}}^{c(+)}(z,t^\prime)\Big\}\;,
\nonumber\\
&&\Big[f_{_{(--)}}^{R,c}(z,\phi)\Big]\Big[f_{_{(--)}}^{R,c}(z,\phi^\prime)\Big]
={\sqrt{tt^\prime}\over N_{_{R_{(uv)}}}^{c(-)}(z)N_{_{R_{(ir)}}}^{c(-)}(z)}
\Big\{\theta(t-t^\prime)\varphi_{_{R_{(uv)}}}^{c(-)}(z,t^\prime)\varphi_{_{R_{(ir)}}}^{c(-)}(z,t)
\nonumber\\
&&\hspace{4.8cm}
+\theta(t^\prime-t)\varphi_{_{R_{(uv)}}}^{c(-)}(z,t)\varphi_{_{R_{(ir)}}}^{c(-)}(z,t^\prime)\Big\}\;.
\nonumber\\
&&\Big[f_{_{(+-)}}^{L,c}(z,\phi)\Big]\Big[f_{_{(+-)}}^{L,c}(z,\phi^\prime)\Big]
={\sqrt{tt^\prime}\over N_{_{L_{(uv)}}}^{c(+)}(z)N_{_{L_{(ir)}}}^{c(-)}(z)}
\Big\{\theta(t-t^\prime)\varphi_{_{L_{(uv)}}}^{c(+)}(z,t^\prime)\varphi_{_{L_{(ir)}}}^{c(-)}(z,t)
\nonumber\\
&&\hspace{4.8cm}
+\theta(t^\prime-t)\varphi_{_{L_{(uv)}}}^{c(+)}(z,t)\varphi_{_{L_{(ir)}}}^{c(-)}(z,t^\prime)\Big\}\;,
\nonumber\\
&&\Big[f_{_{(-+)}}^{R,c}(z,\phi)\Big]\Big[f_{_{(-+)}}^{R,c}(z,\phi^\prime)\Big]
={\sqrt{tt^\prime}\over N_{_{R_{(uv)}}}^{c(-)}(z)N_{_{R_{(ir)}}}^{c(+)}(z)}
\Big\{\theta(t-t^\prime)\varphi_{_{R_{(uv)}}}^{c(-)}(z,t^\prime)\varphi_{_{R_{(ir)}}}^{c(+)}(z,t)
\nonumber\\
&&\hspace{4.8cm}
+\theta(t^\prime-t)\varphi_{_{R_{(uv)}}}^{c(-)}(z,t)\varphi_{_{R_{(ir)}}}^{c(+)}(z,t^\prime)\Big\}\;,
\nonumber\\
&&\Big[f_{_{(-+)}}^{L,c}(z,\phi)\Big]\Big[f_{_{(-+)}}^{L,c}(z,\phi^\prime)\Big]
={\sqrt{tt^\prime}\over N_{_{L_{(uv)}}}^{c(-)}(z)N_{_{L_{(ir)}}}^{c(+)}(z)}
\Big\{\theta(t-t^\prime)\varphi_{_{L_{(uv)}}}^{c(-)}(z,t^\prime)\varphi_{_{L_{(ir)}}}^{c(+)}(z,t)
\nonumber\\
&&\hspace{4.8cm}
+\theta(t^\prime-t)\varphi_{_{L_{(uv)}}}^{c(-)}(z,t)\varphi_{_{L_{(ir)}}}^{c(+)}(z,t^\prime)\Big\}\;,
\nonumber\\
&&\Big[f_{_{(+-)}}^{R,c}(z,\phi)\Big]\Big[f_{_{(+-)}}^{R,c}(z,\phi^\prime)\Big]
={\sqrt{tt^\prime}\over N_{_{R_{(uv)}}}^{c(+)}(z)N_{_{R_{(ir)}}}^{c(-)}(z)}
\Big\{\theta(t-t^\prime)\varphi_{_{R_{(uv)}}}^{c(+)}(z,t^\prime)\varphi_{_{R_{(ir)}}}^{c(-)}(z,t)
\nonumber\\
&&\hspace{4.8cm}
+\theta(t^\prime-t)\varphi_{_{R_{(uv)}}}^{c(+)}(z,t)\varphi_{_{R_{(ir)}}}^{c(-)}(z,t^\prime)\Big\}\;,
\nonumber\\
&&\Big[f_{_{(--)}}^{L,c}(z,\phi)\Big]\Big[f_{_{(--)}}^{L,c}(z,\phi^\prime)\Big]
={\sqrt{tt^\prime}\over N_{_{L_{(uv)}}}^{c(-)}(z)N_{_{L_{(ir)}}}^{c(-)}(z)}
\Big\{\theta(t-t^\prime)\varphi_{_{L_{(uv)}}}^{c(-)}(z,t^\prime)\varphi_{_{L_{(ir)}}}^{c(-)}(z,t)
\nonumber\\
&&\hspace{4.8cm}
+\theta(t^\prime-t)\varphi_{_{L_{(uv)}}}^{c(-)}(z,t)\varphi_{_{L_{(ir)}}}^{c(-)}(z,t^\prime)\Big\}\;,
\nonumber\\
&&\Big[f_{_{(++)}}^{R,c}(z,\phi)\Big]\Big[f_{_{(++)}}^{R,c}(z,\phi^\prime)\Big]
={\sqrt{tt^\prime}\over N_{_{R_{(uv)}}}^{c(+)}(z)N_{_{R_{(ir)}}}^{c(+)}(z)}
\Big\{\theta(t-t^\prime)\varphi_{_{R_{(uv)}}}^{c(+)}(z,t^\prime)\varphi_{_{R_{(ir)}}}^{c(+)}(z,t)
\nonumber\\
&&\hspace{4.8cm}
+\theta(t^\prime-t)\varphi_{_{R_{(uv)}}}^{c(+)}(z,t)\varphi_{_{R_{(ir)}}}^{c(+)}(z,t^\prime)\Big\}\;.
\label{bulkfiles-extension4}
\end{eqnarray}
The above expressions in Eq.(\ref{bulkfiles-extension4}) are valid for $c\neq N+1/2$, one gets the
corresponding expressions for $c=N+1/2$ after replacing $\cos^2c\pi$ with $1$ in the
Eq.(\ref{normalization-z-plane1}).

\section{The four and five dimensional perturbative expansions\label{sec4}}
\indent\indent
The KK excitations affect the theoretical predictions of electroweak scale
although it is very difficult to produce them directly in the colliders running now.
When KK excitations of gauge fields satisfying $(++)$ BCs appear in
the relevant Feynman diagrams as virtual particles in four dimensional effective theory,
the amplitudes possibly contain the factor
\begin{eqnarray}
&&{-i\Big[\chi_{_{(++)}}^G(y_{_{(++)}}^{G(n)},\phi)\Big]\Big[\chi_{_{(++)}}^G
(y_{_{(++)}}^{G(n)},\phi^\prime)\Big]\over p^2-\Lambda_{_{KK}}^2\Big[y_{_{(++)}}^{G(n)}\Big]^2}
\label{propagator-1}
\end{eqnarray}
when we expand them according $\upsilon^2/\Lambda_{_{KK}}^2$, here $\upsilon$ denotes
the nonzero vacuum expectation value (VEV) of neutral Higgs located on IR brane. The denominator
$p^2-\Lambda_{_{KK}}^2\Big[y_{_{(++)}}^{G(n)}\Big]^2$ originates from the four dimensional
propagators of KK excitations for gauge fields in momentum space,
$\Big[\chi_{_{(++)}}^G(y_{_{(++)}}^{G(n)},\phi)\Big]$
and $\Big[\chi_{_{(++)}}^G(y_{_{(++)}}^{G(n)},\phi^\prime)\Big]$  originate from
the neighbor vertices in four dimensional effective theory. Note that
$\pm y_{_{(++)}}^{G(1)}$, $\cdots$, $\pm y_{_{(++)}}^{G(n)}$, $\cdots$ are zeros
of the function $R_{_{(++)}}^{G,\epsilon}(z)$, and the limit
\begin{eqnarray}
&&\lim\limits_{|z|\rightarrow\infty}\Bigg|{-iz\Big[\chi_{_{(++)}}^G(z,\phi)\Big]\Big[\chi_{_{(++)}}^G
(z,\phi^\prime)\Big]\over p^2-\Lambda_{_{KK}}^2z^2}\Bigg|=0
\label{propagator-2}
\end{eqnarray}
when we adopt the analytical extension of the combination of bulk profiles for $(++)$ BCs gauge
fields in Eq.(\ref{bulkfiles-extension3}). Applying Eq.(\ref{lemma3I}), we have
\begin{eqnarray}
&&iD_{_{(++)}}^{G}(p;\phi,\phi^\prime)
=\sum\limits_{n=1}^\infty{-i\Big[\chi_{_{(++)}}^G(y_{_{(++)}}^{G(n)},\phi)\Big]
\Big[\chi_{_{(++)}}^G(y_{_{(++)}}^{G(n)},\phi^\prime)\Big]\over p^2-\Lambda_{_{KK}}^2
\Big[y_{_{(++)}}^{G(n)}\Big]^2}
\nonumber\\
&&\hspace{-0.6cm}=
{i\over p^2}\bigg\{\Big[\chi_{_{(++)}}^G(0,\phi)\Big]
\Big[\chi_{_{(++)}}^G(0,\phi^\prime)\Big]
-\Big[\chi_{_{(++)}}^G({p\over \Lambda_{_{KK}}},\phi)\Big]
\Big[\chi_{_{(++)}}^G({p\over \Lambda_{_{KK}}},\phi^\prime)\Big]\bigg\}
\nonumber\\
&&\hspace{0.0cm}
-{i\over2\Lambda_{_{KK}}p}{\Big[\chi_{_{(++)}}^G({p\over \Lambda_{_{KK}}},\phi)\Big]
\Big[\chi_{_{(++)}}^G({p\over \Lambda_{_{KK}}},\phi^\prime)\Big]\over
R_{_{(++)}}^{G,\epsilon}({p\over \Lambda_{_{KK}}})}
{\partial R_{_{(++)}}^{G,\epsilon}\over\partial z}(z)\bigg|_{z={p\over \Lambda_{_{KK}}}}
\nonumber\\
&&\hspace{-0.6cm}=
-{i\Omega_{_{(++)}}^G(p/\Lambda_{_{KK}})\over\pi\Lambda_{_{KK}}^2}
\bigg\{{\pi tt^\prime\over2R_{_{(++)}}^{G,\epsilon}(p/\Lambda_{_{KK}})}
\Big[\theta(t-t^\prime)\Phi_{_{(uv)}}^{G(+)}(p/\Lambda_{_{KK}},t^\prime)
\Phi_{_{(ir)}}^{G(+)}(p/\Lambda_{_{KK}},t)
\nonumber\\
&&\hspace{0.0cm}
+\theta(t^\prime-t)\Phi_{_{(uv)}}^{G(+)}(p/\Lambda_{_{KK}},t)
\Phi_{_{(ir)}}^{G(+)}(p/\Lambda_{_{KK}},t^\prime)\Big]\bigg\}
+{i\over4\pi p^2}
\label{propagator-3}
\end{eqnarray}
with
\begin{eqnarray}
&&\Omega_{_{(++)}}^G(x)={2R_{_{(++)}}^{G,\epsilon}(x)+x\Big[\Phi_{_{(uv)}}^{G(+)}(x,1)
-\epsilon\Phi_{_{(ir)}}^{G(+)}(x,\epsilon)\Big]\over x^2\Big[N_{_{(uv)}}^{G(+)}(x)\Big]
\Big[N_{_{(ir)}}^{G(+)}(x)\Big]}\;.
\label{propagator-4}
\end{eqnarray}
The factor in the parenthesis $\{\cdots\}$ of the first term in Eq.(\ref{propagator-3})
is the five dimensional mixed position/momentum-space propagator derived in \cite{Randall1},
the factor $\Omega_{_{(++)}}^G/\pi$ originates from the normalization constants of
bulk profiles for gauge fields with $(++)$ BCs in four dimensional effective theory,
the different definitions of couplings between the gauge and fermion fields in five dimensional
full theory and corresponding four dimensional effective theory,
and the difference between the normalization of kinetic terms of relevant fields
in five dimensional full theory and corresponding four dimensional
effective theory, respectively. Additional, $D_{_{(++)}}^{G}(p;\phi,\phi^\prime)$
satisfies the following equation
\begin{eqnarray}
&&\Big\{{1\over r^2}{\partial\over\partial\phi}\Big[e^{-2\sigma(\phi)}
{\partial\over\partial\phi}\Big]+p^2\Big\}D_{_{(++)}}^{G}(p;\phi,\phi^\prime)
=-{i\Omega_{_{(++)}}^G(p/\Lambda_{_{KK}})\over\pi\Lambda_{_{KK}}^2r}\delta(\phi-\phi^\prime)
\label{propagator-5}
\end{eqnarray}
and the corresponding BCs
\begin{eqnarray}
&&{\partial D_{_{(++)}}^{G}\over\partial\phi_{_{(uv)}}}(p;\phi,\phi^\prime)
\bigg|_{\phi_{_{(uv)}}=0}=0\;,
\nonumber\\
&&{\partial D_{_{(++)}}^{G}\over\partial\phi_{_{(ir)}}}(p;\phi,\phi^\prime)
\bigg|_{\phi_{_{(ir)}}=\pi/2}=0\;.
\label{propagator-6}
\end{eqnarray}
For small $z$,
\begin{eqnarray}
&&R_{_{(++)}}^{G,\epsilon}(z)=-{2\ln \epsilon\over\pi}\Big\{1
-{z^2\over4}\Big[1+\epsilon^2+{1-\epsilon^2\over\ln\epsilon}\Big]+{\cal O}(z^4)\Big\}\;,
\nonumber\\
&&{\partial R_{_{(++)}}^{G,\epsilon}\over\partial z}(z)
={z\over\pi}\Big[1-\epsilon^2+(1+\epsilon^2)\ln\epsilon\Big]
\Big\{1-{z^2\over16}{2(1+4\epsilon^2+\epsilon^4)\ln\epsilon+3(1-\epsilon^4)
\over1-\epsilon^2+(1+\epsilon^2)\ln\epsilon}+{\cal O}(z^4)\Big\}\;,
\nonumber\\
&&{t\Phi_{_{(uv)}}^{G(+)}(z,t)\over\Big[N_{_{(uv)}}^{G(+)}
(z)\Big]}={1\over\sqrt{2\pi}}\Big\{1+{z^2\over4}
\Big[t^2(1-2\ln t+2\ln\epsilon)+1+{1-\epsilon^2\over\ln\epsilon}\Big]
\nonumber\\
&&\hspace{2.6cm}
+{z^4\over16}\Big[-{5\over4}t^4+\epsilon^2t^2+(t^4+2\epsilon^2t^2)(\ln t-\ln\epsilon)
\nonumber\\
&&\hspace{2.6cm}
+\Big({1-\epsilon^2\over\ln\epsilon}+1+\epsilon^2\Big)\Big(t^2-2t^2\ln t+2t^2\ln\epsilon\Big)
\nonumber\\
&&\hspace{2.6cm}
+2+{\ln\epsilon\over2}+{47-32\epsilon^2-15\epsilon^4\over16\ln\epsilon}
+{3(1-\epsilon^2)^2\over2\ln^2\epsilon}\Big]+{\cal O}(z^6)\Big\}\;,
\nonumber\\
&&{t\Phi_{_{(ir)}}^{G(+)}(z,t)\over\Big[N_{_{(ir)}}^{G(+)}
(z)\Big]}={1\over\sqrt{2\pi}}\Big\{1+{z^2\over4}
\Big[t^2(1-2\ln t)+\epsilon^2+{1-\epsilon^2\over\ln\epsilon}\Big]
\nonumber\\
&&\hspace{2.6cm}
+{z^4\over16}\Big[-{5\over4}t^4+t^2+(t^4+2t^2)\ln t
+\Big({1-\epsilon^2\over\ln\epsilon}+1+\epsilon^2\Big)\Big(t^2-2t^2\ln t\Big)
\nonumber\\
&&\hspace{2.6cm}
+2\epsilon^4-{\epsilon^4\ln\epsilon\over2}+{15+32\epsilon^2-47\epsilon^4\over16\ln\epsilon}
+{3(1-\epsilon^2)^2\over2\ln^2\epsilon}\Big]+{\cal O}(z^6)\Big\}\;.
\label{propagator-7}
\end{eqnarray}
Inserting the above equations into Eq.(\ref{propagator-3}) and assuming
$p\rightarrow0$, one obtains obviously
\begin{eqnarray}
&&\sum\limits_{n=1}^\infty{\Big[\chi_{_{(++)}}^G(y_{_{(++)}}^{G(n)},\phi)\Big]
\Big[\chi_{_{(++)}}^G(y_{_{(++)}}^{G(n)},\phi^\prime)\Big]\over\Lambda_{_{KK}}^2
\Big[y_{_{(++)}}^{G(n)}\Big]^2}
\nonumber\\
&&\hspace{-0.5cm}=
{1\over8\pi\Lambda_{_{KK}}^2}\Big\{t^2(2\ln t-1)+t^{\prime2}(2\ln t^\prime-1)
\nonumber\\
&&-2\ln\epsilon\Big[t^2\theta(t^\prime-t)+t^{\prime2}\theta(t-t^\prime)\Big]
-{1-\epsilon^2\over\ln\epsilon}\Big\}\;,
\label{propagator-8}
\end{eqnarray}
which is Eq.(34) exactly in Ref.\cite{Casagrande}.
Actually, this result can also be gotten directly by the residue theorem. Applying
Eq.(\ref{lemma3I}) and Eq.(\ref{propagator-7}), we have
\begin{eqnarray}
&&\sum\limits_{n=1}^\infty{\Big[\chi_{_{(++)}}^G(y_{_{(++)}}^{G(n)},\phi)\Big]
\Big[\chi_{_{(++)}}^G(y_{_{(++)}}^{G(n)},\phi^\prime)\Big]\over\Lambda_{_{KK}}^2
\Big[y_{_{(++)}}^{G(n)}\Big]^2}
\nonumber\\
&&\hspace{-0.5cm}=
-{1\over2}{\rm Res}\Big\{\Big[{2\over z^3}+{1\over z^2R_{_{(++)}}^{G,\epsilon}(z)}
{\partial R_{_{(++)}}^{G,\epsilon}\over\partial z}(z)\Big]\Big[\chi_{_{(++)}}^G(z,\phi)\Big]
\Big[\chi_{_{(++)}}^G(z,\phi^\prime)\Big],z=0\Big\}
\nonumber\\
&&\hspace{-0.5cm}=
{1\over8\pi\Lambda_{_{KK}}^2}\Big\{t^2(2\ln t-1)+t^{\prime2}(2\ln t^\prime-1)
\nonumber\\
&&-2\ln\epsilon\Big[t^2\theta(t^\prime-t)+t^{\prime2}\theta(t-t^\prime)\Big]
-{1-\epsilon^2\over\ln\epsilon}\Big\}\;,
\nonumber\\
&&\sum\limits_{n=1}^\infty{\Big[\chi_{_{(++)}}^G(y_{_{(++)}}^{G(n)},\phi)\Big]
\Big[\chi_{_{(++)}}^G(y_{_{(++)}}^{G(n)},\phi^\prime)\Big]\over\Lambda_{_{KK}}^4
\Big[y_{_{(++)}}^{G(n)}\Big]^4}
\nonumber\\
&&\hspace{-0.5cm}=
-{1\over2}{\rm Res}\Big\{\Big[{2\over z^5}+{1\over z^4R_{_{(++)}}^{G,\epsilon}(z)}
{\partial R_{_{(++)}}^{G,\epsilon}\over\partial z}(z)\Big]\Big[\chi_{_{(++)}}^G(z,\phi)\Big]
\Big[\chi_{_{(++)}}^G(z,\phi^\prime)\Big],z=0\Big\}
\nonumber\\
&&\hspace{-0.5cm}=
-{1\over32\pi\Lambda_{_{KK}}^4}\Big\{t^4\Big[\ln t-{5\over4}\Big]+t^2\Big[2+{1-\epsilon^2\over\ln\epsilon}
-{2(1-\epsilon^2)\over\ln\epsilon}\ln t\Big]
\nonumber\\
&&+t^{\prime4}\Big[\ln t^\prime-{5\over4}\Big]+t^{\prime2}\Big[2+{1-\epsilon^2\over\ln\epsilon}
-{2(1-\epsilon^2)\over\ln\epsilon}\ln t^\prime\Big]
\nonumber\\
&&+4t^2t^{\prime2}\Big[\ln t\ln t^\prime-(\ln\epsilon+{1\over2})\ln(tt^\prime)
+{1\over4}+{1\over2}\ln\epsilon\Big]
\nonumber\\
&&-\ln\epsilon\Big[\Big(t^4-4t^2t^{\prime2}\ln t\Big)\theta(t^\prime-t)
+\Big(t^{\prime4}-4t^2t^{\prime2}\ln t^\prime\Big)\theta(t-t^\prime)\Big]
\nonumber\\
&&+{5(1-\epsilon^4)\over8\ln\epsilon}+{(1-\epsilon^2)^2\over\ln^2\epsilon}\Big\}\;,
\label{propagator-9}
\end{eqnarray}
where the second equation is the Eq.(36) exactly in Ref.\cite{Casagrande}.

When the exciting KK modes of the left-handed fields with $(++)$ BCs appear in
relevant Feynman diagrams as virtual particles in four dimensional effective theory,
the amplitudes certainly contain the factor
\begin{eqnarray}
&&{i\Big\{/\!\!\!p\Big[f_{_{(++)}}^{L,c}(y_{_{(\pm\pm)}}^{c(n)},\phi)\Big]
\Big[f_{_{(++)}}^{L,c}(y_{_{(\pm\pm)}}^{c(n)},\phi^\prime)\Big]
+\Lambda_{_{KK}}\Big[y_{_{(\pm\pm)}}^{c(n)}\Big]
\Big[f_{_{(--)}}^{R,c}(y_{_{(\pm\pm)}}^{c(n)},\phi)\Big]
\Big[f_{_{(++)}}^{L,c}(y_{_{(\pm\pm)}}^{c(n)},\phi^\prime)\Big]\Big\}
\over p^2-\Lambda_{_{KK}}^2\Big[y_{_{(\pm\pm)}}^{c(n)}\Big]^2}
\label{propagator-10}
\end{eqnarray}
when we expand them according $\upsilon^2/\Lambda_{_{KK}}^2$. The  limit
\begin{eqnarray}
&&\lim\limits_{|z|\rightarrow\infty}\Bigg|{iz/\!\!\!p\Big[f_{_{(++)}}^{L,c}(z,\phi)\Big]
\Big[f_{_{(++)}}^{L,c}(z,\phi^\prime)\Big]
\over p^2-\Lambda_{_{KK}}^2z^2}\Bigg|=0
\label{propagator-11}
\end{eqnarray}
and the function
\begin{eqnarray}
&&\Bigg|{z^2\Lambda_{_{KK}}
\Big[f_{_{(--)}}^{R,c}(z,\phi)\Big]\Big[f_{_{(++)}}^{L,c}(z,\phi^\prime)\Big]
\over p^2-\Lambda_{_{KK}}^2z^2}\Bigg|
\label{propagator-12}
\end{eqnarray}
is uniformly bounded. Since $\pm y_{_{(\pm\pm)}}^{c(1)}$, $\cdots$,
$\pm y_{_{(\pm\pm)}}^{c(n)}$, $\cdots$ are the zeros of the function $R_{_{(\pm\pm)}}^{c,\epsilon}(z)$,
we have the following equation using Eq.(\ref{lemma3I}):
\begin{eqnarray}
&&iD_{_{L(++)}}^{c}(p;\phi,\phi^\prime)
\nonumber\\
&&\hspace{-0.6cm}=
\sum\limits_{n=1}^\infty{i\over p^2-\Lambda_{_{KK}}^2
\Big[y_{_{(\pm\pm)}}^{c(n)}\Big]^2}\bigg\{/\!\!\!p\Big[f_{_{(++)}}^{L,c}(y_{_{(\pm\pm)}}^{c(n)},\phi)\Big]
\Big[f_{_{(++)}}^{L,c}(y_{_{(\pm\pm)}}^{c(n)},\phi^\prime)\Big]
\nonumber\\
&&\hspace{0.0cm}
+\Lambda_{_{KK}}\Big[y_{_{(\pm\pm)}}^{c(n)}\Big]
\Big[f_{_{(--)}}^{R,c}(y_{_{(\pm\pm)}}^{c(n)},\phi)\Big]
\Big[f_{_{(++)}}^{L,c}(y_{_{(\pm\pm)}}^{c(n)},\phi^\prime)\Big]\bigg\}
\nonumber\\
&&\hspace{-0.6cm}=
\sum\limits_{n=1}^\infty{i/\!\!\!p\Big[f_{_{(++)}}^{L,c}(y_{_{(\pm\pm)}}^{c(n)},\phi)\Big]
\Big[f_{_{(++)}}^{L,c}(y_{_{(\pm\pm)}}^{c(n)},\phi^\prime)\Big]\over p^2-\Lambda_{_{KK}}^2
\Big[y_{_{(\pm\pm)}}^{c(n)}\Big]^2}
\nonumber\\
&&\hspace{-0.6cm}=
{i/\!\!\!p\over p^2}\bigg\{\Big[f_{_{(++)}}^{L,c}(0,\phi)\Big]\Big[f_{_{(++)}}^{L,c}(0,\phi^\prime)\Big]
-\Big[f_{_{(++)}}^{L,c}({p\over \Lambda_{_{KK}}},\phi)\Big]
\Big[f_{_{(++)}}^{L,c}({p\over \Lambda_{_{KK}}},\phi^\prime)\Big]\bigg\}
\nonumber\\
&&\hspace{0.0cm}
-{i/\!\!\!p\over2\Lambda_{_{KK}}p}{\Big[f_{_{(++)}}^{L,c}({p\over \Lambda_{_{KK}}},\phi)\Big]
\Big[f_{_{(++)}}^{L,c}({p\over \Lambda_{_{KK}}},\phi^\prime)\Big]\over
R_{_{(\pm\pm)}}^{c,\epsilon}({p\over \Lambda_{_{KK}}})}
{\partial R_{_{(\pm\pm)}}^{c,\epsilon}\over\partial z}(z)\bigg|_{z={p\over \Lambda_{_{KK}}}}
\nonumber\\
&&\hspace{-0.6cm}=
-{i/\!\!\!p\Omega_{_{(++)}}^{L,c}(p/\Lambda_{_{KK}})\over\pi\Lambda_{_{KK}}^2}
\bigg\{{\pi tt^\prime\over2R_{_{(\pm\pm)}}^{c,\epsilon}(p/\Lambda_{_{KK}})}
\Big[\theta(t-t^\prime)\varphi_{_{L_{(uv)}}}^{c(+)}(p/\Lambda_{_{KK}},t^\prime)
\varphi_{_{L_{(ir)}}}^{c(+)}(p/\Lambda_{_{KK}},t)
\nonumber\\
&&\hspace{0.0cm}
+\theta(t^\prime-t)\varphi_{_{L_{(uv)}}}^{c(+)}(p/\Lambda_{_{KK}},t)
\varphi_{_{L_{(ir)}}}^{c(+)}(p/\Lambda_{_{KK}},t^\prime)\Big]\bigg\}
+{i/\!\!\!p\over p^2}\bigg[{-2(1-2c)\epsilon\ln\epsilon\over\pi(1-\epsilon^{1-2c})}\bigg]
{(tt^\prime)^{-c}\over4}
\label{propagator-13}
\end{eqnarray}
with
\begin{eqnarray}
&&\Omega_{_{(++)}}^{L,c}(x)={2R_{_{(++)}}^{G,\epsilon}(x)+x\Big[\varphi_{_{L_{(uv)}}}^{c(+)}(x,1)
-\epsilon\varphi_{_{L_{(ir)}}}^{c(+)}(x,\epsilon)\Big]\over x^2\Big[N_{_{L_{(uv)}}}^{c(+)}(x)\Big]
\Big[N_{_{L_{(ir)}}}^{c(+)}(x)\Big]}\;.
\label{propagator-14}
\end{eqnarray}
Similarly, we also get the summing over the infinite series of exciting KK modes
for right-handed fermion as
\begin{eqnarray}
&&iD_{_{R(++)}}^{c}(p;\phi,\phi^\prime)
\nonumber\\
&&\hspace{-0.6cm}=
\sum\limits_{n=1}^\infty{i\over p^2-\Lambda_{_{KK}}^2
\Big[y_{_{(\mp\mp)}}^{c(n)}\Big]^2}\bigg\{/\!\!\!p\Big[f_{_{(++)}}^{R,c}(y_{_{(\mp\mp)}}^{c(n)},\phi)\Big]
\Big[f_{_{(++)}}^{R,c}(y_{_{(\mp\mp)}}^{c(n)},\phi^\prime)\Big]
\nonumber\\
&&\hspace{0.0cm}
+\Lambda_{_{KK}}\Big[y_{_{(\mp\mp)}}^{c(n)}\Big]
\Big[f_{_{(--)}}^{L,c}(y_{_{(\mp\mp)}}^{c(n)},\phi)\Big]
\Big[f_{_{(++)}}^{R,c}(y_{_{(\mp\mp)}}^{c(n)},\phi^\prime)\Big]\bigg\}
\nonumber\\
&&\hspace{-0.6cm}=
\sum\limits_{n=1}^\infty{i/\!\!\!p\Big[f_{_{(++)}}^{R,c}(y_{_{(\mp\mp)}}^{c(n)},\phi)\Big]
\Big[f_{_{(++)}}^{R,c}(y_{_{(\mp\mp)}}^{c(n)},\phi^\prime)\Big]\over p^2-\Lambda_{_{KK}}^2
\Big[y_{_{(\mp\mp)}}^{c(n)}\Big]^2}
\nonumber\\
&&\hspace{-0.6cm}=
{i/\!\!\!p\over p^2}\bigg\{\Big[f_{_{(++)}}^{R,c}(0,\phi)\Big]\Big[f_{_{(++)}}^{R,c}(0,\phi^\prime)\Big]
-\Big[f_{_{(++)}}^{R,c}({p\over \Lambda_{_{KK}}},\phi)\Big]
\Big[f_{_{(++)}}^{R,c}({p\over \Lambda_{_{KK}}},\phi^\prime)\Big]\bigg\}
\nonumber\\
&&\hspace{0.0cm}
-{i/\!\!\!p\over2\Lambda_{_{KK}}p}{\Big[f_{_{(++)}}^{R,c}({p\over \Lambda_{_{KK}}},\phi)\Big]
\Big[f_{_{(++)}}^{R,c}({p\over \Lambda_{_{KK}}},\phi^\prime)\Big]\over
R_{_{(\mp\mp)}}^{c,\epsilon}({p\over \Lambda_{_{KK}}})}
{\partial R_{_{(\mp\mp)}}^{c,\epsilon}\over\partial z}(z)\bigg|_{z={p\over \Lambda_{_{KK}}}}
\nonumber\\
&&\hspace{-0.6cm}=
-{i/\!\!\!p\Omega_{_{(++)}}^{R,c}(p/\Lambda_{_{KK}})\over\pi\Lambda_{_{KK}}^2}
\bigg\{{\pi tt^\prime\over2R_{_{(\mp\mp)}}^{c,\epsilon}(p/\Lambda_{_{KK}})}
\Big[\theta(t-t^\prime)\varphi_{_{R_{(uv)}}}^{c(+)}(p/\Lambda_{_{KK}},t^\prime)
\varphi_{_{R_{(ir)}}}^{c(+)}(p/\Lambda_{_{KK}},t)
\nonumber\\
&&\hspace{0.0cm}
+\theta(t^\prime-t)\varphi_{_{R_{(uv)}}}^{c(+)}(p/\Lambda_{_{KK}},t)
\varphi_{_{R_{(ir)}}}^{c(+)}(p/\Lambda_{_{KK}},t^\prime)\Big]\bigg\}
+{i/\!\!\!p\over p^2}\bigg[{-2(1+2c)\epsilon\ln\epsilon\over\pi(1-\epsilon^{1+2c})}\bigg]
{(tt^\prime)^{c}\over4}
\label{propagator-15}
\end{eqnarray}
with
\begin{eqnarray}
&&\Omega_{_{(++)}}^{R,c}(x)={2R_{_{(\mp\mp)}}^{c,\epsilon}(x)+x\Big[\varphi_{_{R_{(uv)}}}^{c(+)}(x,1)
-\epsilon\varphi_{_{R_{(ir)}}}^{c(+)}(x,\epsilon)\Big]\over x^2\Big[N_{_{R_{(uv)}}}^{c(+)}(x)\Big]
\Big[N_{_{R_{(ir)}}}^{c(+)}(x)\Big]}\;.
\label{propagator-16}
\end{eqnarray}
For small $z$,
\begin{eqnarray}
&&R_{_{(\pm\pm)}}^{c,\epsilon}(z)={2\epsilon^{c-{1\over2}}(1-\epsilon^{1-2c})
\over(1-2c)\Gamma({1\over2}-c)\Gamma({1\over2}+c)}\Big\{1
-{z^2\over2(1-\epsilon^{1-2c})}\Big[{1-\epsilon^{3-2c}\over3-2c}
+{\epsilon^2-\epsilon^{1-2c}\over1+2c}\Big]
+{\cal O}(z^4)\Big\}\;,
\nonumber\\
&&{\partial R_{_{(\pm\pm)}}^{c,\epsilon}\over\partial z}(z)
=-{2z\epsilon^{c-{1\over2}}\over(1-2c)\Gamma({1\over2}-c)\Gamma({1\over2}+c)}\Big\{
\Big[{1-\epsilon^{3-2c}\over3-2c}+{\epsilon^2-\epsilon^{1-2c}\over1+2c}\Big]
\nonumber\\
&&\hspace{2.4cm}
-{z^2\over2}\Big[{1-\epsilon^{5-2c}\over(5-2c)(3-2c)}
+{2(\epsilon^2-\epsilon^{3-2c})\over(3-2c)(1+2c)}
+{(\epsilon^4-\epsilon^{1-2c})\over(1+2c)(3+2c)}\Big]
+{\cal O}(z^4)\Big\}\;,
\nonumber\\
&&{\sqrt{t}\varphi_{_{L_{(uv)}}}^{c(+)}(z,t)\over\Big[N_{_{L_{(uv)}}}^{c(+)}
(z)\Big]}={t^{-c}\over2}\sqrt{-2(1-2c)\epsilon\ln\epsilon\over\pi(1-\epsilon^{1-2c})}
\bigg\{1-{z^2\over2}\Big[{\epsilon^2\over1+2c}+{t^2\over1-2c}-{2\epsilon^{1-2c}
t^{1+2c}\over(1-2c)(1+2c)}
\nonumber\\
&&\hspace{2.6cm}
-{(1+2c)(1-\epsilon^{3-2c})+(3-2c)(\epsilon^2-\epsilon^{1-2c})\over
(3-2c)(1+2c)(1-\epsilon^{1-2c})}\Big]+{\cal O}(z^4)\bigg\}\;,
\nonumber\\
&&{\sqrt{t}\varphi_{_{L_{(ir)}}}^{c(+)}(z,t)\over\Big[N_{_{L_{(ir)}}}^{c(+)}
(z)\Big]}={t^{-c}\over2}\sqrt{-2(1-2c)\epsilon\ln\epsilon\over\pi(1-\epsilon^{1-2c})}
\bigg\{1-{z^2\over2}\Big[{1\over1+2c}+{t^2\over1-2c}-{2t^{1+2c}\over(1-2c)(1+2c)}
\nonumber\\
&&\hspace{2.6cm}
-{(1+2c)(1-\epsilon^{3-2c})+(3-2c)(\epsilon^2-\epsilon^{1-2c})\over
(3-2c)(1+2c)(1-\epsilon^{1-2c})}\Big]+{\cal O}(z^4)\bigg\}\;.
\label{propagator-17}
\end{eqnarray}
Using Eq.(\ref{propagator-17}), we have
\begin{eqnarray}
&&\sum\limits_{n=1}^\infty{\Big[f_{_{(++)}}^{L,c}(y_{_{(\pm\pm)}}^{c(n)},\phi)\Big]
\Big[f_{_{(++)}}^{L,c}(y_{_{(\pm\pm)}}^{c(n)},\phi^\prime)\Big]\over\Lambda_{_{KK}}^2
\Big[y_{_{(\pm\pm)}}^{c(n)}\Big]^2}
\nonumber\\
&&\hspace{-0.5cm}=
-{1\over2\Lambda_{_{KK}}^2}{\rm Res}\Big\{\Big[{2\over z^3}+{1\over z^2R_{_{(\pm\pm)}}^{c,\epsilon}(z)}
{\partial R_{_{(\pm\pm)}}^{c,\epsilon}\over\partial z}(z)\Big]\Big[f_{_{(++)}}^{L}(z,\phi)\Big]
\Big[f_{_{(++)}}^{L}(z,\phi^\prime)\Big],z=0\Big\}
\nonumber\\
&&\hspace{-0.5cm}=
-{(1-2c)\epsilon\ln\epsilon\over(1-\epsilon^{1-2c})}{(tt^\prime)^{-c}\over4\pi\Lambda_{_{KK}}^2}
\Big\{{2(1-2c)(1-\epsilon^{3-2c})\over(3-2c)(1+2c)(1-\epsilon^{1-2c})}+{t^2+t^{\prime2}\over1-2c}
\nonumber\\
&&-{2\over(1-2c)(1+2c)}
\Big[\theta(t^\prime-t)\Big(t^{\prime1+2c}+\epsilon^{1-2c}t^{1+2c}\Big)
\nonumber\\
&&+\theta(t-t^\prime)\Big(t^{1+2c}+\epsilon^{1-2c}t^{\prime1+2c}\Big)\Big]\Big\}\;,
\nonumber\\
&&\sum\limits_{n=1}^\infty{\Big[f_{_{(++)}}^{R,c}(y_{_{(\mp\mp)}}^{c(n)},\phi)\Big]
\Big[f_{_{(++)}}^{R,c}(y_{_{(\mp\mp)}}^{c(n)},\phi^\prime)\Big]\over\Lambda_{_{KK}}^2
\Big[y_{_{(\mp\mp)}}^{c(n)}\Big]^2}
\nonumber\\
&&\hspace{-0.5cm}=
-{1\over2\Lambda_{_{KK}}^2}{\rm Res}\Big\{\Big[{2\over z^3}+{1\over z^2R_{_{(\mp\mp)}}^{c,\epsilon}(z)}
{\partial R_{_{(\mp\mp)}}^{c,\epsilon}\over\partial z}(z)\Big]\Big[f_{_{(++)}}^{R}(z,\phi)\Big]
\Big[f_{_{(++)}}^{R}(z,\phi^\prime)\Big],z=0\Big\}
\nonumber\\
&&\hspace{-0.5cm}=
-{(1+2c)\epsilon\ln\epsilon\over(1-\epsilon^{1+2c})}{(tt^\prime)^{c}\over4\pi\Lambda_{_{KK}}^2}
\Big\{{2(1+2c)(1-\epsilon^{3+2c})\over(3+2c)(1-2c)(1-\epsilon^{1+2c})}+{t^2+t^{\prime2}\over1+2c}
\nonumber\\
&&-{2\over(1-2c)(1+2c)}
\Big[\theta(t^\prime-t)\Big(t^{\prime1-2c}+\epsilon^{1+2c}t^{1-2c}\Big)
\nonumber\\
&&+\theta(t-t^\prime)\Big(t^{1-2c}+\epsilon^{1+2c}t^{\prime1-2c}\Big)\Big]\Big\}\;.
\label{propagator-18}
\end{eqnarray}

\section{Summing over infinite series of KK modes in unified extra dimension\label{sec5}}
In this section, we depart from the main line above and discuss how to sum over
the infinite series of KK modes in unified extra dimension. In these models all fields
can propagate in all available dimension, the SM particles correspond to
zero modes in the KK decomposition of five dimensional fields with $(++)$ BCs.
The towers of the SM particle KK partners and additional towers of KK modes of
five dimensional fields with $(--)$ BCs do not correspond to any field
in the SM\cite{Arkani-Hamed,Barbieri}. The simplest model of this type
is proposed by Appelquist, Cheng and Dobrescu\cite{ACD} in which the only
additional free parameter relating to the SM is the compactification scale
$1/r$.

Assuming that the topology of the fifth dimension is the orbifold $S^1/Z_2$
and the coordinate $y\equiv x^5$ runs from $0$ to $2\pi r$, one can write
the KK expansions of five dimensional fields respectively as\cite{Buras7}:
\begin{eqnarray}
&&\psi^+(x,y)={1\over\sqrt{2\pi r}}\psi_{_{R(0)}}(x)+{1\over\sqrt{\pi r}}
\sum\limits_{n=1}^\infty\Big\{\psi_{_{R(n)}}(x)\cos{ny\over r}
+\psi_{_{L(n)}}(x)\sin{ny\over r}\Big\}\;,
\nonumber\\
&&\psi^-(x,y)={1\over\sqrt{2\pi r}}\psi_{_{L(0)}}(x)+{1\over\sqrt{\pi r}}
\sum\limits_{n=1}^\infty\Big\{\psi_{_{L(n)}}(x)\cos{ny\over r}
+\psi_{_{R(n)}}(x)\sin{ny\over r}\Big\}\;,
\nonumber\\
&&A^\mu(x,y)={1\over\sqrt{2\pi r}}A^\mu_{_{(0)}}(x)+{1\over\sqrt{\pi r}}
\sum\limits_{n=1}^\infty A^\mu_{_{(n)}}(x)\cos{ny\over r}\;,
\nonumber\\
&&A^5(x,y)={1\over\sqrt{\pi r}}\sum\limits_{n=1}^\infty A^5_{_{(n)}}(x)\sin{ny\over r}\;,
\nonumber\\
&&\phi^+(x,y)={1\over\sqrt{2\pi r}}\phi_{_{(0)}}(x)+{1\over\sqrt{\pi r}}
\sum\limits_{n=1}^\infty\phi_{_{(n)}}^+(x)\cos{ny\over r}\;,
\nonumber\\
&&\phi^-(x,y)={1\over\sqrt{\pi r}}\sum\limits_{n=1}^\infty\phi_{_{(n)}}^-(x)\sin{ny\over r}\;.
\label{unified-1}
\end{eqnarray}
Here, the Dirac spinors $\psi^\pm=(P_R+P_L)\psi^\pm=\psi_L^\pm+\psi_R^\pm$
satisfy the following BCs
\begin{eqnarray}
&&\left\{\begin{array}{l}\partial_y\psi_R^+\Big|_{y=\{0,\pi r\}}=0,\\
\psi_L^+\Big|_{y=\{0,\pi r\}}=0,\end{array}\right.\;\;\; {\rm or}\;\;\;
\left\{\begin{array}{l}\partial_y\psi_L^-\Big|_{y=\{0,\pi r\}}=0,\\
\psi_R^-\Big|_{y=\{0,\pi r\}}=0,\end{array}\right.
\label{unified-2}
\end{eqnarray}
with the chirality projectors $P_{R/L}=(1\pm\gamma_5)/2$. Additional,
the vector fields satisfy the BCs
\begin{eqnarray}
&&\left\{\begin{array}{l}\partial_y A^\mu\Big|_{y=\{0,\pi r\}}=0,\\
A^5\Big|_{y=\{0,\pi r\}}=0,\end{array}\right.
\label{unified-3}
\end{eqnarray}
and the scalar fields $\phi^\pm$ satisfy the BCs
\begin{eqnarray}
&&\left\{\begin{array}{l}\partial_y\phi^+\Big|_{y=\{0,\pi r\}}=0,\\
\phi^-\Big|_{y=\{0,\pi r\}}=0.\end{array}\right.
\label{unified-4}
\end{eqnarray}

In unified extra dimension, the couplings involving KK excitations do not depend
on the bulk profiles since the integral over the 5th coordinate y
can be integrated out explicitly. Correspondingly, the summing over KK excitations
is simplified drastically because it is unnecessary to extend the square
of bulk profile to the complex plane when we apply the residue theorem.
Actually, the authors of Ref.\cite{Buras7} have applied the relation
\begin{eqnarray}
&&\sum\limits_{n=1}^\infty{b\over n^2+c}={b(\sqrt{c}\pi\coth(\sqrt{c}\pi)-1)\over2c}
\label{unified-5}
\end{eqnarray}
to perform the summing over KK excitations in some lower energy processes.
In unified extra dimension, the KK mass $n/r$ is the zero of order one
of the function $\sin(\pi rz)$, and residue of the function $G(z)=\pi r\cos(\pi rz)/\sin(\pi rz)$
at $z=n/r$ is uniform one. Obviously, the function
\begin{eqnarray}
&&f(z)={B\over z^2+C}
\label{unified-6}
\end{eqnarray}
is uniformly bounded because
\begin{eqnarray}
&&\lim\limits_{|z|\rightarrow\infty}\Big|zf(z)\Big|=0.
\label{unified-7}
\end{eqnarray}
Here,
\begin{eqnarray}
&&B={b\over r^2},\;\;C={c\over r^2}\;.
\label{unified-8}
\end{eqnarray}
Using the residue theorem, one obtains obviously
\begin{eqnarray}
&&\sum\limits_{n=1}^\infty{b\over n^2+c}=\sum\limits_{n=1}^\infty{B\over
({n\over r})^2+C}
\nonumber\\
&&\hspace{-0.5cm}=
-{1\over2}{\rm Res}\Big\{{\pi r\cos(\pi rz)\over\sin\pi rz}{B\over
z^2+C},z=0\Big\}
-{1\over2}{\rm Res}\Big\{{\pi r\cos(\pi rz)\over\sin\pi rz}{B\over
z^2+C},z=i\sqrt{C}\Big\}
\nonumber\\
&&-{1\over2}{\rm Res}\Big\{{\pi r\cos(\pi rz)\over\sin\pi rz}{B\over
z^2+C},z=-i\sqrt{C}\Big\}
\nonumber\\
&&\hspace{-0.5cm}=
-{B\over2C}-{B\pi r\cos(i\sqrt{C}\pi r)\over2\sqrt{C}\sin(i\sqrt{C}\pi r)}
={b(\sqrt{c}\pi\coth(\sqrt{c}\pi)-1)\over2c}\;,
\label{unified-9}
\end{eqnarray}
which is the equation presented in Eq.(\ref{unified-5}). Furthermore, the
residue theorem can be applied to sum over the infinite series of KK excitations
in more complicate forms.

\section{The corrections to $B\rightarrow X_s\gamma$ from neutral Higgs in
warped extra dimension\label{sec6}}
\indent\indent
Within the framework of warped extra dimension, the neutral Higgs located on IR brane induces
the mixing between states of the same electric charge, a consequence
of the mixing is that the FCNC transitions are also mediated by neutral
Higgs, the KK excitations of gluon and photon, and the KK excitations of neutral electroweak
gauge bosons besides the charged electroweak gauge bosons $W^\pm$ together with
their KK partners. In this section, we demonstrate that the radiative
correction to the rare decay $b\rightarrow s+\gamma$ from neutral Higgs contains
the suppression factor $m_b^3m_s/m_{_{\rm w}}^4$ comparing with the corrections from other sector.

The effective Hamilton for $B\rightarrow X_s\gamma$ at scales
$\mu={\cal O}(m_b)$ is given by\cite{Buras6}:
\begin{eqnarray}
&&{\cal H}_{eff}(b\rightarrow s\gamma)=-{G_F\over\sqrt{2}}\Big[
\sum\limits_{i=1}^6C_i(\mu){\cal Q}_i(\mu)+C_{7\gamma}(\mu){\cal Q}_{7\gamma}(\mu)
+C_{8G}(\mu){\cal Q}_{8G}(\mu)
\nonumber\\
&&\hspace{3.0cm}+C_{7\gamma}^N(\mu){\cal Q}_{7\gamma}^N(\mu)
+C_{8G}^N(\mu){\cal Q}_{8G}^N(\mu)\Big]\;,
\label{eff_Hamilton}
\end{eqnarray}
with $G_F$ denoting the Fermi constant. The magnetic penguin operators are
\begin{eqnarray}
&&{\cal Q}_{7\gamma}={e\over8\pi^2}m_b\overline{s}_\alpha\sigma^{\mu\nu}
(1+\gamma_5)b_\alpha F_{\mu\nu}\;,
\nonumber\\
&&{\cal Q}_{8G}={g_s\over8\pi^2}m_b\overline{s}_\alpha T^a_{\alpha\beta}
\sigma^{\mu\nu}(1+\gamma_5)b_\beta G_{\mu\nu}^a\;,
\nonumber\\
&&{\cal Q}_{7\gamma}^N={e\over8\pi^2}m_b\overline{s}_\alpha\sigma^{\mu\nu}
(1-\gamma_5)b_\alpha F_{\mu\nu}\;,
\nonumber\\
&&{\cal Q}_{8G}^N={g_s\over8\pi^2}m_b\overline{s}_\alpha T^a_{\alpha\beta}
\sigma^{\mu\nu}(1-\gamma_5)b_\beta G_{\mu\nu}^a\;,
\label{operators}
\end{eqnarray}
and concrete expressions of other dimension six operators can be found in Ref.\cite{Buras6}.
Here $\alpha,\;\beta=1,\;2,\;3$ denote the color indices of quarks, $F_{\mu\nu}$
and $G_{\mu\nu}^a\;(a=1,\;\cdots,\;8)$ are the electromagnetic and strong field
strength tensors, respectively. The magnetic penguin operators ${\cal Q}_{7\gamma},\;
{\cal Q}_{8G},\;{\cal Q}_{7\gamma}^N,\;{\cal Q}_{8G}^N$ in the effective Hamilton
are induced by the virtual heavy freedoms through the one loop diagrams.
As the internal virtual particles are neutral Higgs and electric charge $-1/3$ quarks,
the corresponding corrections to relevant Wilson coefficients can be written as
\begin{eqnarray}
&&C_{7\gamma}(\mu_{_{\rm EW}})={4s_{_{\rm w}}^2Q_f\over e^2}\bigg\{
\sum\limits_{i=1}^3\Big(\xi_{(0)}\Big)_{s,d_i}
\Big(\xi_{(0)}\Big)_{d_i,b}^\dagger F_1(x_{_h},x_{d_i})
\nonumber\\
&&\hspace{2.2cm}
+\sum\limits_{d_i=1}^3{m_{d_i}\over m_b}\Big(\xi_{(0)}\Big)_{s,d_i}^\dagger
\Big(\xi_{(0)}\Big)_{d_i,b}^\dagger F_2(x_{_h},x_{d_i})
\nonumber\\
&&\hspace{2.2cm}
+\sum\limits_{i=1}^3\sum\limits_{n=1}^\infty\Big(\zeta_{(n)}\Big)_{s,D_i}^\dagger
\Big(\zeta_{(n)}\Big)_{D_i,b}F_1(x_{_h},x_{D_i})
\nonumber\\
&&\hspace{2.2cm}
+\sum\limits_{d_i=1}^3\sum\limits_{n=1}^\infty
{m_{D_i}\over m_b}\Big(\zeta_{(n)}\Big)_{s,D_i}
\Big(\zeta_{(n)}\Big)_{D_i,b}F_2(x_{_h},x_{D_i})\bigg\}\;,
\nonumber\\
&&C_{8G}(\mu_{_{\rm EW}})={1\over Q_{_f}}C_{7\gamma}(\mu_{_{\rm EW}})\;,
\nonumber\\
&&C_{7\gamma}^{N}(\mu_{_{\rm EW}})={4s_{_{\rm w}}^2Q_f\over e^2}\bigg\{
\sum\limits_{i=1}^3\Big(\xi_{(0)}\Big)_{s,d_i}^\dagger
\Big(\xi_{(0)}\Big)_{d_i,b}F_1(x_{_h},x_{d_i})
\nonumber\\
&&\hspace{2.2cm}
+\sum\limits_{d_i=1}^3{m_{d_i}\over m_b}\Big(\xi_{(0)}\Big)_{s,d_i}
\Big(\xi_{(0)}\Big)_{d_i,b}F_2(x_{_h},x_{d_i})
\nonumber\\
&&\hspace{2.2cm}
+\sum\limits_{i=1}^3\sum\limits_{n=1}^\infty\Big(\xi_{(n)}\Big)_{s,D_i}^\dagger
\Big(\xi_{(n)}\Big)_{D_i,b}F_1(x_{_h},x_{D_i})
\nonumber\\
&&\hspace{2.2cm}
+\sum\limits_{d_i=1}^3\sum\limits_{n=1}^\infty{m_{D_i}\over m_b}\Big(\xi_{(n)}\Big)_{s,D_i}
\Big(\xi_{(n)}\Big)_{D_i,b}F_2(x_{_h},x_{D_i})
\nonumber\\
&&\hspace{2.2cm}
+\sum\limits_{i=1}^3\sum\limits_{n=1}^\infty\Big(\eta_{(n)}\Big)_{s,D_i}^\dagger
\Big(\eta_{(n)}\Big)_{D_i,b}F_1(x_{_h},x_{D_i})
\nonumber\\
&&\hspace{2.2cm}
+\sum\limits_{d_i=1}^3\sum\limits_{n=1}^\infty
{m_{D_i}\over m_b}\Big(\eta_{(n)}\Big)_{s,D_i}
\Big(\eta_{(n)}\Big)_{D_i,b}F_2(x_{_h},x_{D_i})\bigg\}\;,
\nonumber\\
&&C_{8G}^N(\mu_{_{\rm EW}})={1\over Q_{_f}}C_{7\gamma}^{N}(\mu_{_{\rm EW}})\;,
\label{magnetic_penguin_S1}
\end{eqnarray}
with $Q_{_f}=-1/3,\;x_{d_i}=m_{d_i}^2/m_{_{\rm w}}^2,\;x_{_h}=m_{_h}^2/m_{_{\rm w}}^2$.
The abbreviation $s_{_{\rm w}}=\sin\theta_{_W}$ where $\theta_{_W}$ is the Weinberg angle,
and the functions $F_{1,2}(x,y)$ are defined as
\begin{eqnarray}
&&F_1(x,y)=\Big\{{1\over24}{\partial^3\varrho_{_{3,1}}\over\partial x^3}
-{1\over4}{\partial^2\varrho_{_{2,1}}\over\partial x^2}
+{1\over 4}{\partial\varrho_{_{1,1}}\over\partial x}\Big\}(x,y)\;,
\nonumber\\
&&F_2(x,y)=\Big\{{1\over 4}{\partial^2\varrho_{_{2,1}}\over\partial x^2}
-{1\over2}{\partial\varrho_{_{1,1}}\over\partial x}
+{1\over2}{\partial\varrho_{_{1,1}}\over\partial y}\Big\}(x,y)\;,
\label{magnetic_functions}
\end{eqnarray}
with
\begin{eqnarray}
&&\varrho_{_{m,n}}(x,y)={x^m\ln^nx-y^m\ln^ny\over x-y}\;.
\label{magnetic_function1}
\end{eqnarray}

The couplings $\xi_{(0)}$ depend on the bulk profiles of zero modes
for charge $-1/3$ quarks, and the mixing between zero modes of charge $-1/3$
quarks and corresponding KK partners. To obtain approximately the mixing
between the zero modes of charge $-1/3$ quarks and exciting KK modes, we write
the infinite dimensional column vectors for down type quarks in the chirality
basis as\cite{Buras5}
\begin{eqnarray}
&&\Psi_L(-1/3)=\Big(q_{_{d_L}}^{i(0)}(++),\cdots,q_{_{d_L}}^{i(n)}(++),
D_L^{i(n)}(+-),d_L^{i(n)}(--),\cdots\Big)^T\;,
\nonumber\\
&&\Psi_R(-1/3)=\Big(d_R^{i(0)}(++),\cdots,q_{_{d_R}}^{i(n)}(--),
D_R^{i(n)}(-+),d_R^{i(n)}(++),\cdots\Big)^T\;.
\label{mass1}
\end{eqnarray}
Here, $i=1,\;2,\;3$ is the index of generation, $n=1,\;2,\;\cdots,\;\infty$
is the index of KK modes, the signs in parentheses denote
the BCs satisfied by corresponding fields on UV and IR branes respectively.
We then formally diagonalize the $-1/3$ charge mass matrix and write the mass
eigenvector as
\begin{eqnarray}
&&\Psi_L^{(m)}(-1/3)={\cal D}_L^\dagger\Psi_L(-1/3)\;,
\nonumber\\
&&\Psi_R^{(m)}(-1/3)={\cal D}_R^\dagger\Psi_R(-1/3)\;.
\label{mass2}
\end{eqnarray}
Using above equations, one writes the couplings as
\begin{eqnarray}
&&\Big(\xi_{(0)}\Big)_{d,b}=\sum\limits_{\alpha,\beta=1}^3
\Big[f_{_{(++)}}^{R,c_{_T}^\alpha}(0,{\pi\over2})
f_{_{(++)}}^{L,c_{_B}^\beta}(0,{\pi\over2})\Big]
\Big({\cal D}_R\Big)_{d,\alpha}^\dagger
\Big(\lambda^{d}\Big)_{\alpha\beta}^\dagger\Big({\cal D}_L\Big)_{\beta,b}\;,
\nonumber\\
&&\Big(\xi_{(n)}\Big)_{D_i,b}=\sum\limits_{\alpha,\beta=1}^3\Big[f_{_{(++)}}^{L,c_{_B}^\beta}(0,{\pi\over2})\Big]
\Big({\cal D}_R\Big)_{D_i,9n+\alpha}^\dagger
\Big[f_{_{(++)}}^{R,c_{_T}^\alpha}(y_{_{(\mp\mp)}}^{c_{_T}^\alpha(n)},{\pi\over2})\Big]
\Big(\lambda^{d}\Big)_{\alpha\beta}^\dagger \Big({\cal D}_L\Big)_{\beta,b}\;,
\nonumber\\
&&\Big(\eta_{(n)}\Big)_{D_i,b}=\sum\limits_{\alpha,\beta=1}^3\Big[f_{_{(++)}}^{L,c_{_B}^\beta}(0,{\pi\over2})\Big]
\Big({\cal D}_R\Big)_{D_i,9n-3+\alpha}^\dagger
\Big[f_{_{(-+)}}^{R,c_{_T}^\alpha}(y_{_{(\pm\mp)}}^{c_{_T}^\alpha(n)},{\pi\over2})\Big]
\Big(\lambda^{d}\Big)_{\alpha\beta}^\dagger \Big({\cal D}_L\Big)_{\beta,b}\;,
\nonumber\\
&&\Big(\zeta_{(n)}\Big)_{D_i,b}=\sum\limits_{\alpha,\beta=1}^3\Big[f_{_{(++)}}^{R,c_{_T}^\beta}(0,{\pi\over2})\Big]
\Big({\cal D}_L\Big)_{D_i,9n-6+\alpha}^\dagger
\Big[f_{_{(++)}}^{L,c_{_B}^\alpha}(y_{_{(\pm\pm)}}^{c_{_B}^\alpha(n)},{\pi\over2})\Big]
\Big(\lambda^{d}\Big)_{\alpha\beta}^\dagger \Big({\cal D}_R\Big)_{\beta,b}\;.
\label{mass3}
\end{eqnarray}

For the infinite dimensional column vectors, we have no means of obtaining
the mixing matrices ${\cal D}_{L,R}$ exactly. Adopting the
effective Lagrangian approach\cite{Buras4,Santiago4}, we expand the left-
and right-handed mixing matrices according $\upsilon^2/\Lambda_{_{KK}}^2$
and then approximate the Wilson coefficients $C_{7\gamma}$
in Eq.(\ref{magnetic_penguin_S1}) as
\begin{eqnarray}
&&C_{7\gamma}(\mu_{_{\rm EW}})
={4s_{_{\rm w}}^2Q_f\over e^2}\sum\limits_{i=1}^3\sum\limits_{\alpha=1}^3
\sum\limits_{\beta=1}^3\sum\limits_{\gamma=1}^3\bigg\{
-\Big(\delta_{b,d_i}{m_b^2\over\Lambda_{_{KK}}^2}
+\delta_{s,d_i}{m_s^2\over\Lambda_{_{KK}}^2}\Big)
\nonumber\\
&&\hspace{2.2cm}\times
\Big[f_{_{(++)}}^{R,c_{_T}^\alpha}(0,{\pi\over2})f_{_{(++)}}^{R,c_{_T}^\gamma}(0,{\pi\over2})\Big]
\Big({\cal D}_R^{(0)}\Big)_{s,\alpha}^\dagger\Big(\lambda^d\Big)^\dagger_{\alpha\beta}
\Big(\lambda^d\Big)_{\beta\gamma}\Big({\cal D}_R^{(0)}\Big)_{\gamma,b}
\nonumber\\
&&\hspace{2.2cm}\times
\sum\limits_{n=1}^\infty{1\over\Big[y_{_{(\pm\pm)}}^{c_{_B}^\beta(n)}\Big]^2}
\Big[f_{_{(++)}}^{L,c_{_B}^\beta}(y_{_{(\pm\pm)}}^{c_{_B}^\beta(n)},{\pi\over2})
f_{_{(++)}}^{L,c_{_B}^\beta}(y_{_{(\pm\pm)}}^{c_{_B}^\beta(n)},{\pi\over2})\Big]
\nonumber\\
&&\hspace{2.2cm}
-{m_bm_s\over\Lambda_{_{KK}}^2}\Big(\delta_{b,d_i}+\delta_{s,d_i}\Big)
\Big[f_{_{(++)}}^{L,c_{_B}^\alpha}(0,{\pi\over2})f_{_{(++)}}^{L,c_{_B}^\gamma}(0,{\pi\over2})\Big]
\Big({\cal D}_L^{(0)}\Big)_{s,\alpha}^\dagger\Big(\lambda^d\Big)_{\alpha\beta}
\nonumber\\
&&\hspace{2.2cm}\times
\Big(\lambda^d\Big)_{\beta\gamma}^\dagger
\Big({\cal D}_L^{(0)}\Big)_{\gamma,b}
\sum\limits_{n=1}^\infty\bigg({1\over\Big[y_{_{(\pm\mp)}}^{c_{_T}^\beta(n)}\Big]^2}
\Big[f_{_{(-+)}}^{R,c_{_T}^\beta}(y_{_{(\pm\mp)}}^{c_{_T}^\beta(n)},{\pi\over2})
f_{_{(-+)}}^{R,c_{_T}^\beta}(y_{_{(\pm\mp)}}^{c_{_T}^\beta(n)},{\pi\over2})\Big]
\nonumber\\
&&\hspace{2.2cm}
+{1\over\Big[y_{_{(\mp\mp)}}^{c_{_T}^\beta(n)}\Big]^2}
\Big[f_{_{(++)}}^{R,c_{_T}^\beta}(y_{_{(\mp\mp)}}^{c_{_T}^\beta(n)},{\pi\over2})
f_{_{(++)}}^{R,c_{_T}^\beta}(y_{_{(\mp\mp)}}^{c_{_T}^\beta(n)},{\pi\over2})\Big]\bigg)
\bigg\}F_1(x_{_h},x_{d_i})
\nonumber\\
&&\hspace{2.2cm}
+{4s_{_{\rm w}}^2Q_f\over e^2}\sum\limits_{i=1}^3\sum\limits_{\alpha=1}^3
\sum\limits_{\beta=1}^3\sum\limits_{\gamma=1}^3\bigg\{
-\Big(\delta_{b,d_i}{m_b^2\over\Lambda_{_{KK}}^2}
+\delta_{s,d_i}{m_s^2\over\Lambda_{_{KK}}^2}\Big)
\nonumber\\
&&\hspace{2.2cm}\times
\Big[f_{_{(++)}}^{L,c_{_B}^\alpha}(0,{\pi\over2})f_{_{(++)}}^{L,c_{_B}^\gamma}(0,{\pi\over2})\Big]
\Big({\cal D}_L^{(0)}\Big)_{s,\alpha}^\dagger\Big(\lambda^d\Big)^\dagger_{\alpha\beta}
\Big(\lambda^d\Big)_{\beta\gamma}\Big({\cal D}_L^{(0)}\Big)_{\gamma,b}
\nonumber\\
&&\hspace{2.2cm}\times
\sum\limits_{n=1}^\infty\bigg({1\over\Big[y_{_{(\pm\mp)}}^{c_{_T}^\beta(n)}\Big]^2}
\Big[f_{_{(-+)}}^{R,c_{_T}^\beta}(y_{_{(\pm\mp)}}^{c_{_T}^\beta(n)},{\pi\over2})
f_{_{(-+)}}^{R,c_{_T}^\beta}(y_{_{(\pm\mp)}}^{c_{_T}^\beta(n)},{\pi\over2})\Big]
\nonumber\\
&&\hspace{2.2cm}
+{1\over\Big[y_{_{(\mp\mp)}}^{c_{_T}^\beta(n)}\Big]^2}
\Big[f_{_{(++)}}^{R,c_{_T}^\beta}(y_{_{(\mp\mp)}}^{c_{_T}^\beta(n)},{\pi\over2})
f_{_{(++)}}^{R,c_{_T}^\beta}(y_{_{(\mp\mp)}}^{c_{_T}^\beta(n)},{\pi\over2})\Big]\bigg)
\nonumber\\
&&\hspace{2.2cm}
-\Big(\delta_{b,d_i}{m_bm_s\over\Lambda_{_{KK}}^2}
+\delta_{s,d_i}{m_s^3\over\Lambda_{_{KK}}^2m_b}\Big)\Big[f_{_{(++)}}^{R,c_{_T}^\alpha}(0,{\pi\over2})
f_{_{(++)}}^{R,c_{_T}^\gamma}(0,{\pi\over2})\Big]\Big({\cal D}_R^{(0)}\Big)_{s,\alpha}^\dagger
\Big(\lambda^d\Big)_{\alpha\beta}^\dagger
\nonumber\\
&&\hspace{2.2cm}\times
\Big(\lambda^d\Big)_{\beta\gamma}\Big({\cal D}_R^{(0)}\Big)_{\gamma,b}
\sum\limits_{n=1}^\infty{1\over\Big[y_{_{(\pm\pm)}}^{c_{_B}^\beta(n)}\Big]^2}
\Big[f_{_{(++)}}^{L,c_{_B}^\beta}(y_{_{(\pm\pm)}}^{c_{_B}^\beta(n)},{\pi\over2})
f_{_{(++)}}^{L,c_{_B}^\beta}(y_{_{(\pm\pm)}}^{c_{_B}^\beta(n)},{\pi\over2})\Big]\bigg\}
\nonumber\\
&&\hspace{2.2cm}\times
F_2(x_{_h},x_{d_i})
\nonumber\\
&&\hspace{2.2cm}
+{2s_{_{\rm w}}^2m_{_{\rm w}}^2Q_f\over e^2\Lambda_{_{KK}}^2}\sum\limits_{\alpha=1}^3
\sum\limits_{\beta=1}^3\sum\limits_{\gamma=1}^3\Big[f_{_{(++)}}^{R,c_{_T}^\alpha}(0,{\pi\over2})
f_{_{(++)}}^{R,c_{_T}^\gamma}(0,{\pi\over2})\Big]\Big({\cal D}_R^{(0)}\Big)_{s,\alpha}^\dagger
\Big(\lambda^d\Big)_{\alpha\beta}^\dagger
\nonumber\\
&&\hspace{2.2cm}\times
\Big(\lambda^d\Big)_{\beta\gamma}\Big({\cal D}_R^{(0)}\Big)_{\gamma,b}
\sum\limits_{n=1}^\infty\Big[f_{_{(++)}}^{L,c_{_B}^\beta}(y_{_{(\pm\pm)}}^{c_{_B}^\beta(n)},{\pi\over2})
f_{_{(++)}}^{L,c_{_B}^\beta}(y_{_{(\pm\pm)}}^{c_{_B}^\beta(n)},{\pi\over2})\Big]
\nonumber\\
&&\hspace{2.2cm}\times
F_1({m_{_{\rm w}}^2\over\Lambda_{_{KK}}^2}x_{_h},\Big[y_{_{(\pm\pm)}}^{c_{_B}^\beta(n)}
\Big]^2)+{\cal O}(\upsilon^4/\Lambda_{_{KK}}^4)
\nonumber\\
&&\hspace{1.6cm}=
-{Q_fm_bm_s\over m_{_{\rm w}}^2\Lambda_{_{KK}}^2}\sum\limits_{i=1}^3
\sum\limits_{\beta=1}^3\bigg\{\Big(\delta_{b,d_i}m_b^2+\delta_{s,d_i}m_s^2\Big)
\Big({\cal D}_L^{(0)}\Big)_{s,\beta}^\dagger\Theta_1(c_{_B}^\beta)
\Big({\cal D}_L^{(0)}\Big)_{\beta,b}
\nonumber\\
&&\hspace{2.2cm}
+m_bm_s\Big(\delta_{b,d_i}+\delta_{s,d_i}\Big)
\Big({\cal D}_R^{(0)}\Big)_{s,\beta}^\dagger\Big[\Theta_2(c_{_T}^\beta)
+\Theta_1(-c_{_T}^\beta)\Big]
\Big({\cal D}_R^{(0)}\Big)_{\beta,b}\bigg\}
\nonumber\\
&&\hspace{2.2cm}\times
F_1(x_{_h},x_{d_i})
\nonumber\\
&&\hspace{2.2cm}
-{Q_fm_bm_s\over m_{_{\rm w}}^2\Lambda_{_{KK}}^2}\sum\limits_{i=1}^3\bigg\{
m_bm_s\Big(\delta_{b,d_i}+\delta_{s,d_i}\Big)
\Big({\cal D}_L^{(0)}\Big)_{s,\beta}^\dagger\Theta_1(c_{_B}^\beta)
\Big({\cal D}_L^{(0)}\Big)_{\beta,b}
\nonumber\\
&&\hspace{2.2cm}
+\Big(\delta_{b,d_i}m_b^2+\delta_{s,d_i}m_s^2\Big)
\Big({\cal D}_R^{(0)}\Big)_{s,\beta}^\dagger\Big[\Theta_2(c_{_T}^\beta)
+\Theta_1(-c_{_T}^\beta)\Big]\Big({\cal D}_R^{(0)}\Big)_{\beta,b}\bigg\}
\nonumber\\
&&\hspace{2.2cm}\times
F_2(x_{_h},x_{d_i})+{\cal O}(\upsilon^4/\Lambda_{_{KK}}^4)\;,
\label{mass4}
\end{eqnarray}
Here, ${\cal D}_{L,R}^{(0)}$ are the rotations of zero modes of charge $-1/3$
quarks from chirality eigenvectors to mass eigenvectors in the absence
of the mixing between zero modes and exciting KK modes of quarks.
The nonnegative functions are defined as
\begin{eqnarray}
&&\Theta_1(c)={(1-2c)(1-\epsilon^{3-2c})\over(3-2c)(1+2c)(1-\epsilon^{1-2c})}
+{1\over1-2c}-{1+\epsilon^{1-2c}\over(1-2c)(1+2c)}\;
\nonumber\\
&&\Theta_2(c)=
{1-\epsilon^{3+2c}\over2(3+2c)(1-\epsilon^{1+2c})}
+{\epsilon^2-\epsilon^{1+2c}\over2(1-2c)(1-\epsilon^{1+2c})}
\nonumber\\
&&\hspace{1.6cm}
+{1\over2(1+2c)}+{\epsilon^2\over2(1-2c)}-{\epsilon^{1+2c}\over(1-2c)(1+2c)}\;.
\label{mass5}
\end{eqnarray}
Analogously, we find that the corrections from neutral Higgs to the Wilson coefficients
$C_{8G}(\mu_{_{\rm EW}})$, $C_{7\gamma}^{N}(\mu_{_{\rm EW}})$ and $C_{8G}^N(\mu_{_{\rm EW}})$
all contain the suppression factor $m_b^3m_s/m_{_{\rm w}}^4$
comparing with the corrections from other sectors. The detailed analysis on the radiative
corrections to $b\rightarrow s\gamma$, $(g-2)_\mu$ etc. are presented elsewhere\cite{Feng}.

\section{Summary\label{sec7}}
\indent\indent
In this work, we verify that the eigenvalues of KK excitations in warped extra dimension
are real, and are symmetrically distributed contrasting to the origin in complex plane.
We also present the sufficient condition to judge if the infinite series of KK excitations
is convergent. Applying the residue theorem, we sum over the infinitely series
of KK modes, and analyze the possible relation between summation of the product of
KK mode propagator with the corresponding bulk profiles in four dimensional
effective theory and the propagator of field in five dimensional full theory.
Additional, we also sum over the infinitely series of KK
modes for the gauge boson satisfying the $(++)$ BCs, and recover the results
in literature which are obtained through the equation of motion and completeness
relation of all KK modes. We extend this method to sum over the KK modes
in unified extra dimension, and obtain the equation applied extensively in literature.
As an example, we demonstrate that the radiative correction from the
penguin diagram composed by neutral Higgs and charge $-1/3$ quarks
contains the suppression factor $m_b^3m_s/m_{_{\rm w}}^4$
comparing with the corrections from other sectors.

\begin{acknowledgments}
\indent\indent
The work has been supported by the National Natural Science Foundation of China (NNSFC)
with Grant No. 10975027.
\end{acknowledgments}


\begin{thebibliography}{99}
\bibitem{RS}L.~Randall and R.~Sundrum, Phys.~Rev.~Lett.~{\bf 83}(1999)3370;
{\it ibid.}~{\bf 83}(1999)4690.
\bibitem{Chang}S.~Chang, J.~Hisano, H.~Nakano, N.~Okada and M.~Yamaguchi,
Phys.~Rev.~D{\bf 62}(2000)084025.
\bibitem{Csaki}C.~Csaki, J.~Hubisz, and P.~Meade, hep-ph/0510275.
\bibitem{Gherghetta1}T.~Gherghetta, hep-ph/0601213.
\bibitem{Grossman}Y.~Grossman and M.~Neubert, Phys.~Lett.~B{\bf 474}(2000)361.
\bibitem{Gherghetta2}T.~Gherghetta and A.~Pomarol, Nucl.~Phys.~B{\bf 586}(2000)141.
\bibitem{Huber1}S.~J.~Huber, Nucl.~Phys.~B{\bf 666}(2003)269.
\bibitem{Agashe1}K.~Agashe, G.~Perez, and A.~Soni, Phys.~Rev.~D{\bf 71}(2005)016002.
\bibitem{Agashe2}K.~Agashe, A.~Delgado, M.~J.~May, And R.~Sundrum,
JHEP{\bf 0308}(2003)050.
\bibitem{Csaki2}C.~Csaki, C.~Grojean, L.~Pilo, and J.~Terning,
Phys.~Rev.~Lett.~{\bf 92}(2004)101802.
\bibitem{Agashe3}K.~Agashe, R.~Contino, and A.~Pomarol, Nucl.~Phys.~B{\bf 719}(2005)165.
\bibitem{Csaki3}G.~Cacciapaglia, C.~Csaki, G.~Marandella, and J.~Terning,
Phys.~Rev.~D{\bf 75}(2007)015003.
\bibitem{Contino}R.~Contino, L.~Da~Rold, and A.~Pomarol, Phys.~Rev.~D{\bf 75}(2007)055014.
\bibitem{Carena}M.~S.~Carena, E.~Ponton, J.~Santiago, and C.~E.~M.~Wagner,
Nucl.~Phys.~B{\bf 759}(2006)202.
\bibitem{Agashe4}K.~Agashe, A.~Delgado, and R.~Sundrum, Ann.~Phys.~{\bf 304}(2003)145.
\bibitem{Agashe5}K.~Agashe, R.~Contino, and R.~Sundrum, Phys.~Rev.~Lett.~{\bf 95}(2005)171804.
\bibitem{Agashe6}K.~Agashe, R.~Contino, Nucl.~Phys.~B{\bf 742}(2006)59.
\bibitem{Casagrande}S.~Casagrande, F.~Goertz, U.~Haisch, M.~Neubert and T.~Prof,
JHEP{\bf 0810}(2008)094.
\bibitem{Bauer1}M.~Bauer, S.~Casagrande, L.~Grunder, U.~Haisch, M.~Neubert,
Phys.~Rev.~D{\bf 79}(2009)076001.
\bibitem{Bauer2}M.~Bauer, S.~Casagrande, U.~Haisch, M.~Neubert, arXiv:0912.1625[hep-ph].
\bibitem{Agashe7}K.~Agashe, R.~Contino, L.~Da~Rold, and A.~Pomarol,
Phys.~Lett.~B{\bf 641}(2006)62.
\bibitem{Santiago1}M.~S.~Carena, E.~Ponton, J.~Santiago, and C.~E.~M.~Wagner,
Phys.~Rev.~D{\bf 76}(2007)035006.
\bibitem{Djouadi}A.~Djouadi, G.~Moreau, and F.~Richard, Nucl.~Phys.~B{\bf 773}(2007)43.
\bibitem{Bouchart}C.~Bouchart and G.~Moreau, Nucl.~Phys.~B{\bf 810}(2009)66.
\bibitem{Csaki4}C.~Csaki, A.~Falkowski, and A.~Weiler, JHEP{\bf 0809}(2008)008.
\bibitem{Buras1}M.~Blanke, A.~J.~Buras, B.~Duling, S.~Gori, and A.~Weiler,
JHEP{\bf 0903}(2009)001.
\bibitem{Agashe8}K.~Agashe, A.~E.~Blechman, and F.~Petriello,
Phys.~Rev.~D{\bf 74}(2006)053011.
\bibitem{Davidson}S.~Davidson, G.~Isidori, and S.~Uhlig, Phys.~Lett.~B{\bf 663}(2008)73.
\bibitem{Iltan}E.~O.~Iltan, Eur.~Phys.~J.~C{\bf 54}(2008)583.
\bibitem{Agashe9}K.~Agashe, A.~Azatov, and L.~Zhu, Phys.~Rev.~D{\bf 79}(2009)056006.
\bibitem{Cacciapaglia}G.~Cacciapaglia, C.~Csaki, J.~Galloway, G.~Marandella,
J.~Terning and A.~Weiler, JHEP{\bf 0804}(2008)006.
\bibitem{Santiago2}J.~Santiago, JHEP{\bf 0812}(2008)046.
\bibitem{Chen1}M.-C.~Chen, and H.-B.~Yu, Phys.~Lett.~B{\bf 672}(2009)253.
\bibitem{Csaki5}C.~Csaki, A.~Falkowski, and A.~Weiler, Phys.~Rev.~D{\bf 80}(2009)016001.
\bibitem{Csaki6}C.~Csaki, C.~Delaunay, C.~Grojean, and Y.~Grossman, JHEP{\bf 0810}(2008)055.
\bibitem{Kadosh}A.~Kadosh and E.~Pallante, arXiv:1004.0321[hep-ph].
\bibitem{Perez1}A.~L.~Fitzpatrick, G.~Perez, and L.~Randall, arXiv:0710.1869.
\bibitem{Perez2}G.~Perez, and L.~Randall, JHEP{\bf 0901}(2009)077.
\bibitem{Falkowski1}A.~Falkowski, and M.~P\'erez-Victoria, JHEP{\bf 0812}(2008)107.
\bibitem{Gherghetta3}B.~Batell, T.~Gherghetta, and D.~Sword,
Phys.~Rev.~D{\bf 78}(2008)116011.
\bibitem{Delgado}A.~Delgado, and D.~Diego, Phys.~Rev.~D{\bf 80}(2009)024030.
\bibitem{Santiago3}S.~M.~Aybat, and J.~Santiago, Phys.~Rev.~D{\bf 80}(2009)035005.
\bibitem{Gherghetta4}T.~Gherghetta, and D.~Sword, Phys.~Rev.~D{\bf 80}(2009)065015.
\bibitem{Cabrer1}J.~A.~Cabrer, G.~Gersdorff, and M.~Quiros, arXiv:0907.5361[hep-ph].
\bibitem{Huber2}M.~Atkins, and S.~J.~Huber, arXiv:1002.5044[hep-ph].
\bibitem{Buras2}M.~Blanke, A.~J.~Buras, B.~Duling, S.~Gori, and A.~ Weiler, JHEP{\bf 0903}(2009)001.
\bibitem{Buras3}M.~Blanke, A.~J.~Buras, B.~Duling, K.~Gemmler, and S.~Gori, JHEP{\bf 0903}(2009)108.
\bibitem{Buras4}A.~J.~Buras, B.~Duling, and S.~Gori, JHEP{\bf 0909}(2009)076.
\bibitem{Azatov}A.~Azatov, M.~Toharia, L.~Zhu, Phys.~Rev.~D{\bf 80}(2009)035016.
\bibitem{Agashe10}K.~Agashe, H.~Davoudiasl, G.~Perez, and A.~Soni, Phys.~Rev.~D{\bf 76}(2007)036006.
\bibitem{Agashe11}K.~Agashe, H.~Davoudiasl, S.~Gopalakrisshna, T.~Han, G.~Huang,
G.~Perez, Z.~Si, and A.~Soni, Phys.~Rev.~D{\bf 76}(2007)115015;
K.~Agashe, S.~Gopalakrisshna, T.~Han, G.~Huang, and A.~Soni,
Phys.~Rev.~D{\bf 80}(2009)075007;
K.~Agashe, A.~Belyaev, T.~Krupovnickas, G.~Perez, and J.~Virzi,
Phys.~Rev.~D{\bf 77}(2008)015003.
\bibitem{Csaki7}C.~Csaki, Y.~Grossman, P.~Tanedo and Y.~Tsai, arXiv: 1004.2037[hep-ph].
\bibitem{Carena1}M.~S.~Carena, A.~Delgado, E.~Ponton, T.~M.~Tait, and
C.~E.~M.~Wagner,  Phys.~Rev.~D{\bf 71}(2005)015010.
\bibitem{Hirn}J.~Hirn and V.~Sanz, Phys.~Rev.~D{\bf 76}(2007)044022.
\bibitem{Azatov}A.~Azatov, M.~Toharia, L.~Zhu, Phys.~Rev.~D{\bf 82}(2010)056004.
\bibitem{Agashe12}K.~Agashe, G.~Perez, and A.~Soni, Phys.~Rev.~Lett.~{\bf 93}(2004)201804.
\bibitem{Buras5}M.~E.~Albrecht, M.~Blanke, A.~J.~Buras, B.~Duling, and K.~Gemmler,
JHEP{\bf 0909}(2009)064.
\bibitem{Buras6}G.~Buchalla, A.~J.~Buras, M.~E.~Lautenbacher,
Rev.~Mod.~Phys{\bf 68}(1996)1125.
\bibitem{Randall1}L.~Randall, M.~D.~Schwartz, JHEP{\bf 0111}(2001)003;
Phys.~Rev.~Lett.~{\bf 88}(2002)081801; R.~Contino, A.~Pomarol, JHEP{\bf 0411}(2004)058.
\bibitem{textbook}John~B.~Conway, {\it Functions of one complex variable},
Springer-Verlag, 1978.
\bibitem{Wangzx}Zhu-Xi~Wang and Dun-Ren~Guo, {\it An introduction to special functions},
Science Press, Beijing, China 1965 (in Chinese).
\bibitem{Arkani-Hamed}N.~Arkani-Hamed, H.~Cheng, B.~A.~Dobrescu, L.~J.~Hall,
Phys.~Rev.~D{\bf 62}(2000)096006.
\bibitem{Barbieri}R.~Barbieri, L.~J.~Hall, Y.~Nomura, Phys.~Rev.~D{\bf 63}(2001)105007.
\bibitem{ACD}T.~Appeiquist, H.-C.~Cheng, B.~A.~Dobrescu, Phys.~Rev.~D{\bf 64}(2001)035002.
\bibitem{Buras7}A.~J.Buras, M.~Spranger, A.~Weiler,  Nucl.~Phys.~B{\bf 660}(2003)225.
\bibitem{Santiago4}F.~A.~Aguila, J.~Santiago, M.~P\'erez-Victoria, JHEP{\bf 0009}(2000)011;
F.~A.~Aguila, M.~P\'erez-Victoria, J.~Santiago, Phys.~Lett.~B{\bf 492}(2000)98.
\bibitem{Feng}Tai-Fu Feng {\it et.al}, in preparation.
\end{thebibliography}
\end{document}